\documentclass[useAMS,usenatbib]{mnras}

\usepackage{epsfig}
\usepackage{aas_macros}
\usepackage{natbib}
\usepackage{times}
\usepackage{graphicx, bm, amssymb, xcolor, colortbl}
\usepackage{verbatim}
\usepackage[all]{hypcap}
\usepackage{xspace}
\usepackage{multirow}
\usepackage{pgf,tikz}
\usepackage{mathrsfs}
\usetikzlibrary{arrows}
\usepackage{amsmath}
\usepackage{subfigure}
\usepackage{array}

\usepackage{environ}
\makeatletter
\newsavebox{\measure@tikzpicture}
\NewEnviron{scaletikzpicturetowidth}[1]{%
  \def\tikz@width{#1}%
  \begin{lrbox}{\measure@tikzpicture}%
  \BODY
  \end{lrbox}%
  \pgfmathparse{#1/\wd\measure@tikzpicture}%
  \BODY
}
\makeatother

\usepackage{etoolbox}
\makeatletter
\patchcmd\@combinedblfloats{\box\@outputbox}{\unvbox\@outputbox}{}{%
    \errmessage{\noexpand\@combinedblfloats could not be patched}%
}%
\makeatother

\newcommand{\msun}{\mbox{M$_\odot$}}

\newcommand{\pc}{\mbox{${\rm pc}$}}

\newcommand{\be}{\begin{equation}}
\newcommand{\ee}{\end{equation}}
\newcommand{\bea}{\begin{eqnarray}}
\newcommand{\eea}{\end{eqnarray}}
\newcommand{\dd}{\mathrm{d}}

\title{The role of galactic dynamics in shaping the physical properties of giant molecular clouds in Milky Way-like galaxies}

\author[]{Sarah~M.~R.~Jeffreson\thanks{s.jeffreson@uni-heidelberg.de}$^{1}$,  J.~M.~Diederik~Kruijssen$^{1}$, Benjamin~W.~Keller$^{1}$,
\newauthor
M{\'e}lanie~Chevance$^{1}$ and Simon~C.~O.~Glover$^{2}$ \\
\\
$^{1}$ Astronomisches Rechen-Institut, Zentrum f\"{u}r Astronomie der Universit\"{a}t Heidelberg, M\"{o}nchhofstra\ss e 12-14, 69120 Heidelberg, Germany \\
$^{2}$ Institut f\"{u}r Theoretische Astrophysik, Zentrum f\"{u}r Astronomie der Universit\"{a}t Heidelberg, Albert-Ueberle-Str.~2, 69120 Heidelberg, Germany \\
}

\begin{document}

\date{Accepted X{\sevensize xxxx} XX. Received 2020 July 1; in original form 2020 May 8.}

\pagerange{\pageref{firstpage}--\pageref{lastpage}} \pubyear{2020}

\maketitle

\label{firstpage}

\begin{abstract}
We examine the role of the large-scale galactic-dynamical environment in setting the properties of giant molecular clouds in Milky Way-like galaxies. We perform three high-resolution simulations of Milky Way-like discs with the moving-mesh hydrodynamics code {\sc Arepo}, yielding a statistical sample of $\sim 80,000$ giant molecular clouds and $\sim 55,000$ HI clouds. We account for the self-gravity of the gas, momentum and thermal energy injection from supernovae and HII regions, mass injection from stellar winds, and the non-equilibrium chemistry of hydrogen, carbon and oxygen. By varying the external gravitational potential, we probe galactic-dynamical environments spanning an order of magnitude in the orbital angular velocity, gravitational stability, mid-plane pressure and the gradient of the galactic rotation curve. The simulated molecular clouds are highly overdense ($\sim 100 \times$) and over-pressured ($\sim 25 \times$) relative to the ambient interstellar medium. Their gravo-turbulent and star-forming properties are decoupled from the dynamics of the galactic mid-plane, so that the kpc-scale star formation rate surface density is related only to the number of molecular clouds per unit area of the galactic mid-plane. Despite this, the clouds display clear, statistically-significant correlations of their rotational properties with the rates of galactic shearing and gravitational free-fall. We find that galactic rotation and gravitational instability can influence their elongation, angular momenta, and tangential velocity dispersions. The lower pressures and densities of the HI clouds allow for a greater range of significant dynamical correlations, mirroring the rotational properties of the molecular clouds, while also displaying a coupling of their gravitational and turbulent properties to the galactic-dynamical environment.
\end{abstract}


\section{Introduction}
\label{Sec::Introduction}
Within the hierarchical structure of the interstellar medium, giant molecular clouds (GMCs) correspond to the spatial scales and densities at which the vast majority of star formation occurs~\citep[e.g.][]{Kennicutt+Evans12}. The physics that drive the formation, evolution and destruction of molecular clouds are therefore the physics that control galaxy-scale observables such as the star formation relation of~\cite{Kennicutt98}. In particular, the large scatter in the observed gas depletion times of nearby galaxies on sub-kpc scales~\citep{Bigiel08,Leroy08,Blanc09,Liu11,Rahman+11,Schruba10,Schruba11,Rahman+12,Leroy+13} indicates that star formation is not controlled exclusively by the quantity of molecular gas available on kpc-scales, but is governed to a great extent by cloud-scale ($\sim 5$-$200$~pc) processes~\citep[e.g.][]{Feldmann2011,Calzetti12,Kruijssen2014,Kruijssen18a}.

The set of possible physical mechanisms governing the properties of GMCs is wide and varied~\citep[e.g.][]{ChevanceReview20}. Large-scale dynamical processes such as galactic shear~\citep{Luna06,Leroy08,Suwannajak14,Colombo18}, interactions with spiral arms~\citep{Koda09,Meidt13}, and gravitational instability on the Toomre scale~\citep{Freeman17,Marchuk18} are observed to vary the star-forming properties of the ISM on cloud scales. Epicyclic motions driven by the galactic rotation curve~\citep{Meidt18,Meidt20,Kruijssen2019,Utreras20}, accretion flows from galactic scales down to GMC scales~\citep{KlessenHennebelle10}, and collisions between clouds~\citep{Tan00,TaskerTan09,Tasker11} can drive turbulence in GMCs and so explain their large non-thermal line-widths~\citep{Fukui01,Engargiola03,RosolowskyBlitz05}. Stellar feedback processes such as supernovae~\citep{AvillezBreitschwerdt05,Joung09,WalchNaab15,Martizzi15,IffrigHennebelle15,KimCG&Ostriker15a,KimCG&Ostriker15b,Padoan16,Padoan17}, photoionisation by massive stars~\citep[e.g.][]{Matzner+02,Krumholz&Matzner09,Gritschneder09,Walch12,KimJG&Ostriker18}, and stellar winds~\citep{Haid2016,Haid2018,Rahner2017} are seen to destroy GMCs in simulations, and to inject turbulence into the wider ISM, providing support against gravitational collapse on large scales~\citep{Ostriker+10,OstrikerShetty2011}. These theoretical results are borne out in observations, which show a correlation between the mid-plane hydrostatic pressures of galaxies and their star formation rates~\citep{Sun2020}, and which host clouds that are typically destroyed within a dynamical time, by the feedback from massive stars~\citep{Kruijssen2019,Chevance20,ChevanceReview20}. Cloud collapse can also be slowed~\citep{Tassis04,Mouschovias06} or its onset delayed~\citep{Girichidis18,Hennebelle&Inutsuka19} by the presence of magnetic fields, which are observed to penetrate into star-forming regions and can provide pressure support~\citep[see the review by][]{Crutcher12}. While simulations by~\cite{Su16} show that magnetic fields have a much smaller effect on the galactic-averaged SFR than does stellar feedback,~\cite{Nixon18} argue against the implication that magnetic fields are irrelevant in GMC evolution, instead explaining this result in terms of local magnetic dissipation in the highly star-forming regions within clouds.

With the recent advent of large interferometric instruments such as the Atacama Large Millimeter/Submillimeter Array (ALMA), it has become possible to resolve cloud-scale observables outside the Milky Way. These signatures of cloud-scale physics can now be retrieved across the local galaxy population~\citep{Schinnerer13,Elmegreen17,Faesi18}, permitting a statistical sample of GMC properties across a wide range of galactic environments. Measurements of the star formation efficiency per free-fall time~\citep{Utomo18}, as well as the GMC turbulent velocity dispersion, turbulent pressure, surface density and spatial extent~\citep{Leroy17b,Sun18,Sun2020} have already revealed a strong correlation with the galaxy-scale properties. A systematic variation in the dense gas fraction across galaxy discs has similarly been observed by~\cite{Usero15,Bigiel16}, and is correlated with the molecular gas surface density, the stellar surface density, and the dynamical equilibrium pressure~\citep{Gallagher18}. Within the Milky Way, a large scatter in the `size-linewidth' relation of~\cite{Larson1981} is observed between the Central Molecular Zone~\citep{Oka+01,Shetty12,KruijssenLongmore13,Kauffmann17}, the outer disc~\citep{Heyer+01}, and the collective GMC population of the entire Galaxy~\citep{Heyer+09,Roman-Duval+10,Rice+16,Miville-Deschenes17,Colombo+19}.

Combined with these observations, a systematic theoretical study of observable cloud properties across different galactic environments is necessary to discern the dominant physical mechanisms controlling the GMC lifecycle. Given that each distinct galactic environment hosts a unique set of galactic-dynamical processes~\citep[e.g.][]{Jeffreson+Kruijssen18}, we conduct in this work a systematic examination of the correlations between the average physical properties of GMCs and the galactic-dynamical time-scales of their host galaxies. We use a spatially-resolved, statistical sample of $\sim 80,000$ GMCs, drawn from three numerical simulations of Milky Way-pressured disc galaxies that we perform using the moving-mesh hydrodynamics code {\sc Arepo}~\citep{Springel10}. In this context, `Milky Way-pressured' refers to the fact that our simulated galaxies have a Milky Way-like division of mass between the galactic disc, bulge and halo, as well as between stars and atomic and molecular gas, leading to Milky Way-like values of the mid-plane pressure. To obtain further insight into the origin of each dynamical correlation, we also examine the sample of $\sim 55,000$ HI clouds across the three galaxies, in addition to the GMCs. The HI clouds represent a level up in the hierarchical structure of the ISM, i.e.~they comprise the atomic gas out of which GMCs condense on time-scales of around $30$~Myr in Milky Way-like environments~\citep{Larson94,Goldsmith07,Ward20}.

The remainder of the paper is structured as follows. In Section~\ref{Sec::numerical-methods}, we describe the three isolated galaxy simulations used in this work to explore the properties of GMCs across different galactic-dynamical environments. We explain the set of numerical methods that we have employed and the basic analysis methods that we have used to identify GMCs and HI clouds. In Section~\ref{Sec::galaxy-properties} we compare our simulations to key observable properties of Milky Way-like galaxies and their GMCs from the literature, demonstrating an acceptable level of agreement. Section~\ref{Sec::Theory} reviews the analytic theory of~\cite{Jeffreson+Kruijssen18} for GMC evolution under the influence of galactic dynamics, and maps the simulation data into the environmental parameter space spanned by the theory to reveal galactic-dynamical trends in the properties of our simulated GMCs. Our results and their implications are explored in Section~\ref{Sec::results-properties}. In Section~\ref{Sec::discussion}, we discuss our results in the context of existing ISM models and simulations of GMCs, and additionally examine the caveats of our simulations. Finally, we present a summary of our conclusions in Section~\ref{Sec::conclusion}.

\section{Simulations} \label{Sec::numerical-methods}
We consider three simulated galaxy discs, spanning a range of galactic-dynamical environments at Milky Way ISM mid-plane pressures. The simulations are set up as isolated gaseous discs in an external gravitational potential that models the dark matter halo, the stellar disc, and the stellar bulge. Subsequent star formation produces live stellar particles.

\begin{table*}
\begin{center}
\label{Tab::params}
  \caption{Physical parameters for the disc galaxies modelled in this work, including the values of the analytic constants for the background potentials in Section~\ref{Sec::ext-ptnl}. Properties of the gas disc ($M_{\rm gas}$, $R_{\rm gas}$ and $z_{\rm gas}$) are quoted at the fiducial simulation time of $\sim 600$~Myr, at which point molecular cloud properties are measured. All masses are given in units of $10^{10} {\rm M}_\odot$, while all length-scales are given in units of kpc. The columns report: (1) Halo mass, (2) halo scale-radius, (3) core cut-off radius, (4) bulge mass, (5) bulge turnover radius, (6) stellar disc mass, (7) stellar disc scale-radius, (8) stellar disc scale-height, (9) final gas disc mass, (10) gas disc scale-radius, (11) gas disc scale-height.}
  \begin{tabular}{@{}l c c c c c c c c c c c@{}}
  \hline
   Sim. & $M_{\rm h}$ & $a_{\rm h}$ & $a_{\rm c}$ & $M_{\rm b}$ & $a_{\rm b}$ & $M_{\rm d}$ & $a_{\rm d}$ & $b_{\rm d}$ & $M_{\rm gas}$ & $R_{\rm gas}$ & $z_{\rm gas}$ \\
    & (1) & (2) & (3) & (4) & (5) & (6) & (7) & (8) & (9) & (10) & (11) \\
  \hline
  \hline
   FLAT & 116 & 47 & - & 1.5 & 0.4 & 3.5 & 2.6 & 0.4 & 5.8 & 7.4 & 0.38 \\
   SLOPED & 130 & 49 & - & 0.5 & 2 & 3.5 & 5.4 & 0.3 & 5.9 & 7.7 & 0.28 \\
   CORED & 150 & 51.5 & 5 & - & - & 3.5 & 6 & 1 & 6 & 7.4 & 0.25 \\
  \hline
\end{tabular}
\end{center}
\end{table*}

\subsection{Initial conditions} \label{Sec::ICs}
The initial conditions for each galaxy are generated using {\sc MakeNewDisk}~\citep{Springel2005} with $9 \times 10^6$ gas cells each. The velocity of each cell centroid is determined by its acceleration due to the external potential described in Section~\ref{Sec::ext-ptnl}, with all parameters given in Table~\ref{Tab::params}. The density distribution of the gas disc follows an exponential profile of the form
\begin{equation}
\label{Eqn::gas-disc}
\rho_{\rm gas} (R,z) = \frac{M_{\rm gas}}{4\pi R_{\rm gas} z_{\rm gas}} \exp{\Big(-\frac{R}{R_{\rm gas}}\Big)} \exp{\Big(-\frac{|z|}{z_{\rm gas}}\Big)},
\end{equation}
where $R$ is the galactocentric radius, $z$ is the height above the galactic mid-plane, and $M_{\rm gas}$ is the total gas mass of the disc. The disc scale-length $R_{\rm gas}$ is fully-determined by the external potential and the disc scale-height $z_{\rm gas}$ is set by the condition of hydrostatic equilibrium.

To each initial condition, we add a background grid of side-length $500$~kpc, composed of $32$ cells per side. The large size of this box ensures no interaction between the simulation boundaries and the gas cells in the isolated disc galaxy. We set an upper limit of $1.25 \times 10^5$~kpc$^3$ on the gas cell volume for each simulation, limiting the size of the background grid cells under the adaptive mesh refinement scheme described in Section~\ref{Sec::adaptive-mesh-refinement}.

Finally, we `warm up' the initial condition for $500$~Myr by injecting kinetic and thermal energy into all cells above a hydrogen number density threshold of $100$~${\rm cm}^{-3}$. We use the supernova feedback prescription outlined in Section~\ref{Sec::feedback}, but circumvent the creation of star particles to inject this feedback instantaneously into the relevant gas cells. This period of evolution allows for the dispersal of resonant rings formed within the galactic mid-plane, as the gas cells refine and re-distribute within the potential well. It produces a flocculent spiral-arm structure, and so significantly reduces the run-time required for the simulation to settle into a state of dynamical equilibrium.

\subsection{External potential} \label{Sec::ext-ptnl}
\begin{figure}
  \label{Fig::rotcurve-components}
    \includegraphics[width=\linewidth]{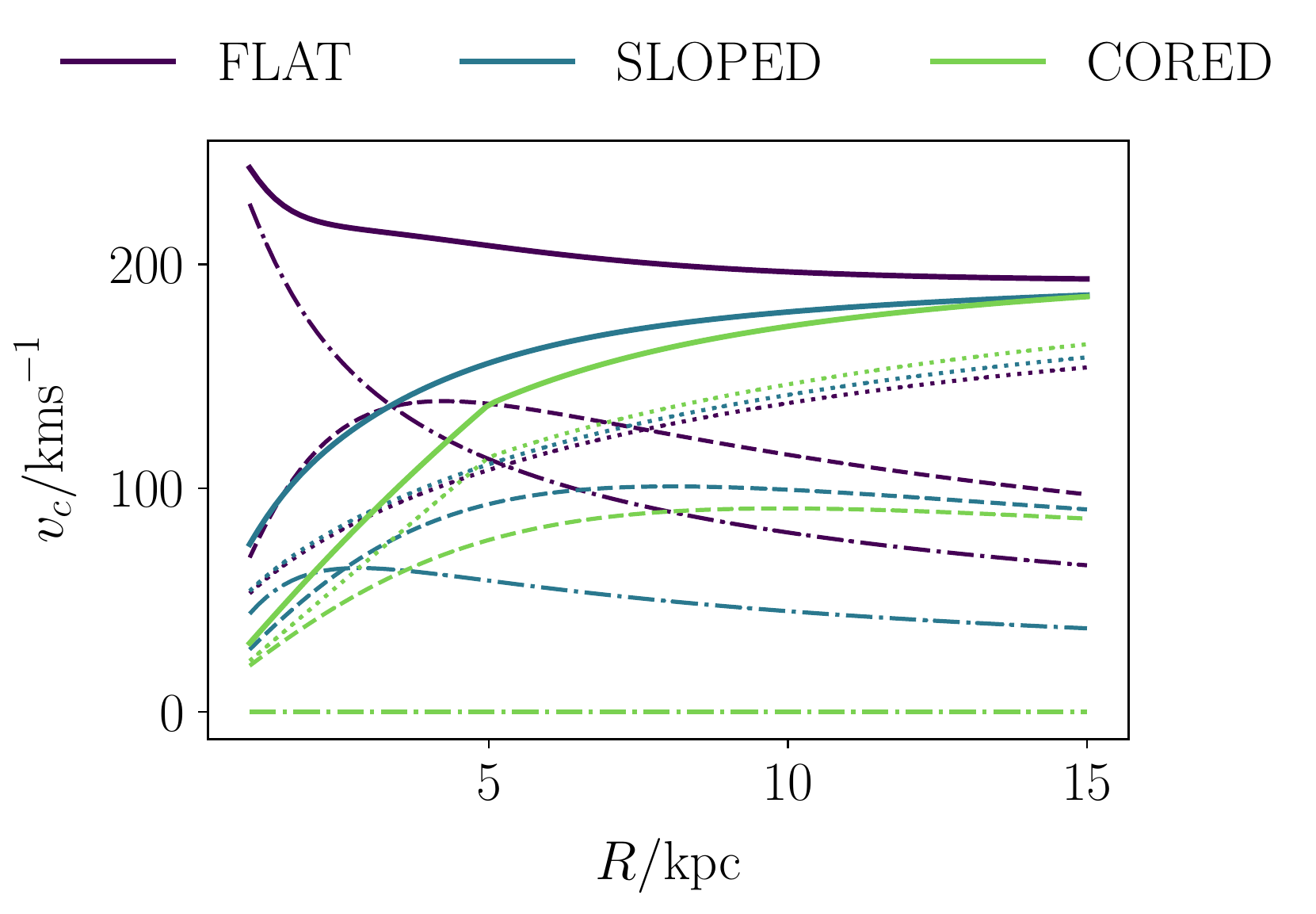}
    \caption{Contribution of the disc (dashed lines, Equation~\ref{Eqn::d-rotcurve}), the bulge (dash-dotted lines, Equation~\ref{Eqn::b-rotcurve}) and the halo (dotted lines, Equation~\ref{Eqn::h-rotcurve}) components of the analytic external gravitational potential to the galactic circular velocity (bold lines) of each simulated galaxy.}
\end{figure}

We consider two different classes of external potential, which span values of the galactic shear parameter $\beta$ from the high-shear case of $\beta = 0$ for a flat rotation curve up to $\beta = 1$ for solid-body rotation. The FLAT $(\beta \sim 0)$ and SLOPED $(\beta \la 0.5)$ initial conditions follow a Milky Way-like external potential consisting of a stellar bulge, a (thick) stellar disc and a cusped dark matter halo~\citep[e.g.][]{Bland-Hawthorn&Gerhard16}. In this initial study we do not consider the influence of the Galactic bar, which is likely to create a higher-pressure environment with higher levels of star formation than the Galactic disc~\citep[e.g.][]{Sun2020}. The CORED $(0 \la \beta \la 1)$ initial condition follows an M33-like potential profile with a stellar disc, a cored dark matter halo, and no stellar bulge~\cite[e.g.][]{Corbelli03}. All analytic parameters are presented in Table~\ref{Tab::params}, and have been chosen to achieve a maximum rotational velocity within the gas disc of $\sim 220$ kms$^{-1}$. The contribution of the halo, bulge and disc components to the galactic circular velocity of each simulation is shown as a function of the galactocentric radius in Figure~\ref{Fig::rotcurve-components}.

\subsubsection{Stellar disc}
The stellar disc component is modelled using a~\cite{Miyamoto&Nagai75} potential of the form
\begin{equation}
\label{Eqn::d-ptnl}
\Phi_{\rm d} = \frac{G M_{\rm d}}{\sqrt{R^2 + [a_{\rm d} + (z^2 + b_{\rm d}^2)^{1/2}]^2}},
\end{equation}
where $G$ is the gravitational constant, $R$ is the galactocentric radius within the galactic mid-plane and $z$ is the perpendicular distance from this plane, such that $r = \sqrt{R^2 + z^2}$ for any distance $r$ away from the disc centre. The parameters $M_{\rm d}$, $a_{\rm d}$ and $b_{\rm d}$ are the mass, scale-length and scale-height, respectively, of the stellar disc. The corresponding rotation curve is given by
\begin{equation}
\label{Eqn::d-rotcurve}
v_{\rm c, d}(R) = \frac{\sqrt{GM_{\rm d}}R}{([a_{\rm d} + b_{\rm d}]^2 + R^2)^{3/4}}.
\end{equation}
Since $a_{\rm d}$ and $b_{\rm d}$ both fall within the gas disc, the stellar disc potential contributes a solid-body ($v_{\rm c} \propto R$) component to the rotation curve for very small galactocentric radii $R \ll a_{\rm d} + b_{\rm d}$, which turns over at large radii to follow the Hernquist profile ($v_{\rm c} \propto \sqrt{R}$) for $R \gg a_{\rm d} + b_{\rm d}$. For each disc, this rotation profile is given by the dashed lines in Figure~\ref{Fig::rotcurve-components}.

\subsubsection{Stellar bulge}
We model the stellar bulge component in the FLAT and SLOPED simulations using a~\cite{Plummer1911} potential of the form
\begin{equation}
\label{Eqn::b-ptnl}
\Phi_{\rm b}(r) = -\frac{GM_{\rm b}}{\sqrt{r^2 + a_{\rm b}^2}},
\end{equation}
where $M_{\rm b}$ is the mass of the bulge and $a_{\rm b}$ is the turnover radius of the density core. This potential gives a rotation curve of
\begin{equation}
\label{Eqn::b-rotcurve}
v_{\rm c, b} = \frac{\sqrt{GM_{\rm b}} R}{(R^2 + a_{\rm b}^2)^{3/4}}
\end{equation}
in the galactic plane, which is identical in form to the stellar disc rotation curve, but with its peak at the bulge turnover radius $a_{\rm b}$. These profiles correspond to the dash-dotted lines in Figure~\ref{Fig::rotcurve-components}.

\subsubsection{Spherical dark matter halo}
The cuspy dark matter halo for the FLAT and SLOPED simulations follows a~\cite{Hernquist90} potential of the form
\begin{equation}
\label{Eqn::h-ptnl}
\Phi_{\rm h}(r) = -\frac{GM_{\rm h}}{r+a_{\rm h}},
\end{equation}
where $M_{\rm h}$ is the halo mass and $a_{\rm h}$ is the halo scale radius. The corresponding rotation profile in the galactic plane is given by
\begin{equation}
\label{Eqn::h-rotcurve}
v_{\rm c, h}(R) = \sqrt{R\frac{\dd \Phi_{\rm h}}{\dd R}} = \frac{\sqrt{GM_{\rm h}R}}{R+a_{\rm h}},
\end{equation}
which peaks at $R \sim a_{\rm h}$, far outside the star-forming disc. The dark matter halo therefore contributes a component $v_{\rm c} \propto \sqrt{R}$ to the rotation curve within the star-forming disc, corresponding to a shear parameter of $\beta = 1/2$. These profiles are given by the purple (FLAT) and blue (SLOPED) dotted lines in Figure~\ref{Fig::rotcurve-components}.

The dark matter potential for the CORED simulation combines the above Hernquist profile with a uniform-density core in a piece-wise fashion, such that
\begin{equation}
\Phi_{\rm h, core}(r) = 
\begin{cases} 
      \frac{G M_{\rm h} R^2}{2 a_{\rm c} (a_{\rm c} + a_{\rm h})^2} & 0 \leq r \leq a_{\rm c} \\
      \Phi_{\rm h}(r) & a_{\rm c} \leq r \\
   \end{cases},
\end{equation}
where $a_{\rm c}$ is the cut-off radius of the core. The resulting rotation curve is solid-body for $R < a_{\rm c}$ and follows Equation~(\ref{Eqn::h-rotcurve}) for $R \geq a_{\rm c}$. It is shown as the green dotted line in Figure~\ref{Fig::rotcurve-components}.

\subsection{Adaptive mesh refinement} \label{Sec::adaptive-mesh-refinement}
\begin{figure}
  \label{Fig::cell-dstbns}
  \includegraphics[width=\linewidth]{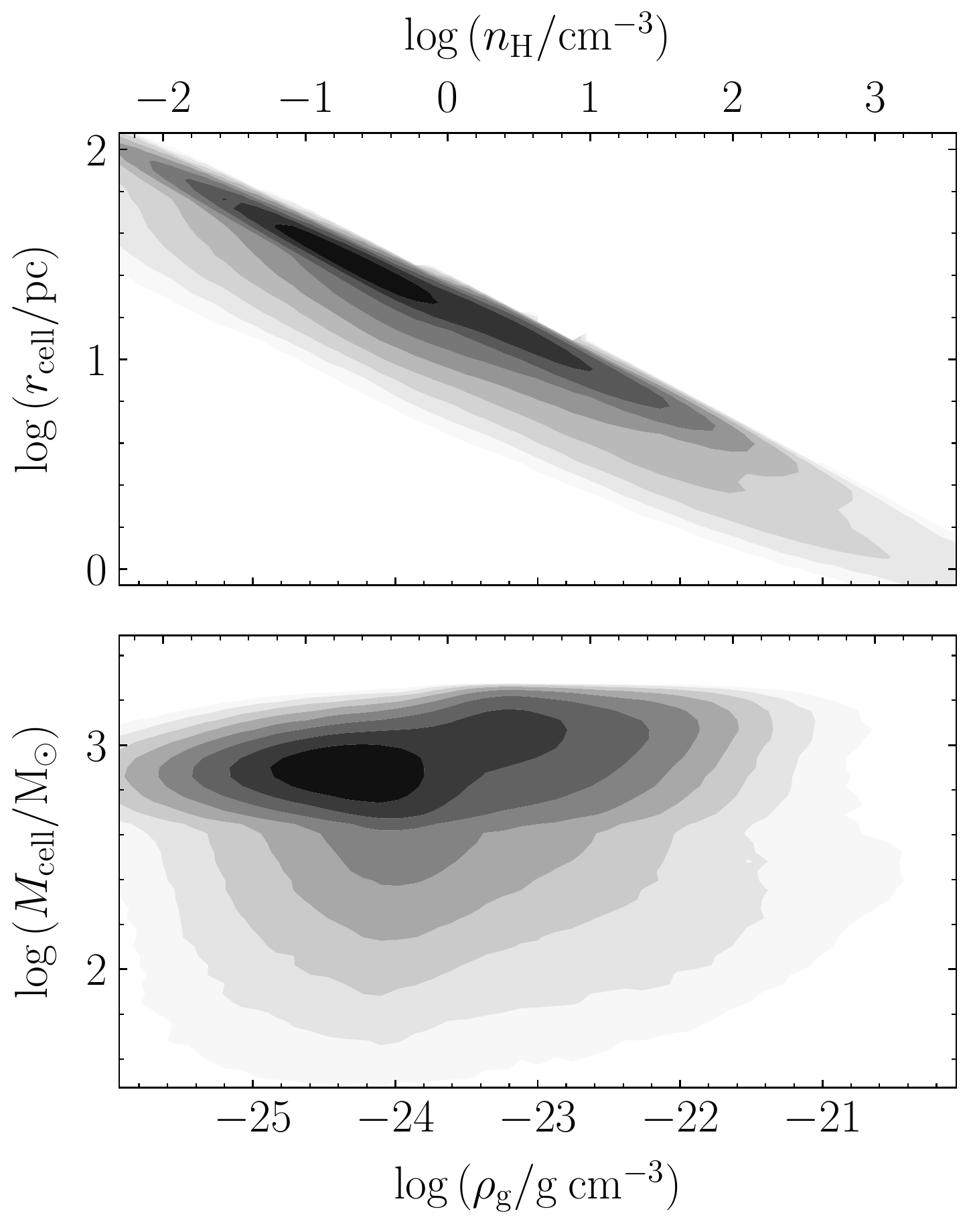}
  \caption{Distribution of gas cell radii $r_{\rm cell}$ (top) and gas cell masses $M_{\rm cell}$ (bottom) as a function of mass density for the FLAT simulation after $600$~Myr of evolution. The cells are not perfectly-spherical, so $r_{\rm cell}$ corresponds to the radius of a sphere with the same volume. The colours are logarithmically-related to the total number of gas cells, at a spacing of $0.3$~dex.}
\end{figure}
The adaptive refinement and de-refinement of gas cells in {\sc Arepo} is determined according to the mass and density aggregated at each grid point. In contrast to Eulerian codes, the mesh moves along with the gas flow, reducing the number of gas cells that must be refined and de-refined during each time-step~\citep{Springel10}. As such, we need only to set a `target' mass resolution for the Voronoi cells, corresponding to the mode of the distribution of cell masses. We use a value of $900$~M$_\odot$, so that the spatial resolution of each simulation extends down to cell diameters of $\sim 3$~pc at our star formation threshold of $n_{\rm thresh} = 2000$~${\rm cm}^{-3}$ (see Section~\ref{Sec::starformation}). The distribution of cell masses and sizes is shown in Figure~\ref{Fig::cell-dstbns}. We do not impose the non-thermal pressure floor that is often used to prevent gravitational fragmentation in regions for which the Jeans length $\lambda_{\rm J} = (\pi c_s^2/G\rho)^{1/2}$ is unresolved. According to the criterion of~\cite{Truelove97}, such fragmentation is a numerical artefact that can be avoided if $\lambda_J$ is sampled by four or more gas cells at all times. However, this would require us to inflate the Jeans length inside our simulated molecular clouds to an unphysical value of $\lambda_J \sim 10$~pc, preventing the physical (but unresolved) gravitational fragmentation required to attain densities close to our star formation threshold, as discussed by~\cite{Teyssier15,Hopkins18}. Instead, we fulfil the three requirements tested by~\cite{Nelson06} for thin isolated discs using Lagrangian methods: namely that (1) the Toomre mass is resolved, (2) the scale-height of the disc is resolved, and (3) fully-adaptive gravitational softening is used up to density threshold for star formation. The first requirement can be formulated as a maximum resolvable surface density $\Sigma_{\rm max}$, given in Equation (11) of~\cite{Nelson06} as $\Sigma_{\rm max} = (\pi/G) \cdot (c_s^4/m_{\rm H} N_{\rm reso})$, where $c_s$ is the sound speed, $m_{\rm H}$ is the proton mass and $N_{\rm reso}$ is the SPH neighbour number. For use with {\sc Arepo}, we set $N_{\rm reso} = 9$ for the linear stencil used to reconstruct the hydrodynamical gradients (the central cell plus eight adjacent cells). Using the average sound speed in our simulations\footnote{Both the sound speed and the turbulent velocity dispersion in the highest-density (molecular) gas are much lower than is the average ISM sound speed. However, the Toomre length in this cold gas is also much longer, ballooning out to kpc-scales for $n_{\rm H} \sim 100$~${\rm cm}^{-3}$. Even in the densest gas, the average ISM sound speed is therefore the appropriate quantity to use in our calculation of $\Sigma_{\rm max}$.} ($c_s \sim 7$~${\rm kms}^{-1}$), we obtain $\Sigma_{\rm max} \sim 3000$~${\rm M}_\odot \: {\rm pc}^{-2}$: larger than the maximum surface density of $\sim 1000$~${\rm M}_\odot \: {\rm pc}^{-2}$ attained in our galaxies. The second requirement is manifestly fulfilled for our gas disc scale-heights of several hundreds of parsec (see Table~\ref{Tab::params}). To fulfil the third requirement, we employ the adaptive gravitational softening scheme in {\sc Arepo} with a softening length of $1.5$ times the cell diameter and a minimum value of $3$~pc to match the spatial resolution at $n_{\rm thresh} = 2000$~${\rm cm}^{-3}$.

\subsection{Star formation prescription} \label{Sec::starformation}
We follow a simple prescription for the star formation rate volume density that reproduces the observed relationship between the star formation rate surface density and the gas surface density~\citep{Kennicutt98}. For a gas cell $i$ with volume density $\rho_i$, the volume density of the star formation rate is computed as
\begin{equation}
\label{Eqn::starformation}
\frac{\dd \rho_{*,i}}{\dd t} = 
\begin{cases}
      \frac{\epsilon \rho_i}{t_{{\rm ff},i}}, \; n_i \geq n_{\rm thresh} \\
      0, \; n_i < n_{\rm thresh}\\
   \end{cases},
\end{equation}
where $t_{{\rm ff},i} = \sqrt{3\pi/(32G \rho_i)}$ is the local free-fall time, $n_i$ is the local hydrogen number density, and $n_{\rm thresh} = 2000$~cm$^{-3}$ is the threshold above which star formation is allowed to occur. The star formation threshold is chosen to ensure that the star-forming gas in our simulations is Jeans-unstable, provided that the gas temperature does not exceed $100$~K, a constraint that is satisfied by all of the dense gas in our simulation (see Section~\ref{Sec::galaxy-properties}). We assign a star formation efficiency per free-fall time of $\epsilon = 0.01$ in accordance with observations of the gas depletion time in nearby galaxies~\citep{Leroy17b,Krumholz&Tan07,Utomo18,Krumholz18}. In practice, Equation~(\ref{Eqn::starformation}) is fulfilled on average for a large number of gas cells by stochastically generating star particles from the set of cells with $n > n_{\rm thresh}$, at a probability of $P_i = 1 - \exp{(-\dd \rho_{*,i}/\dd t \cdot \Delta t/\rho_i)}$. Gas cells with masses larger than twice the simulation mass resolution ($900$~M$_\odot$ here) `spawn' a star particle of mass equal to the mass resolution, and the gas cell mass is reduced by the corresponding amount. Smaller gas cells are deleted entirely and replaced by star particles of an equal mass. In both cases, the velocity of the new star particle is equal to the velocity of the parent gas cell. As for the gas particles, we use a gravitational force softening of $3$~pc for the star particles in our simulations.

One significant concern with the prescription outlined above, which relies solely on a gas density threshold to determine where stars form, is that it may lead to star formation in gas that is not gravitationally-bound. This may occur in gas flows with a high Mach numbers, in which the gas is Jeans unstable with $|U_{\rm grav}| > U_{\rm therm}$ but the ram pressure is high, such that $|U_{\rm grav}| < U_{\rm kin} = 1/2 \: m v^2$. This is a common occurrence in radiative shocks, where the material is cool and at high density, but still has sufficient kinetic energy to prevent it from undergoing gravitational collapse~\citep[see e.g.~the discussion in][]{Federrath10,Gensior20}. To check whether this is a problem for the gas in our simulations, we have computed the total energy $U_{\rm grav} + U_{\rm therm} + U_{\rm kin} < 0$ for the gas cells in our simulations that fall above the star formation threshold, $n > n_{\rm thresh}$. The gravitational potential for each gas cell of mass $M_{\rm cell}$ and size $r_{\rm cell}$ is defined at its edge, such that $U_{\rm grav}=-G M_{\rm cell}^2/r_{\rm cell}$. We find that only $\la 0.04$~per~cent of these cells are not self-gravitating at a simulation time of $600$~Myr.

\subsection{Stochastic stellar population synthesis} \label{Sec::stochastic-stars}
We synthesise a stellar population for every star particle in our simulations using the Stochastically Lighting Up Galaxies ({\sc SLUG}) model~\citep{daSilva12,daSilva14,Krumholz15}. Here we briefly describe the methods used within {\rm SLUG} to track the evolution of each stellar population, but refer to the reader to the cited works for a complete and detailed explanation. The stellar population for a star particle of birth mass $M_{\rm birth}$ is formed via the Monte-Carlo sampling of $N$ stars from a~\cite{Chabrier03} initial stellar mass function (IMF). The integer $N$ is chosen by drawing from the Poisson distribution with an expectation value of $M_{\rm birth}/\overline{M}$, where $\overline{M}$ is the expectation value for the mass of a single star. Averaged over a large number of assignments, this procedure ensures that the assigned masses of the stellar populations converge to the birth masses of the star particles. Each stellar population evolves as a function of the simulation time according to Padova solar metallicity tracks~\citep{Fagotto94a,Fagotto94b,VazquezLeitherer05} with Starburst99-like spectral synthesis~\citep{Leitherer99}. As such, SLUG provides the number of supernovae $N_{\rm SN}$, the ejected mass $\Delta m$ and the ionising luminosity $S$ for each star particle at every time-step, all of which are used in our numerical prescription for stellar feedback. By basing our feedback on the stochastic sampling of the IMF, we avoid arbitrary (but important) choices regarding the time interval over and delay with which stellar feedback acts, which have a qualitative effect on the structure of the ISM \citep{Keller20}.

\subsection{Stellar feedback} \label{Sec::feedback}
Here we describe in detail the numerical methods used to inject stellar feedback from supernovae, stellar winds and HII regions into the simulated ISM. We provide convergence tests for each of the components of our stellar feedback model in a separate paper,~\cite{Jeffreson20b}.

\subsubsection{Supernovae and stellar winds} \label{Sec::SN-and-winds}
For each star particle $i$ in our simulations, we use SLUG to calculate the mass $\Delta m_i$ lost during each numerical time-step, along with the number of supernovae $N_{i, {\rm SN}}$ that have occurred. If $N_{i, {\rm SN}} = 0$, then we assume that the mass loss results from stellar winds, and deposit the mass into the star's nearest-neighbour (NN) gas cell. We do not account for the thermal energy and momentum injected by stellar winds, and we discuss the possible consequences of this in Section~\ref{Sec::caveats}. If $N_{i, {\rm SN}} > 0$, then we assume that all mass loss results from Type II supernovae, and we inject mass, energy and momentum according to the prescription described in~\cite{Keller19}. We give a brief overview of this algorithm below.
\begin{enumerate}
  \item For each star particle $i$, find the NN gas particle $j$.
  \item Determine the total mass $\Delta m_j$, momentum $\Delta \mathbf{p}_{j, {\rm SN}}$ and energy $\Delta E_{j, {\rm SN}}$ delivered by all of the star particles for which $j$ is the NN, such that
\begin{align}
\label{Eqn::centralcell_mass}
\Delta m_j &= \sum_i {\Delta m_i} \\
\label{Eqn::centralcell_mom}
\Delta \mathbf{p}_{j, {\rm SN}} & = \sum_i {\Delta \mathbf{p}_{i, {\rm SN}}} \\
\begin{split} \label{Eqn::centralcell_energy}
\Delta E_{j, {\rm SN}} &= \sum_i{\Delta E_{i, {\rm SN}}} + \sum_i{\frac{|\Delta \mathbf{p}_{i, {\rm SN}}|^2}{2\Delta m_i}} \\
-& \frac{|\sum_i \Delta \mathbf{p}_{i, {\rm SN}}|^2}{2\sum_i \Delta m_i}
\end{split}
\end{align}
where $\Delta \mathbf{p}_{i, {\rm SN}} = \Delta m_i \mathbf{v}_i$ and $\Delta E_{i, {\rm SN}} = N_{i, {\rm SN}} \times 10^{51} {\rm erg}$. The total energy $\Delta E_{j, {\rm SN}}$ received by gas cell $j$ via Equation~(\ref{Eqn::centralcell_energy}) is a combination of the blast-wave energies of the individual SN ejecta (first term on the LHS) and the energy dissipated in the inelastic collisions between these ejecta (second and third terms on the LHS).
  \item For each gas cell $j$ that has received feedback mass, momentum and energy, find the set of neighbouring gas cells $k$ with which it shares a Voronoi face. Compute the radial terminal momentum $p_{t, k}$ for the blast-wave as it passes through each cell $k$, using the (unclustered) parametrization of~\cite{Gentry17}, as
\begin{equation}
\label{Eqn::gentry-momentum}
\frac{p_{{\rm t}, k}}{{\rm M}_\odot {\rm kms}^{-1}} = 4.249 \times 10^5 \: \Big(\frac{w_k \Delta E_{j, {\rm SN}}}{10^{51} {\rm erg}}\Big) \Big(\frac{n_k}{{\rm cm}^{-3}}\Big)^{-0.06},
\end{equation}
where the weight factor $w_k$ is the fractional Voronoi face area shared between cells $j$ and $k$, such that
\begin{equation}
\label{Eqn::weight-fn}
w_k = \frac{A_{j \rightarrow k}}{\sum_k{A_{j \rightarrow k}}},
\end{equation}
ensuring isotropic energy injection. In the above, $n_k$ is the gas number density in cell $k$, and solar metallicity is assumed. Equation~(\ref{Eqn::gentry-momentum}) approximates the mechanical ($PdV$) work done by the blast-wave on the surrounding gas during the Sedov-Taylor (energy-conserving, momentum-generating) phase of its expansion. As we do not resolve this phase of the blast-wave expansion, the momentum given by Equation~(\ref{Eqn::gentry-momentum}) would not be retrieved by simply dumping the energy $w_k \Delta E_{j, {\rm SN}}$ into cell $k$ as thermal energy~\citep{Katz92,Slyz2005,Smith2018}.
  \item Using the terminal momentum $p_{t, k}$, calculate the final momentum $p_{k, {\rm new}}$ of cell $k$ following the energy-conserving procedure of~\cite{Hopkins18b}, as
\begin{equation} \label{Eqn::k-momentum}
\begin{split}
\mathbf{p}_{k, {\rm new}} &= \mathbf{p}_k + \Delta \mathbf{p}_{k, {\rm SN}} \\
\Delta \mathbf{p}_{k, {\rm SN}} &= p_{{\rm fb},k} \hat{\mathbf{r}}_{j \rightarrow k} + w_k \Delta \mathbf{p}_{j, {\rm SN}},
\end{split}
\end{equation}
where $p_{{\rm fb}, k}$ is the smallest of the terminal and energy-conserving momenta in cell $k$, such that
\begin{equation} \label{Eqn::p-fb}
p_{{\rm fb}, k} = \min{\{w_k p_{t, k}, \: \sqrt{2(m_k+w_k \Delta m_j) w_k \Delta E_{j, {\rm SN}}}\}}.
\end{equation}
  \item Calculate the final mass $m_{k, {\rm new}}$ and final energy $E_{k, {\rm new}}$ of the cell $k$, as
\begin{equation} \label{Eqn::k-mass}
\begin{split}
m_{k, {\rm new}} &= m_k + \Delta m_k \\
\Delta m_k &= w_k \Delta m_j,
\end{split}
\end{equation}
and
\begin{equation} \label{Eqn::k-energy}
\begin{split}
E_{k, {\rm new}} &= E_k + \Delta E_k \\
\Delta E_k &= \frac{|\mathbf{p}_k + \Delta \mathbf{p}_{k, {\rm SN}}|^2}{2(m_k + \Delta m_k)} - \frac{|\mathbf{p}_k|^2}{2m_k}.
\end{split}
\end{equation}
  \item Finally, ensuring linear momentum conservation to machine precision requires that the new momentum $\mathbf{p}_{j, {\rm new}}$ of the central cell $j$ is given by
\begin{equation}
\mathbf{p}_{j, {\rm new}} =  \mathbf{p}_j - \sum_k \Delta \mathbf{p}_{k, {\rm SN}}.
\end{equation}
Similarly, to ensure energy conservation to machine precision requires that the updated total energy $E_{j, {\rm new}}$ of the central cell $j$ be given by
\begin{equation}
\begin{split}
 \\
E_{j, {\rm new}} = E_j +  E_{j, {\rm SN}} - \sum_k \Big(\Delta E_{k, {\rm SN}} - \frac{w_k |\Delta \mathbf{p}_{j, {\rm SN}}|^2}{2\Delta m_j} \Big),
\end{split}
\end{equation}
where the final term accounts for the frame-change from the SN frame to the frame of gas cell $j$.
\end{enumerate}
In the above, we do not adjust the chemical state of the cells into which SN mass, momentum and thermal energy is injected. The evolution of chemistry, heating and cooling via {\sc SGChem} (see Section~\ref{Sec::chemistry}) occurs immediately after the injection of feedback, and deals with the change in the ionisation state of the gas cells caused by the injection of thermal energy. Aside from this, we do not evolve metal abundances in our simulations.

\begin{figure*}
  \label{Fig::fof-rsln-test}
  \includegraphics[width=\linewidth]{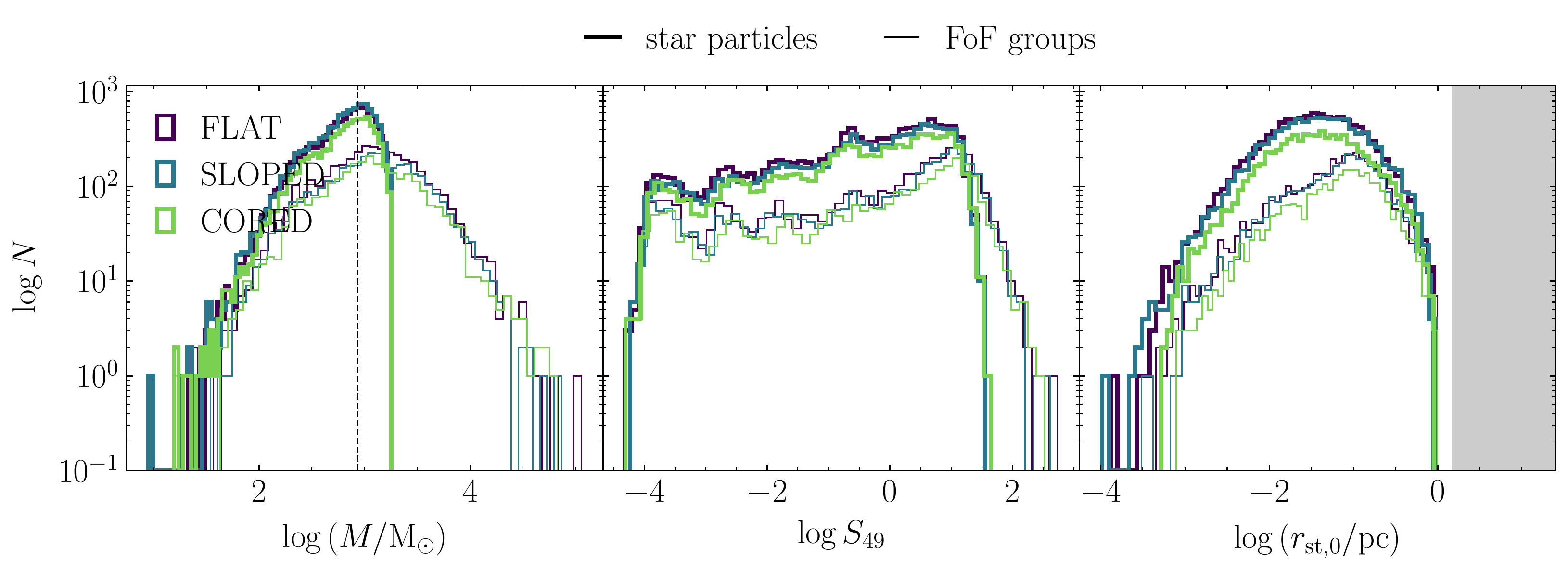}
  \caption{Distribution of cell masses (left), ionising luminosities (centre) and Str{\"o}mgren radii (right) for individual star particles (bold lines) and for FoF groups with overlapping Str{\"o}mgren spheres (thin lines). The vertical dashed line in the left-hand panel represents the target gas cell resolution of $900$~M$_\odot$. The grey shaded region in the right-hand panel represents the range $r_{\rm st, 0}>1.5$~pc of Str{\"o}mgren radii that would be resolved in our simulations. It is clear that all Str{\"o}mgren spheres are unresolved, even at the peak gas cell resolution.}
\end{figure*}

\subsubsection{HII region momentum} \label{Sec::HII-momentum}
We inject thermal and kinetic energy from HII region feedback according to the model of~\cite{Jeffreson20b}. The momentum provided by a hemispherical `blister-type' HII region to the surrounding ISM is given by the momentum of the thin bounding shell at the ionisation front, swept up in its initial rapid expansion to the Str{\"o}mgren radius~\citep{Matzner+02,Krumholz&Matzner09}. The momentum equation for the shell of an HII region with ionising luminosity $S$ and age $t$ can be solved to give a momentum per unit time of
\begin{equation} \label{Eqn::HII-momentum-eqn}
\begin{split}
\frac{\dd p}{\dd t} \sim 1.2 \times 10^3 &M_\odot \: {\rm kms}^{-1} {\rm Myr}^{-1} \times \\
&S_{49} \Big\{1+\Big[\frac{3}{2} \frac{t^2}{t_{\rm ch}^2} + \Big(\frac{25}{28} \frac{t^2}{t_{\rm ch}^2}\Big)^{6/5}\Big]^{1/6}\Big\},
\end{split}
\end{equation}
where $S_{49} = S/10^{49} {\rm s}^{-1}$. The characteristic time $t_{\rm ch}$ at which radiation pressure and gas pressure make equal contributions to the momentum delivered is given approximately by
\begin{equation}
\label{Eqn::t_ch}
t_{\rm ch} \sim 45 \: \overline{n}_{\rm H,2}^{1/6} S_{49}^{7/6} \: {\rm yr},
\end{equation}
where $\overline{n}_{\rm H,2} = n_{\rm H}/100 {\rm cm}^{-2}$ with $n_{\rm H}$ the birth number density of the star particle. The full derivations of Equations~(\ref{Eqn::HII-momentum-eqn}) and~(\ref{Eqn::t_ch}) are given in~\cite{Jeffreson20b}. Following~\cite{Krumholz&Matzner09}, the enhancement of the radiation pressure made by photon trapping (via stellar winds, infrared photons and Lyman-$\alpha$ photons) contributes a factor of $f_{\rm trap} \sim 2$ to the first term on the right-hand side of Equation~(\ref{Eqn::HII-momentum-eqn}), relative to the case of direct radiation pressure. We calculate the physical momentum delivered by each {\sc Arepo} star particle $i$ by grouping together all star particles that have overlapping ionisation front radii, given by
\begin{equation} \label{Eqn::i-front-radius}
r_{\rm II,i}(t) \sim 0.5 \times^{-2} S_{49,i} \: {\rm pc} \Big\{\frac{3}{2} \Big(\frac{t_i}{t_{\rm ch,i}}\Big)^2 + \Big[\frac{25}{28} \Big(\frac{t_i}{t_{\rm ch,i}}\Big)^2\Big]^{6/5}\Big\}^{1/3}.
\end{equation}
This ensures that the amount of momentum injected varies with the size of the physical HII regions in our simulations, and not with the masses of the individual star particles, which in turn depend on the simulation resolution. In practice, we form Friends-of-Friends (FoF) groups of star particles, where two particles are linked together if either falls within the ionisation front of the other. The FoF linking length is then given by $\max{(r_{\rm II,1}, r_{\rm II,2})}$ for star particles with ionisation fronts $r_{\rm II,1}$ and $r_{\rm II,2}$. The entire FoF group injects a momentum per unit time that is given by the sum over the group members, as
\begin{equation}
\begin{split}
\label{Eqn::momentumequation_fof}
\Big(\frac{\dd p}{\dd t} \Big)_{\rm FoF} &= \; 1.2 \times 10^3 \: {\rm M}_\odot \: {\rm km} \: {\rm s}^{-1} {\rm Myr}^{-1} \times \\
 &\sum_{i=1}^N{S_{49,i}} \left\{1 + \Big[\frac{3}{2} \Big(\frac{\langle t \rangle_S}{t_{\rm ch, FoF}}\Big)^2 + \Big(\frac{25}{28} \frac{\langle t \rangle_S}{t_{\rm ch,FoF}}\Big)^{6/5}\Big]^{1/6} \right\},
\end{split}
\end{equation}
where $\langle ... \rangle_S$ denotes the luminosity-weighted average over the star particles $i=1...N$ in the group, and the characteristic time is given by
\begin{equation}
\label{Eqn::t_ch_fof}
t_{\rm ch,FoF} = \sqrt{0.6 \: {\rm s}^2 \Big(\sum^N_{i=1}{S_{49,i}}\Big)^{7/3} \langle \overline{n}_{\rm H,2} \rangle_S^{1/3} \: {\rm pc}}.
\end{equation}
The group injects the momentum $\Delta p_{\rm FoF} = (\dd p/\dd t)_{\rm FoF} \Delta t$ at its luminosity-weighted centre, given by
\begin{equation}
\label{Eqn::centre_fof}
\langle \mathbf{x} \rangle_S = \frac{\sum_{i=1}^N{S_{49,i}\mathbf{x}_i}}{\sum_{i=1}^N{S_{49,i}}}.
\end{equation}
We ensure that the values of $(\dd p/\dd t)_{\rm FoF}$ and $\langle \mathbf{x} \rangle_S$ are consistent across all FoF group members at every time, by updating the FoF groups on global time-steps only. In~\cite{Jeffreson20b}, we argue that this procedure incurs a maximum positional error of $1$~pc on the star particle members that are included in a group: one third the size of the smallest Voronoi cell in our simulations. The numerical time-step $\Delta t$ of momentum injection between updates is set to the numerical time-step of an arbitrary group member. All star particles in each FoF group have comparable time-steps, which are determined according to their instantaneous accelerations.

The injection procedure for the HII region momentum is identical to that employed for supernovae, as described in Section~\ref{Sec::SN-and-winds}. The nearest-neighbour gas cell $j$ to the centre $\langle \mathbf{x} \rangle_S$ of the FoF group accumulates the net radial momenta of all the FoF groups that it hosts, then distributes the momentum to its facing neighbour cells $k$ according to
\begin{equation}
\label{Eqn::j-to-k-inj}
\Delta p_{k, {\rm HII}} = w_k(\theta_k, A_k) \hat{\mathbf{r}}_{j \rightarrow k} \Delta p_{j, {\rm HII}},
\end{equation}
where $\hat{\mathbf{r}}_{j \rightarrow k}$ is the unit vector connecting the centroid of cell $j$ to that of cell $k$, and the weight factor $w_k(\theta_k, A_k)$ is the fractional Voronoi face area shared between these cells, rescaled to account for the directionality of momentum injection from a blister-type HII region, such that
\begin{equation}
\label{Eqn::jet-profile}
\begin{split}
w(\theta_k, A_k) &= \frac{A_{j \rightarrow k} f(\theta_k)}{\sum_k{A_{j \rightarrow k} f(\theta_k)}} \\
f(\theta_k) &= \Big[\log{\Big(\frac{2}{\Theta}\Big)(1+\Theta^2-\cos^2{\theta_k})}\Big]^{-1}.
\end{split}
\end{equation}
Here, $\Theta=\pi/12$ controls the width of the directed momentum `beam' and $\theta_k$ is the angle between the beam-axis and the unit vector $\hat{\mathbf{r}}_{j \rightarrow k}$ connecting cells $j$ and $k$, defined by
\begin{equation}
\cos{\theta_k} = \frac{\hat{\mathbf{r}}_{j \rightarrow k} \cdot \hat{\mathbf{z}}_{\rm FoF}}{|\hat{\mathbf{z}}_{\rm FoF}|}.
\end{equation}
For each star particle, the vector $\hat{\mathbf{z}}_i$ defining the beam-axis is drawn randomly from a uniform distribution over the spherical polar angles about the star's position at birth, $\phi_i$ and $\theta_i$. This value is fixed throughout the lifetime of the HII region, and the beam-axis $\hat{\mathbf{z}}_{\rm FoF}$ of each FoF group is calculated as a luminosity-weighted average of $\hat{\mathbf{z}}_i$ across the constituent star particles.

\subsubsection{HII region heating} \label{Sec::HII-heating}
We inject enough thermal energy from each HII region to heat the gas inside the Str{\"o}mgren radius of each FoF group to a temperature of $7000$~K, in accordance with~\cite{Ho19}. We do this via an approximate photon-counting procedure that assumes all Str{\"o}mgren radii are either completely unresolved (smaller than the radius of a single Voronoi gas cell) or marginally-resolved (extending into the first layer of neighbouring gas cells). We demonstrate in Figure~\ref{Fig::fof-rsln-test} that this approximation holds for the Str{\"o}mgren radii in all three simulations. As such, we need only inject photons into the Voronoi gas cells that share a face with the nearest-neighbour cell of the FoF group, and so we use the same injection procedure as for the HII region momentum. We count the photons to be injected via a technique similar to that of~\cite{Hopkins18}, described below, and explained fully in~\cite{Jeffreson20b}.
\begin{enumerate}
  \item For each FoF group, find the NN (host) gas particle $j$ for the luminosity-weighted centre of mass.
  \item Increment the total number of photons $S_{\rm in}$ per unit time delivered to this gas cell by its enclosed set of $N$ group centres, so that the final value is given by $S_{j, {\rm in}} = \sum_{{\rm FoF} = 1}^N{S_{\rm FoF}}$, where $S_{\rm FoF}$ is the total ionising luminosity of all stars in the FoF group.
  \item Calculate the number of photons that can be consumed per unit time by ionising the material inside gas cell $j$, given by $S_{j, {\rm cons}} = \alpha_B N_{j, {\rm H}} n_{j, e}$, with $N_{j, {\rm H}}$ the number of hydrogen atoms in the cell and $n_{j, e}$ the number density of electrons.
  \item If $S_{j, {\rm in}} < S_{j, {\rm cons}}$, flag cell $j$ as `ionised' with a probability of $S_{j, {\rm in}}/S_{j, {\rm cons}}$. Over a large number of gas cells, the number of injected photons converges to $S_{j, {\rm in}}$.
  \item If $S_{j, {\rm in}} > S_{j, {\rm cons}}$, flag cell $j$ as `ionised' and calculate the residual ionisation rate $S_{j, {\rm res}} = S_{j, {\rm in}} - S_{j, {\rm cons}}$ that will now be spread over its set of facing Voronoi cells $k$.
\end{enumerate}
If the Str{\"o}mgren radius is completely unresolved, i.e.~$S_{j, {\rm in}} < S_{j, {\rm cons}}$, then the algorithm ends here. If it is marginally-resolved, then we continue as follows.
\begin{enumerate}
  \setcounter{enumi}{6}
  \item For each gas cell $j$ with $S_{j, {\rm res}} > 0$, find the set of neighbouring cells $k$ with which it shares a Voronoi face, which have already been identified for the purpose of injecting feedback from supernovae and momentum from HII regions (see Sections~\ref{Sec::SN-and-winds} and~\ref{Sec::HII-momentum}). Compute the fraction of momentum received by each of these cells $k$ according to
\begin{equation}
S_{k, {\rm in}} = w_k S_{j, {\rm res}},
\end{equation}
where the weight factor $w_k$ is identical to the weight factor used for the injection of HII region momentum.
  \item Ionise each facing cell $k$ with a probability of $S_{k, {\rm in}}/S_{k, {\rm cons}}$. Summed over the set of facing cells for many HII regions, this ensures that the expectation value for the rate of ionisation converges to $S_{j, {\rm res}}$.
\end{enumerate}
The injection of thermal energy via the procedure outlined above is immediately followed by the computation of chemistry and cooling for each gas cell using {\sc SGChem}, as described in Section~\ref{Sec::chemistry}. As such, we do not explicitly adjust the chemical state of the gas cells, relying instead on the chemical network to ionise the gas in accordance with the injection of heating. We set a temperature floor of $7000$~K during this computation.

\subsubsection{HII region stalling} \label{Sec::HII-stalling}
The momentum and thermal energy injected by each HII region in Sections~\ref{Sec::HII-momentum} and~\ref{Sec::HII-heating} is shut off when the rate of HII region expansion drops below the velocity dispersion of the host cloud, such that the ionised and neutral gas are able to mix and the expanding shell loses its coherence~\citep{Matzner+02}. Once this transition has occurred, the shell ceases to expand, and its radius and internal energy are no longer well-defined, such that it no longer transfers momentum to the surrounding gas. Equation~(\ref{Eqn::i-front-radius}) is no longer valid, and we ensure that such `stalled' HII regions are removed from the computation of FoF groups, so that they do not link together two active HII regions and distort the position of their centre of luminosity. In practice, the value of the ionising luminosity $S$ also falls steeply at this time, so that it is safe to ignore the thermal energy that is deposited after stalling has occurred. Prior to FoF group computation, we therefore calculate the numerical rate of HII region expansion $\dot{r}_{{\rm II},i}=\Delta r_{{\rm II}, i}/\Delta t_i$ for each star particle $i$, where $\Delta r_{{\rm II},i}$ is the increment in the ionisation front radius during the particle's time-step $\Delta t_i$. We compare this value to the cloud velocity dispersion $\sigma_{\rm cl}$, approximated according to~\cite{Krumholz&Matzner09} for a blister-type HII region centred at the origin of a cloud with an average density of $\overline{\rho}(r) = 3/(3-k_\rho) \rho_0 (r/r_0)^{-k_\rho}$. Assuming that the cloud is in approximate virial balance with $\alpha_{\rm vir} = 1$ on the scale of the HII region, this gives
\begin{equation}
\label{Eqn::sigma}
\begin{split}
\sigma_{\rm cl}(r_{\rm II}) &= \sqrt{\frac{\alpha_{\rm vir} G M(<r_{\rm II})}{5r_{\rm II}}} \\
&= \sqrt{\frac{2\pi}{15} \alpha_{\rm vir} G \overline{\rho}(r_{\rm st,0}) r_{\rm ch}^{2-k_\rho} r_{\rm st,0}^{k_\rho} x_{\rm II}^{2-k_\rho}},
\end{split}
\end{equation}
where we again take $k_{\rho} = 1$. If we find that $\dot{r}_{{\rm II},i} < \sigma_{\rm cl}$, then we flag the star particle as `stalled' and shut off its HII region feedback.

\subsection{ISM Chemistry, heating and cooling} \label{Sec::chemistry}
The chemical evolution of the gas in our simulations is tracked via a simplified set of reactions involving hydrogen, carbon and oxygen, according to the chemical network of~\cite{GloverMacLow07a,GloverMacLow07b} and~\cite{NelsonLanger97}. This chemical network is interfaced with {\sc Arepo} via the package called {\sc SGChem}, and will be referred to as such throughout this paper. The network follows the fractional abundances of ${\rm H}, {\rm H}_2, {\rm H}^+, {\rm He}, {\rm C}^+, {\rm CO}, {\rm O}$ and ${\rm e}^-$, which are related by the equalities
\begin{equation}
\begin{split}
&x_{\rm H} = 1 - x_{{\rm H}^+} - 2x_{{\rm H}_2} \\
&x_{\rm e^-} = x_{\rm H^+} + x_{\rm C^+} + x_{\rm Si^+} \\
&x_{\rm C^+} = \max(0, x_{\rm C, tot} - x_{\rm CO}) \\
&x_{\rm O} = \max(0, x_{\rm O, tot} - x_{\rm CO}), \\
\end{split}
\end{equation}
with the abundance of helium set to its standard cosmic value of $x_{\rm He} = 0.1$, and the abundances of silicon, carbon and oxygen set in accordance with~\cite{Sembach00}, to values consistent with the local warm neutral medium: $x_{\rm Si, tot} = 1.5 \times 10^{-5}$, $x_{\rm C, tot} = 1.4 \times 10^{-4}$ and $x_{\rm O, tot} = 3.2 \times 10^{-4}$. Silicon is assumed to be singly-ionised throughout the simulation, as is any carbon that is not locked up in CO molecules. The evolution of all chemical species is coupled to the heating, cooling, and dynamical evolution of the gas, via the atomic and molecular cooling function presented in~\cite{Glover10}. This includes chemical cooling due to fine structure-emission from C$^+$, O and Si$^+$, Lyman $\alpha$ emission from atomic hydrogen, ${\rm H}_2$ line emission, gas-grain cooling, and electron recombination on grain surfaces and in reaction with polycyclic aromatic hydrocarbons (PAHs). In hot gas, cooling may also occur via the collisional processes of ${\rm H}_2$ dissociation, Bremsstrahlung, the ionisation of atomic hydrogen, as well as via permitted and semi-forbidden transitions of metal atoms and ions. To treat the contribution from metals, we assume collisional ionization equilibrium and use the cooling rates tabulated by~\cite{GnatFerland12}. The dominant heating mechanism is photoelectric emission from dust grains and PAHs, with lesser contributions from cosmic ray ionisation and H$_2$ photodissociation. We assign a value of $1.7$~\cite{Habing68} units for the interstellar radiation field (ISRF) strength according to~\cite{Mathis83}, and a value of $2 \times 10^{-16}$~s$^{-1}$ to the cosmic ray ionisation rate~\citep[e.g.][]{Indriolo&McCall12}. The dust grain number density is computed by assuming the solar value for the dust-to-gas ratio, and the dust temperature is obtained according to the procedure described in Appendix A of~\cite{Glover&Clark12}. The full list of heating and cooling processes is given in Table 1 of~\cite{Glover10}.

\subsection{Thermal and chemical post-processing} \label{Sec::chem-postproc}
To calculate the mass fractions of ${\rm H}$ and ${\rm H}_2$ for each Voronoi cell in our simulations, we post-process each snapshot using the astrochemistry and radiative transfer model {\sc Despotic}~\citep{Krumholz13a}.\footnote{We use the non-equilibrium molecular fraction from the on-the-fly chemistry for cooling, but due to the limitations of our resolution, we cannot accurately compute the self-shielding of molecular hydrogen from the UV radiation field during run-time. In addition, because we do not resolve all of the dense substructures in the clouds that are created by the turbulent velocity field, we tend to underestimate the ${\rm H}_2$ formation rate within the clouds. This means that the on-the-fly ${\rm H}_2$ fractions are always too low by a factor of around $2$, and so we re-calculate an equilibrium molecular fraction in post-processing.} We follow the method used in~\cite{Fujimoto19} and treat each Voronoi gas cell as a separate one-zone, spherical `cloud' model, characterised by its hydrogen number density $n_{\rm H}$, column density $N_{\rm H}$, and its virial parameter $\alpha_{\rm vir}$. Within {\sc Despotic}, the line emission from each cloud is computed via the escape probability formalism, which is coupled self-consistently to the chemical and thermal evolution of the gas. The carbon and oxygen chemistry follows the chemical network of~\cite{Gong17}, modified by the addition of cosmic rays and the grain photoelectric effect, subject to dust- and self-shielding for each component, line cooling due to ${\rm C}^+$, ${\rm C}^+$, ${\rm O}$ and ${\rm CO}$, as well as thermal exchange between dust and gas. The ISRF strength and the cosmic ionisation rate are matched to the values used by the live chemistry during our simulations, and the rate of photoelectric heating is held fixed, both spatially and temporally. For each one-zone model, this system of coupled rate equations is converged to a state of chemical and thermal equilibrium.

\begin{figure}
  \label{Fig::DESPOTIC-threshold}
  \includegraphics[width=\linewidth]{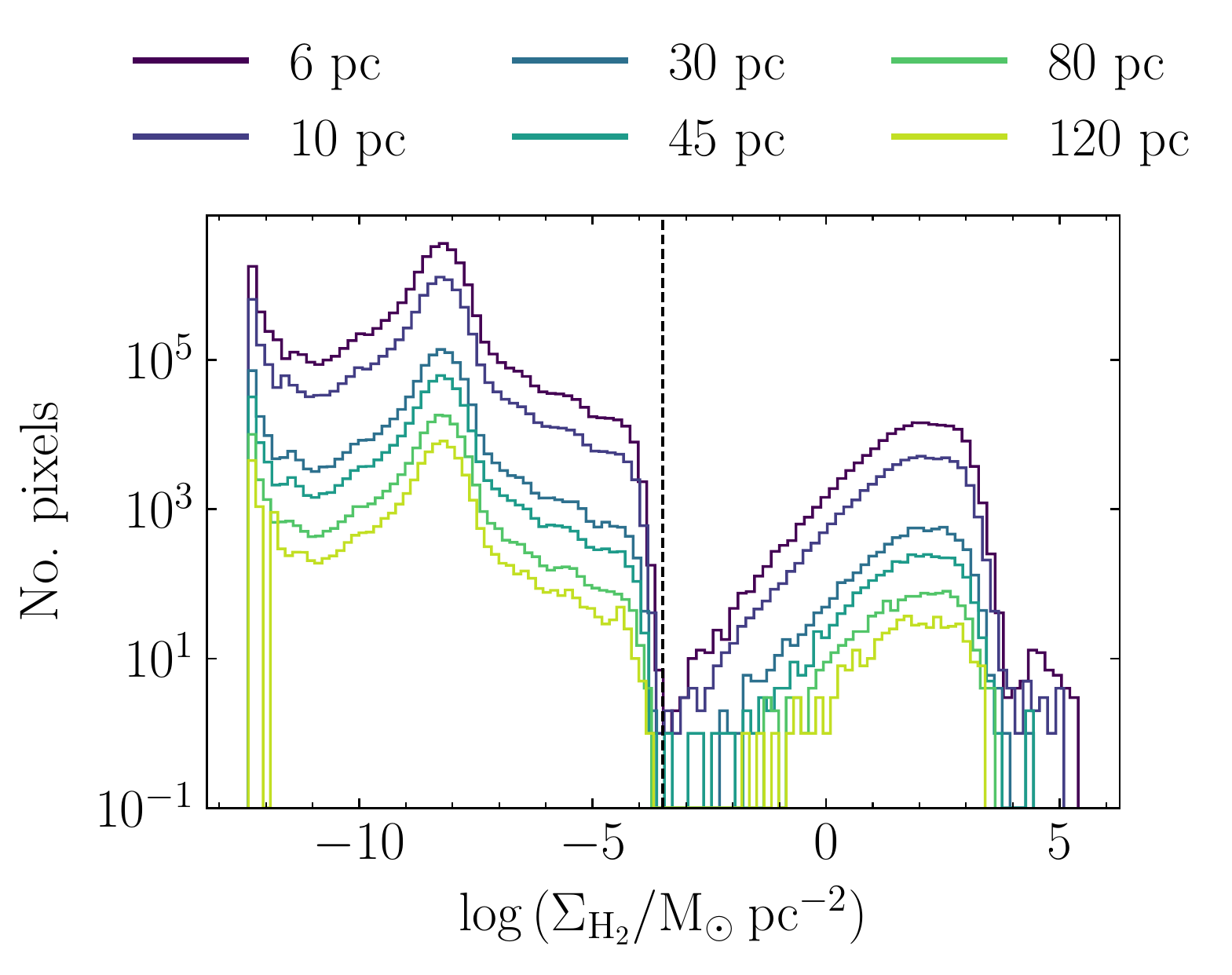}
  \caption{Distributions of the molecular hydrogen column density $\Sigma_{\rm H_2}$ for six different ray-tracing maps of the FLAT simulation at resolutions of $6$~pc, $10$~pc, $30$~pc, $45$~pc, $80$~pc and $120$~pc. When normalised by the total number of pixels, all six histograms are identical. The snapshot is taken at a simulation time of $600$~Myr. The vertical dashed line shows the cutoff used for cloud identification.}
\end{figure}

\begin{figure*}
  \label{Fig::cloud-ID}
  \includegraphics[width=\linewidth]{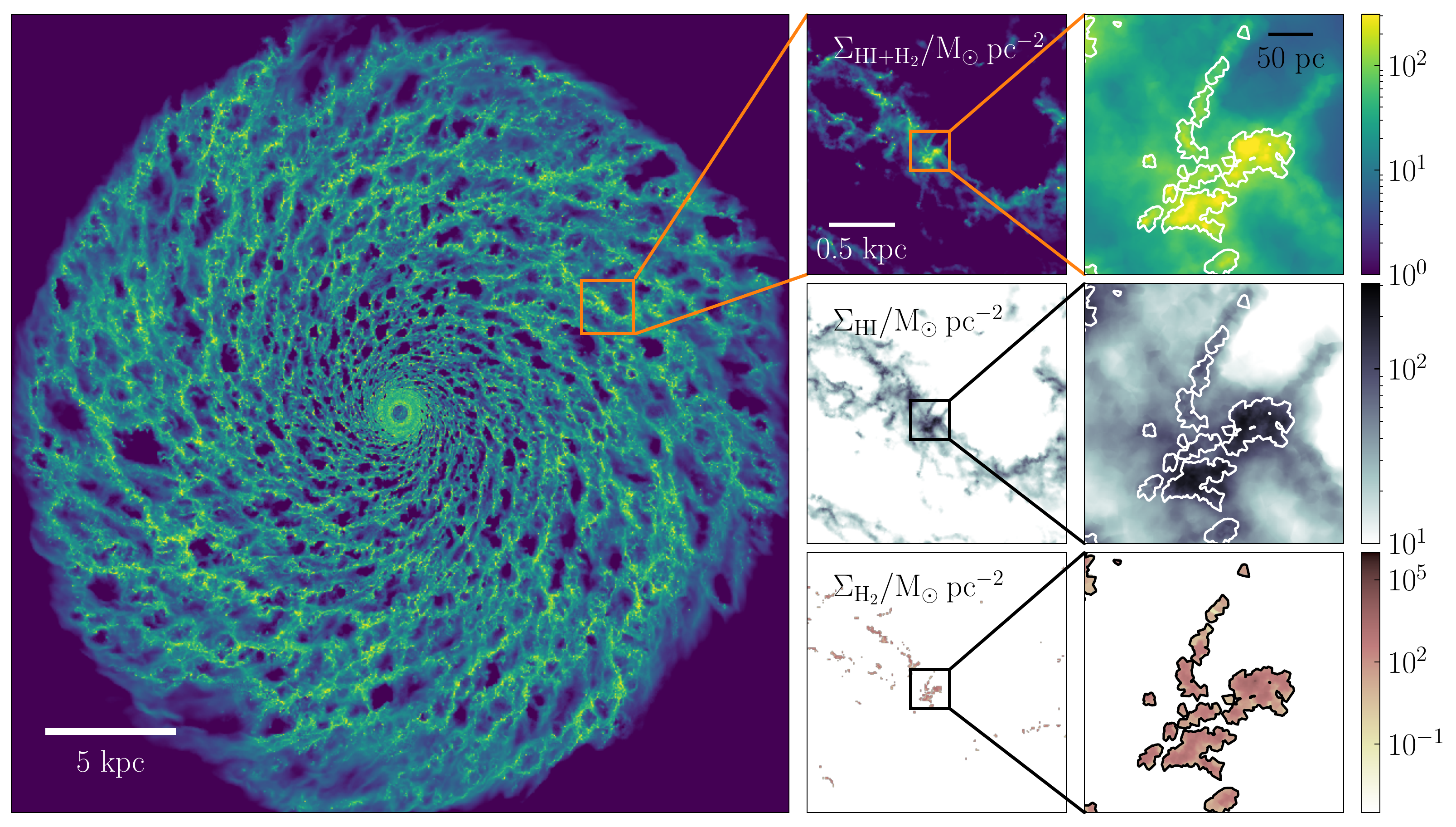}
  \caption{Molecular cloud identification in the FLAT simulation at a time of $600$~Myr. {\it Left:} Column density of the combined HI and H$_2$ gas distribution. {\it Centre:} Zoomed $2$-kpc section of the disc (white box in the left-hand panel), viewed in total cold gas (top), HI gas only (centre) and H$_2$ gas only (bottom). {\it Right:} Zoomed $300$-pc section of the disc, viewed in total cold gas (top), HI gas only (centre) and H$_2$ gas only (bottom). The contours in the right-hand panels indicate the boundaries of GMCs identified via the method described in Section~\ref{Sec::cloud-ID}.}
\end{figure*}

It would be prohibitively computationally-expensive to perform the above convergence calculation for every gas cell in our simulations, so we instead interpolate over a table of pre-calculated cloud models, spaced at regular logarithmic intervals in $n_{\rm H}$, $N_{\rm H}$ and $\alpha_{\rm vir}$. The hydrogen volume density can be straight-forwardly calculated for each Voronoi cell as $n_{\rm H} = \rho/\mu m_{\rm H}$, where $\rho$ is the mass volume density field for the gas and $\mu \sim 1.4$. Following~\cite{Fujimoto19}, the hydrogen column density is computed according to the local approximation of~\cite{Safranek-Shrader+17}, given as
\begin{equation}
N_{\rm H} = \lambda_{\rm J} n_{\rm H},
\end{equation}
where $\lambda_{\rm J} = (\pi c_s^2/G\rho)^{1/2}$ is the Jeans length, computed using an upper limit of $T=40$~K on the gas cell temperature. We define the virial parameter as in~\cite{MacLaren88,BertoldiMcKee1992}, such that
\begin{equation}
\alpha_{\rm vir} = \frac{5\sigma_{\rm g}^2}{\pi G \rho L^2},
\end{equation}
where $\sigma_{\rm g}$ is the turbulent gas velocity dispersion as calculated in~\cite{Gensior20} and $L$ is the smoothing length over which this velocity dispersion is calculated (see Appendix~\ref{Sec::analysis-gas-veldisp}). Together with the assumption of equilibrium and the rates of heating and cooling listed above, these three parameters constrain the abundance of atomic hydrogen $f_{\rm HI}$ and the $^{12}{\rm CO}$ line luminosity $L_{\rm CO}$ for the the $1 \rightarrow 0$ transition. To mimic observations, we use the latter to compute the molecular hydrogen surface density, as described in Section~\ref{Sec::H2-cloud-ID}.

We have also tested an alternative approach to that described above, by applying the {\sc TreeCol} algorithm~\citep{Clark2012} to attenuate the ISRF in the immediate vicinity of each Voronoi gas cell. This would allow us to account for the dust- and self-shielding of molecular hydrogen during run-time, and to self-consistently couple the resulting abundances to the live, non-equilibrium chemical network described in Section~\ref{Sec::chemistry}. However, we have found that at the spatial resolution of our simulations, the resulting molecular hydrogen surface density has a maximum value of $2$~M$_\odot$~${\rm pc}^{-2}$: half of the observed value for the Milky Way~\citep{Wolfire03,Kennicutt+Evans12}. By contrast, the molecular hydrogen abundances obtained in post-processing fall between values of $1$ and $4$~M$_\odot$~${\rm pc}^{-2}$, in agreement with observations (see Section~\ref{Sec::MW-comparison-disc-stab}). This behaviour is in keeping with the resolution requirements reported by~\cite{Seifried17} and~\cite{Joshi19}, who show that the spatial resolution should reach $0.1$~pc in the densest gas, in order to accurately model the ${\rm H}_2$ and CO fractions there. We therefore opt to use the {\sc Despotic} model instead, as we expect that the resolution requirements of this method are less severe than for the computation of abundances via on-the-fly chemistry. That is, at our mass resolution of $900$~M$_\odot$, the lack of sub-structure at high gas densities will have a large impact on the time-scale required for the chemical abundances to reach equilibrium during the non-equilibrium {\sc SGChem} chemistry computation. It will have a lesser impact on the equilibrium abundances themselves, as calculated during post-processing.

\begin{figure*}
\label{Fig::disc-morphology}
  \includegraphics[width=\linewidth]{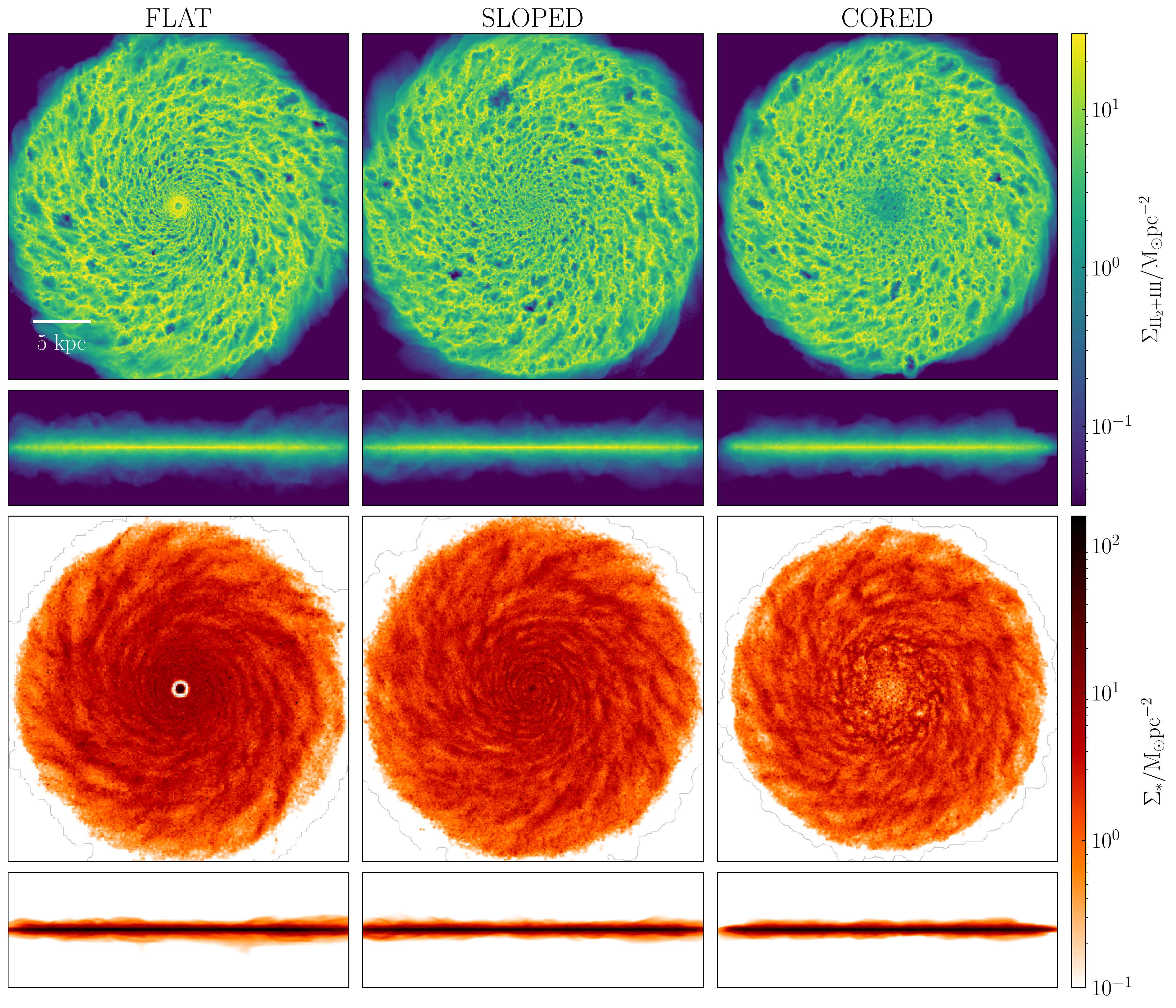}
  \vspace{-0.5cm} \caption{Face-on and edge-on column density maps of the total gas (top two rows) and the live stellar component (bottom two rows) for each galactic disc, at a simulation time of $600$~Myr.}
\end{figure*}

\subsection{Cloud identification} \label{Sec::cloud-ID}
\subsubsection{Molecular clouds} \label{Sec::H2-cloud-ID}
We identify GMCs in our simulations as peaks in the molecular hydrogen column density that are traced by CO, in order to provide the best possible comparison to the properties of observed clouds. We convert the $^{12}{\rm CO} \: 1 \rightarrow 0$ line luminosity ($L_{\rm CO}[{\rm erg} \: {\rm s}^{-1}]$ per hydrogen atom) from our {\sc Despotic} calculation in Section~\ref{Sec::chem-postproc} to a CO-bright molecular hydrogen surface density using
\begin{equation} \label{Eqn::LCO-to-SigmaH2}
\begin{split}
\Sigma_{{\rm H}_2} [{\rm M}_\odot{\rm pc}^{-2}] = &\frac{2.3 \times 10^{-29} [{\rm M}_\odot ({\rm erg} \: {\rm s}^{-1})^{-1}]}{m_{\rm H}[{\rm M}_\odot]} \\
&\times  \Sigma_{\rm g}[{\rm M}_\odot {\rm pc}^{-2}] \\
&\times \frac{\int^{\infty}_{-\infty} \dd z^\prime {\rho_{\rm g}(z^\prime)} \: L_{\rm CO}[{\rm erg}\:{\rm s}^{-1}]}{\int^{\infty}_{-\infty}{\dd z^\prime \rho_{\rm g}(z^\prime)}},
\end{split}
\end{equation}
where $\rho_{\rm g}(z)$ is the total gas volume density as a function of distance $z$ away from the galactic mid-plane, $\Sigma_{\rm g}$ is the total gas surface density, and $m_{\rm H}$ is the proton mass. The factor $2.3 \times 10^{-29} \: ({\rm erg} \: {\rm s}^{-1})^{-1}$ combines the mass-to-luminosity conversion factor $\alpha_{\rm CO} = 4.3 \: {\rm M}_\odot ({\rm K} \: {\rm kms}^{-1} \: {\rm pc}^{-2})^{-1}$ from~\cite{Bolatto13} and the line-luminosity unit conversion factor $5.31 \times 10^{-30} \: ({\rm K} \: {\rm kms}^{-1} \: {\rm pc}^2)/({\rm erg} \: {\rm s}^{-1})$ from~\cite{SolomonVandenBout05}, using the observed frequency of $115.3$~GHz for the CO $J = 1 \rightarrow 0$ transition at a redshift of $z = 0$. The integral ratio is the 2D density-weighted projection map of the CO line-luminosity per hydrogen atom, computed via the method described in Appendix~\ref{App::analysis}. In Section~\ref{Sec::MW-comparison-phases}, we show that the mass of molecular hydrogen identified in this way makes up approximately one third of the combined mass of the warm and neutral media, in accordance with observed galaxies of a similar mass~\citep[e.g.][]{Saintonge+11}.

For each simulation snapshot, we use Equation~(\ref{Eqn::LCO-to-SigmaH2}) to compute a 2D projection map of $\Sigma_{{\rm H}_2}$ with a side-length of $30$~kpc and a resolution of $6$~pc per pixel, equal to the radius of a typical Voronoi gas cell at our mass resolution of $900$~M$_\odot$ and at the minimum molecular cloud hydrogen number density of $\ga 30$~${\rm cm}^{-3}$. We therefore ensure that inside GMCs, each pixel contains at least one Voronoi cell centroid. Using the {\sc Astrodendro} package for Python, we identify all closed contours at $\log_{10}{(\Sigma_{{\rm H}_2}/{\rm M}_\odot \: {\rm pc}^{-2})} = -3.5$, as indicated by the dashed line in Figure~\ref{Fig::DESPOTIC-threshold}.\footnote{This very conservative lower limit makes use of the range of molecular hydrogen surface densties calculated within the {\sc Despotic} model (spanning over $15$ orders of magnitude, as shown in Figure~\ref{Fig::DESPOTIC-threshold}). We save computation time by ignoring the bulk of the pixels in each map (left-hand side of the dashed line), but avoid taking an arbitrary cut on the value of $\Sigma_{\rm H_2}$. The lowest-density gas on the right-hand side of the dashed line will have little influence on the GMC properties computed in Section~\ref{Sec::results-properties}, as the contribution made by each Voronoi cell is weighted by its ${\rm H}_2$ mass. The assumptions associated with our GMC identification procedure are therefore limited to the assumptions made within the {\sc Despotic} model itself.} We discard contours that enclose fewer than nine pixels in total, allowing us to identify clouds of diameter $\sim 3 \times 6$~pc $ = 18$~pc, oversampled by a factor of three. In the right-hand panels of Figure~\ref{Fig::cloud-ID}, we show a zoom-in of the total gas column density (top), the HI column density (centre) and the ${\rm H}_2$ column density (bottom) for a $250$-pc patch of the ISM, overlaid with the corresponding set of contours. These correspond to the regions outlined by squares in the central panels, which in turn correspond to the outlined region in the left-most panel, showing the total gas column density of the entire galaxy.

To obtain the gas cell population of each molecular cloud, we apply the two-dimensional pixel mask for each {\sc Astrodendro} contour to the field of gas cell positions in each snapshot. Any gas cell with a temperature $T<10^4$~K whose centroid falls inside the contour is considered to be a member of the molecular cloud. We accept only those identified structures with $20$ Voronoi gas cells or more, to ensure that the properties of the clouds (e.g.~velocity dispersions, angular momenta) are resolved. On top of the $18$~pc minimum cloud diameter, this imposes a minimum cloud mass of $18,000$~${\rm M}_\odot$. The temperature threshold ensures that we do not include gas cells that fall far from the galactic mid-plane, but we still expect to include many gas cells along the line of sight that have small molecular gas fractions. This is not a concern, as the properties of each molecular cloud are computed as $L_{\rm CO}$-weighted averages.

\subsubsection{HI clouds} \label{Sec::HI-cloud-ID}
The HI clouds in our simulations are identified similarly to the GMCs. The only difference is that we consider the HI gas column density derived from the HI abundance $f_{\rm HI}$, such that
\begin{equation} \label{Eqn::fHI-ID}
\Sigma_{\rm HI} [{\rm M}_\odot {\rm pc}^{-2}] = \Sigma_{\rm g}[{\rm M}_\odot {\rm pc}^{-2}] \frac{\int^\infty_{-\infty} \dd z^\prime \rho_{\rm g}(z^\prime) f_{\rm HI}}{\int^\infty_{-\infty} \dd z^\prime \rho_{\rm g}(z^\prime)},
\end{equation}
where the fraction on the right-hand side is obtained via the ray-tracing procedure described in Appendix~\ref{App::analysis}. We do not distinguish between atomic and ionised gas during post-processing, in the sense that we take $1 = f_{\rm HI} + 2f_{{\rm H}_2}$, where $f_{\rm HI}$ and $f_{{\rm H}_2}$ are the abundances of atomic and molecular hydrogen, respectively. The lowest-density gas in our simulations is assigned an atomic hydrogen abundance of $f_{\rm HI} \sim 1$. We therefore simply choose a lower limit of $\Sigma_{\rm HI} = 10$~M$_\odot$~${\rm pc}^{-2}$ on the HI cloud surface density.

\begin{figure}
\label{Fig::stellar-veldisp}
  \includegraphics[width=\linewidth]{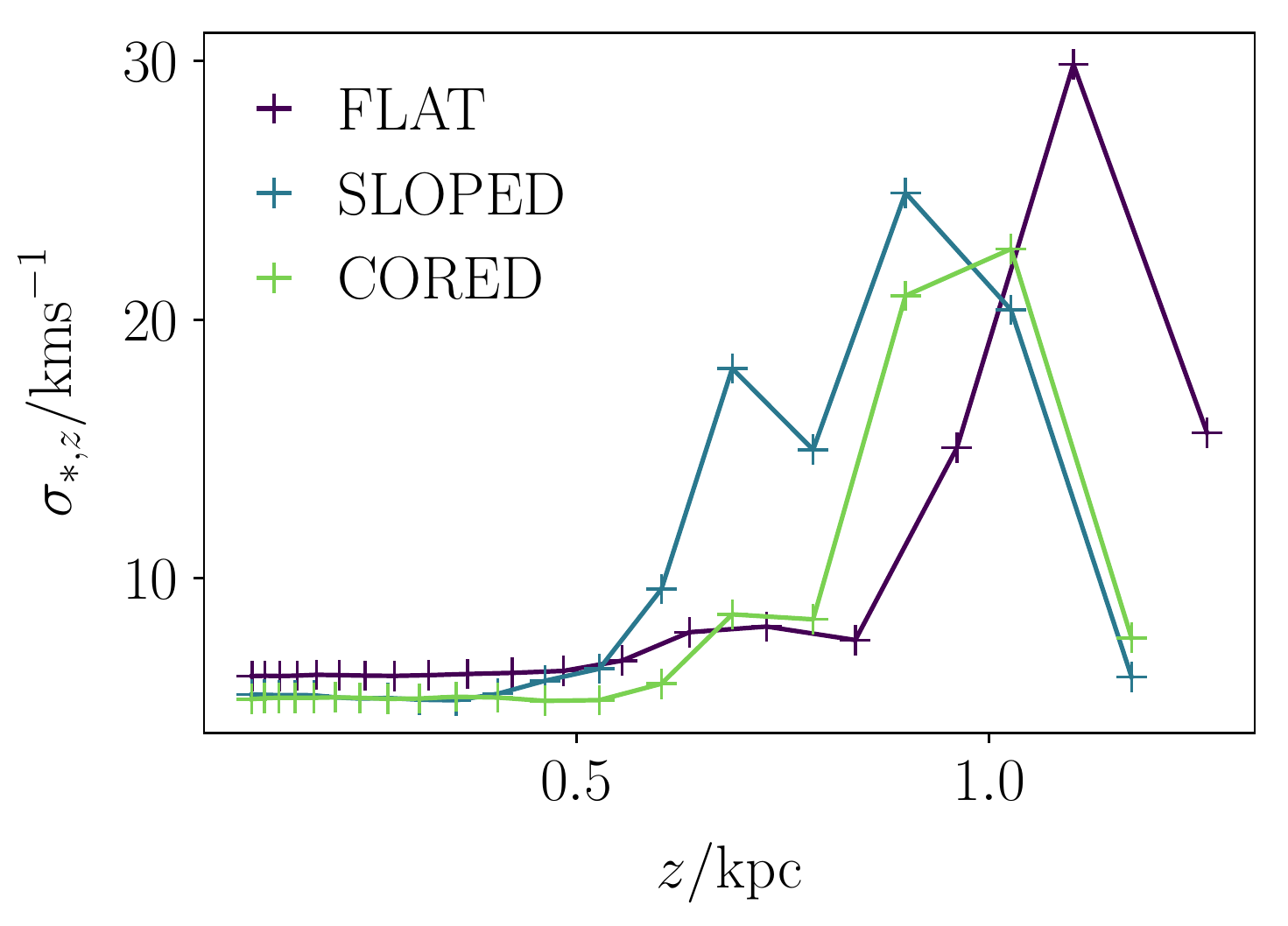}
  \vspace{-0.5cm} \caption{Stellar velocity dispersion as a function of height above/below the galactic mid-plane, over the radial range from $1$ to $13$~kpc. The thin stellar disc ($\la 0.5$~kpc) has a velocity dispersion of $5$-$6$~kms$^{-1}$, while the thick stellar disc ($\ga 0.5$~kpc) has a velocity dispersion of $20$-$30$~kms$^{-1}$.}
\end{figure}

\begin{figure}
\label{Fig::tuning-forks}
  \includegraphics[width=\linewidth]{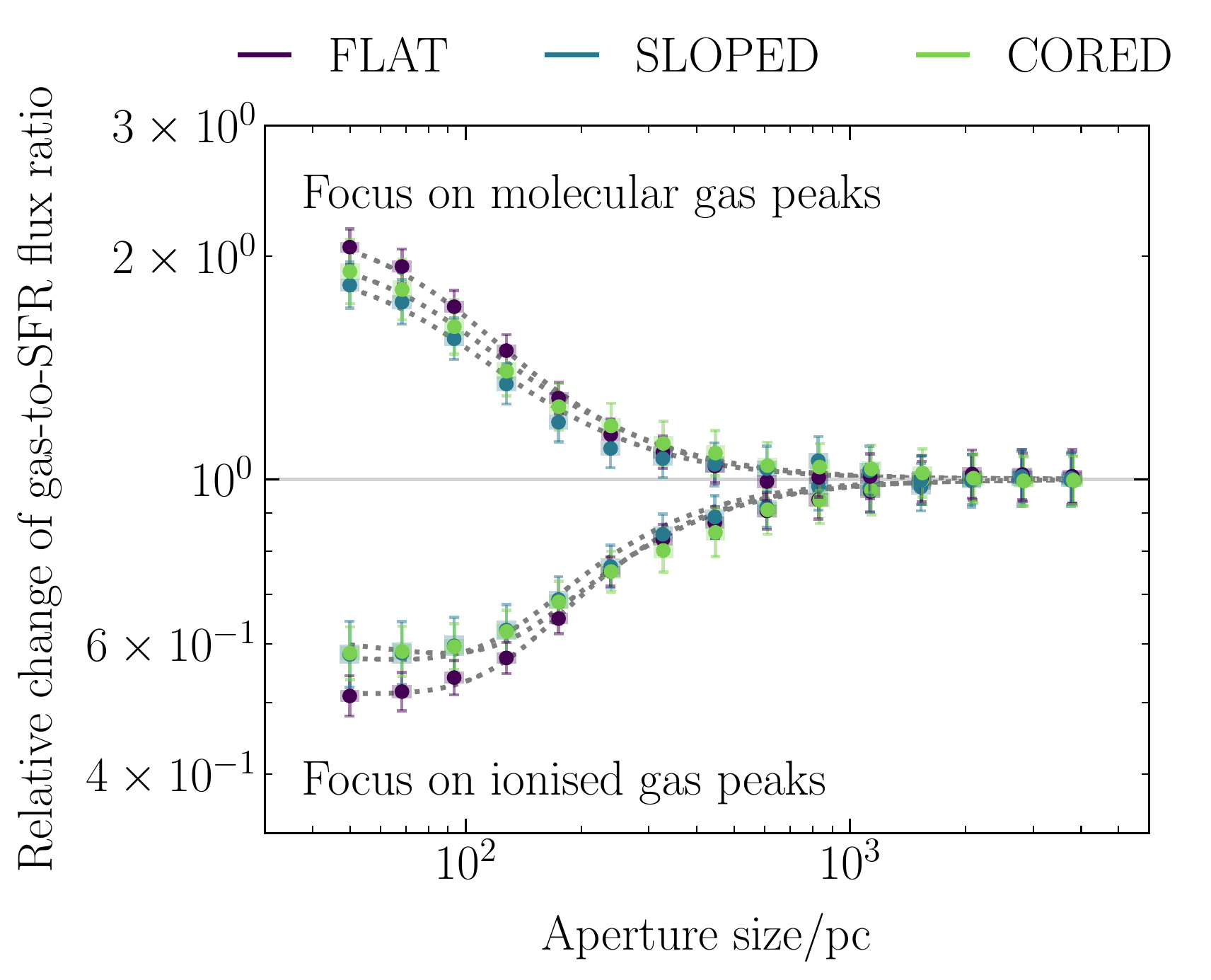}
  \caption{The gas-to-SFR flux ratio relative to the galactic average value as a function of the aperture size, for each of our simulated galaxies at $t=600$~Myr. The upper branch represents apertures focused on molecular gas peaks, while the lower branch represents apertures focused on young stellar surface density (stars with ages $0$-$5$~Myr). The errorbars on each data point represent the $1\sigma$ uncertainty on the value of the gas-to-SFR flux ratio. the dotted lines show the best-fitting models constraining the separation length, GMC lifetime and stellar feedback time-scale~\protect\citep{Kruijssen18a}.}
\end{figure}

\begin{figure*}
\label{Fig::KS-rln}
  \includegraphics[width=\linewidth]{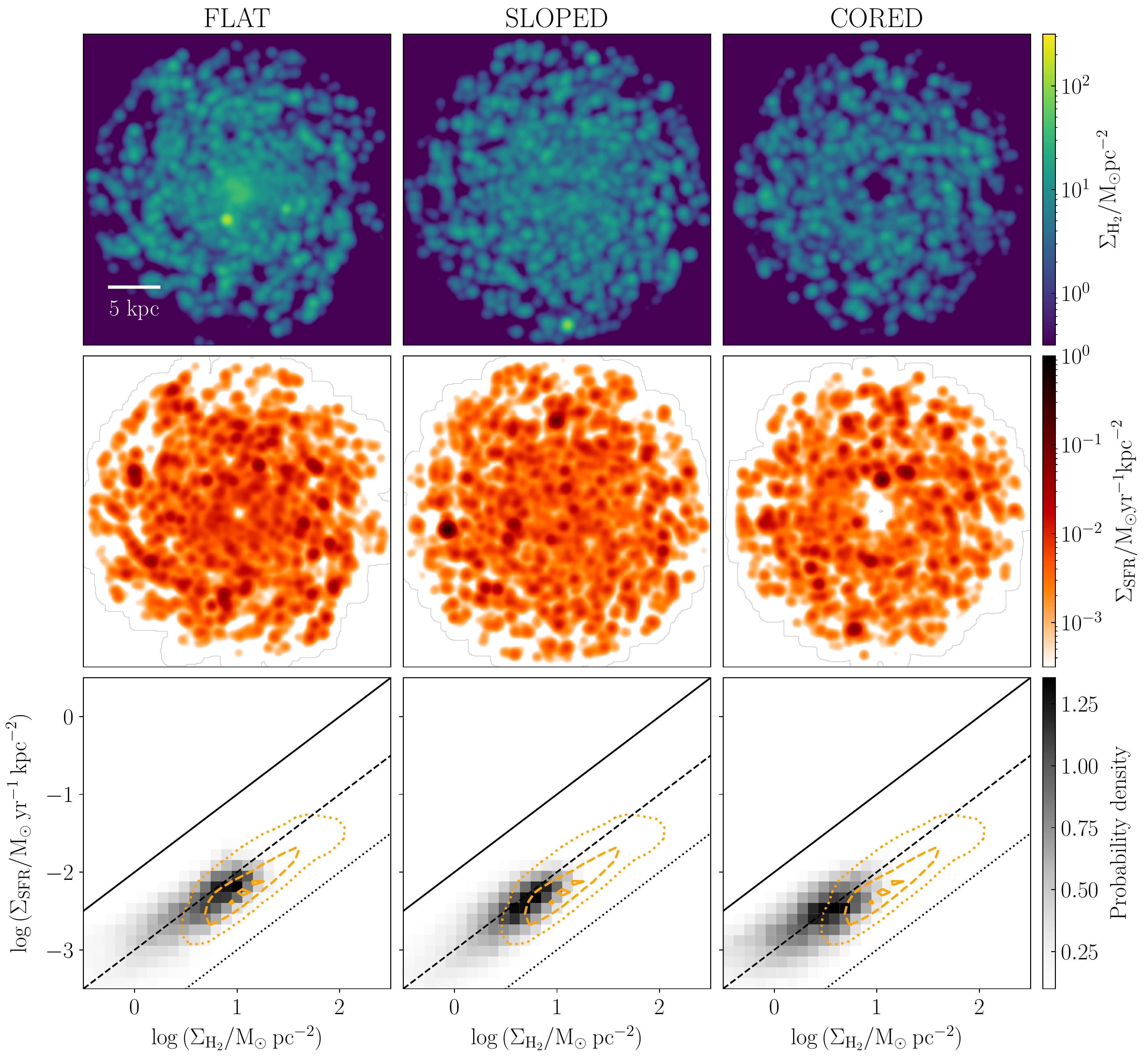}
  \vspace{-0.5cm} \caption{{\it Top row:} CO-bright molecular gas surface density $\Sigma_{{\rm H}_2}$ for each galactic disc, degraded to a spatial resolution of $750$~pc. {\it Middle row:} Star formation rate surface density $\Sigma_{{\rm SFR}}$ for each galactic disc, degraded to a spatial resolution of $750$~pc. {\it Bottom row:} Pixel density as a function of $\Sigma_{{\rm H}_2}$ and $\Sigma_{{\rm SFR}}$ for each disc, corresponding to the resolved molecular star-formation relation of~\protect\cite{Kennicutt98}. Gas depletion times of $10^8$, $10^9$ and $10^{10}$~Myr are given by the black solid, dashed and dotted lines respectively. The orange contours encircle $90$~per~cent (dotted), $50$~per~cent (dashed) and $10$~per~cent (solid) of the observational data for nearby galaxies from~\protect\cite{Bigiel08}. All maps are computed at a simulation time of $600$~Myr.}
\end{figure*}

\begin{figure}
\label{Fig::SFR-vs-time}
  \includegraphics[width=\linewidth]{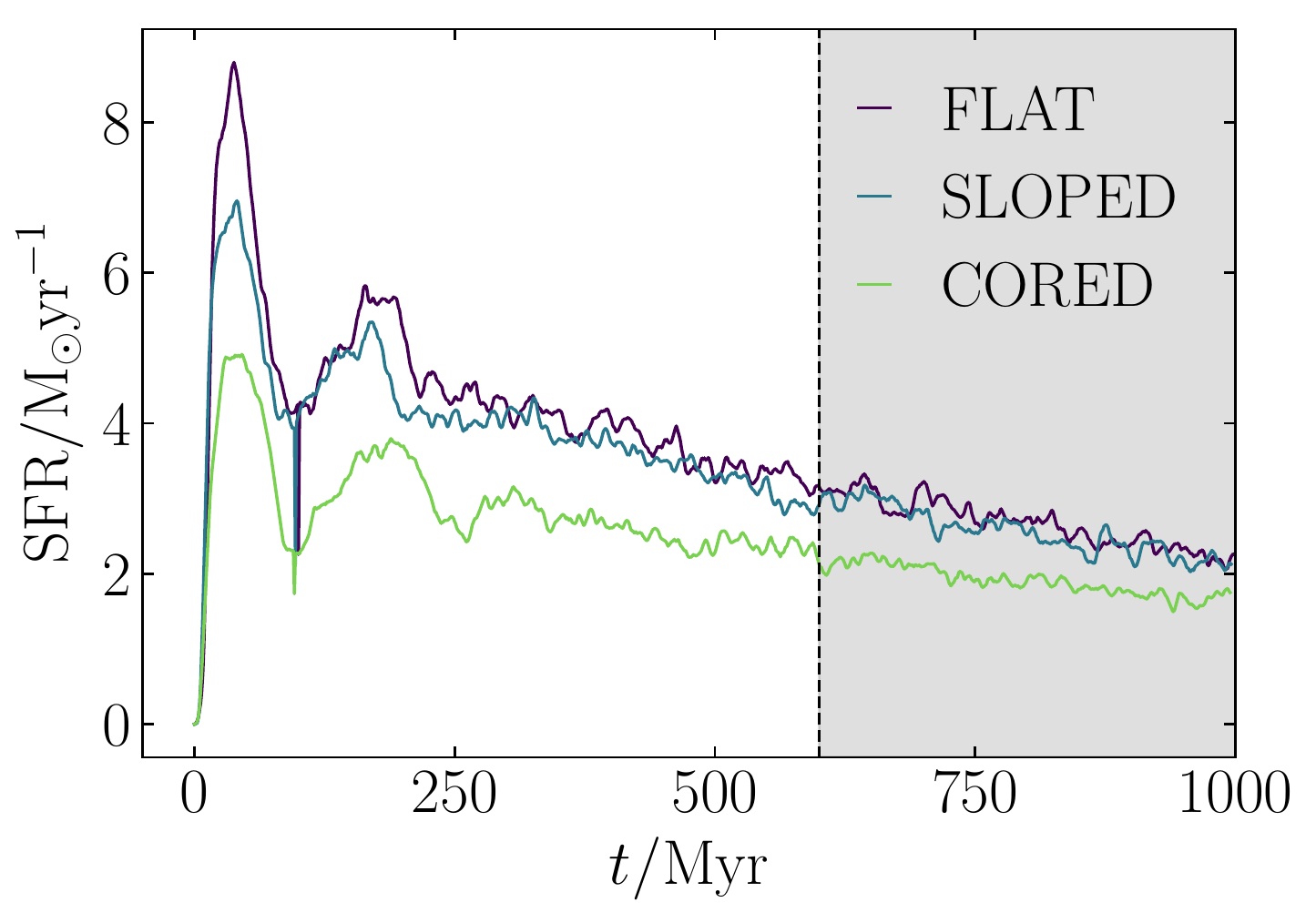}
  \vspace{-0.5cm} \caption{Global galactic star formation rate as a function of simulation time for each galactic disc. The grey shaded region indicates the simulation times $600$~Myr-$1$~Gyr at which we sample the cloud population.}
\end{figure}

\begin{figure}
\label{Fig::radial-profiles}
  \includegraphics[width=\linewidth]{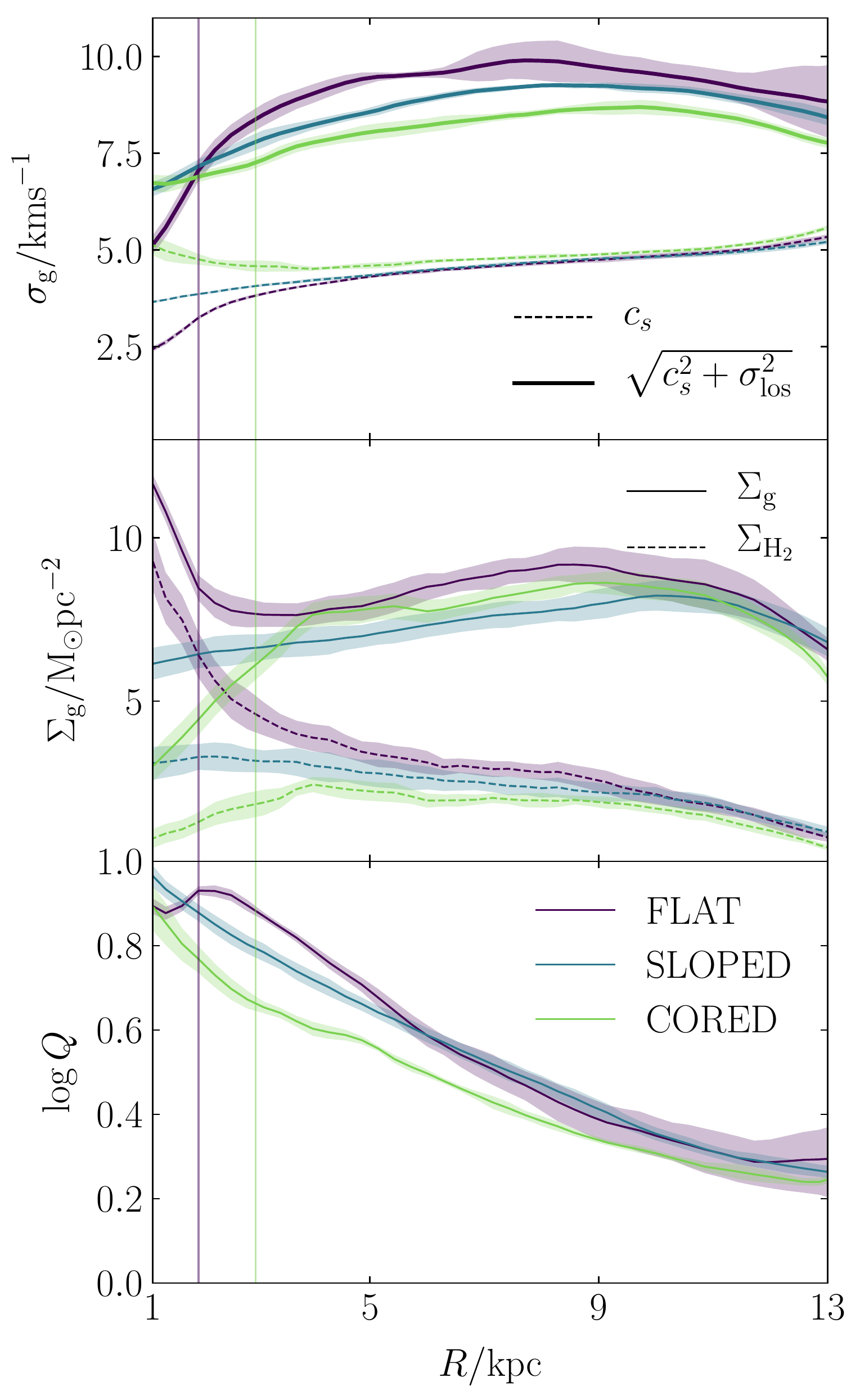}
  \vspace{-0.5cm} \caption{Azimuthally- and temporally-averaged cold-gas ($\leq 10^4$~K) velocity dispersion (upper panel), column density (centre panel) and Toomre $Q$ parameter (lower panel) as a function of the galactocentric radius for each simulated galaxy, across the simulation time interval from $600$~Myr to $1$~Gyr. The thermal $c_s$ and total (thermal plus turbulent) velocity dispersions are denoted by the dashed and solid lines, respectively. The vertical lines denote the minimum radii for cloud identification for the FLAT and CORED simulations (see Section~\ref{Sec::dynamical-span}).}
\end{figure}

\begin{figure}
\label{Fig::ISM-phases}
  \includegraphics[width=.9\linewidth]{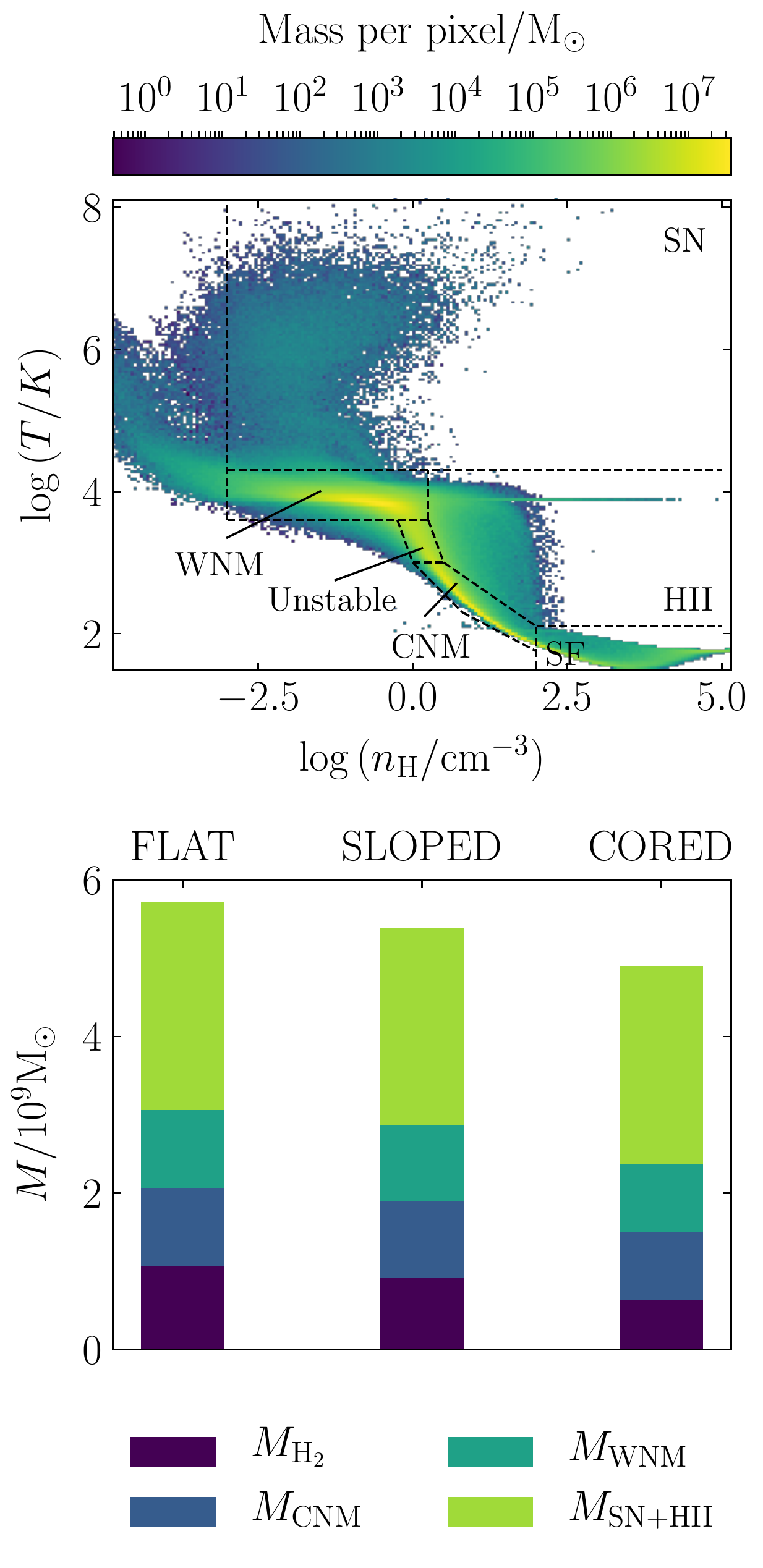}
  \caption{{\it Upper panel:} Density-temperature phase diagram for the FLAT simulation run at a simulation time of $600$~Myr. Dashed lines delineate the regions of phase-space corresponding to the warm neutral medium (WNM), the thermally-unstable phase (Unstable), the cold neutral medium (CNM), gas heated by supernovae (SN), gas heated by HII regions (HII) and star-forming gas (SF). {\it Lower panel:} Partitioning of the gas mass in each galaxy simulation into four ISM phases from warmest to coolest: hot gas that has received thermal energy from stellar feedback ($M_{\rm SN + HII}$), the warm neutral medium ($M_{\rm WNM}$), the cold neutral medium ($M_{\rm CNM}$), and the star-forming gas in the molecular phase ($M_{{\rm H}_2}$).}
\end{figure}

\section{Properties of simulated galaxies} \label{Sec::galaxy-properties}
The initial conditions for our galaxies are Milky Way-like in their masses, sizes and geometries, so it is a necessary (but not sufficient) test of the input physics that they also reproduce the observable properties of the Milky Way on sub-galactic scales.\footnote{We do not expect to reproduce the properties of the central few kpc of the Milky Way, as we do not model the Galactic bar in our simulations. That is, differences between the simulations and observations in this region are expected, and do not raise concerns about the general validity of our model.} In this section, we compare the properties of our simulated discs to observations of Milky Way-like galaxies from the literature, and demonstrate an acceptable level of agreement.

\subsection{Disc morphology} \label{Sec::MW-comparison-morphology}
\subsubsection{Disc structure on kpc-scales} \label{Sec::kpc-scale-disc-structure}
The face-on and edge-on total gas and stellar column densities for each simulated galaxy are displayed in Figure~\ref{Fig::disc-morphology}. As reported in Table~\ref{Tab::params}, stellar feedback inflates the HI gas disc to a scale-height of $\sim 250$-$380$~pc, in agreement with the value observed by~\cite{SavageWakker2009}. In Figure~\ref{Fig::stellar-veldisp} we show the stellar velocity dispersion as a function of the height above the galactic mid-plane. The thin stellar disc has a vertical velocity dispersion of $5$-$6 \: {\rm kms}^{-1}$ and a scale-height of $400$-$500$~pc, while the thick stellar disc has a velocity dispersion of $20$-$30 \: {\rm kms}^{-1}$ and a scale-height of $\sim 1000$~pc. Each of these measured parameters is in approximate agreement with the values observed in the Milky Way~\citep{RixBovy2013}.

High-resolution studies of the nearby Milky Way-like flocculent spiral galaxies NGC 628 and NGC 4254 provide a qualitative point of comparison for the morphology of our HI gas~\citep[see][]{Walter2008}, molecular gas~\citep[see][]{Sun18} and young stars~\citep[see][]{Kreckel2018}. In particular, we find that our maps of the molecular gas surface density (top row of Figure~\ref{Fig::disc-morphology}) are similar in structure to the sub kpc-scale observations of the CO emission in both NGC 628 and NGC 4254. Our maps of the young stellar surface density (central row of Figure~\ref{Fig::disc-morphology}) may likewise be compared to to the sub kpc-scale structure of the H-$\alpha$ emission in NGC 628, while its observed HI gas profile resembles the structure of the HI gas disc for the FLAT simulation in particular (top left-hand panel of Figure~\ref{Fig::disc-morphology}). 

\subsubsection{Spatial decorrelation between molecular gas and stars}
By contrast with the tight correlation observed between molecular gas and tracers of star formation on galaxy scales~\citep[e.g.][]{Kennicutt98}, a spatial decorrelation between CO clouds and HII regions is observed in nearby galaxies~\citep[e.g.][]{Schruba10,Kreckel2018,Kruijssen2019,Schinnerer19,Chevance20}. This spatial decorrelation has been explained by the fact that individual regions in galaxies follow evolutionary lifecycles independent from those of their neighbours, during which clouds assemble, collapse, form stars, and are disrupted by stellar feedback~\citep[e.g.][]{Feldmann2011,Calzetti12,Kruijssen2014,Kruijssen18a}. Although in Section~\ref{Sec::kpc-scale-disc-structure} we have found qualitative similarities between our simulations and observations, we note that~\cite{Fujimoto19} also reproduce the observed morphologies of the ${\rm H}_2$, HI and stellar components in galaxies like the Milky Way and NGC 628, but fail to correctly capture the spatial decorrelation between young stellar regions and dense molecular gas on small scales. We apply the same analysis as~\cite{Fujimoto19} to the maps of $\Sigma_{{\rm H}_2}$ and $\Sigma_{\rm SFR}$ for our galaxies, degraded in spatial resolution via convolution with a Gaussian kernel of ${\rm FWHM}=50$~pc. The result is illustrated in Figure~\ref{Fig::tuning-forks}. In summary, we measure the gas-to-stellar flux ratio enclosed in apertures centred on ${\rm H}_2$ peaks (top branch) and SFR peaks (bottom branch), for aperture sizes ranging between the native resolution of the convolved maps ($50$~pc) and large scales ($4000$~pc). More details about the method can be found in~\cite{Kruijssen2014} and~\citet{Kruijssen18a}. Figure~\ref{Fig::tuning-forks} shows that our galaxies span approximately a factor $2$ in the gas-to-SFR flux ratio. Therefore, by contrast with~\cite{Fujimoto19}, we find a similar gas-to-stellar decorrelation as is observed in several nearby disc galaxies by~\cite{Chevance20}, which show gas-to-SFR flux ratios in the range $[0.3, 0.5]$ for cloud-scale apertures ($\sim 30-150$~pc) centred on stellar peaks and in the range $[1.3, 5.5]$ for cloud-scale apertures centred on gas peaks. Following the formalism of~\cite{Kruijssen18a}, this spatial decorrelation can be used to probe the evolutionary timeline of GMCs and star-forming regions. This will be investigated in more detail in~\cite{Jeffreson20b}.

\subsection{Star formation} \label{Sec::MW-comparison-SFR}
In Figure~\ref{Fig::SFR-vs-time}, we show the total galactic star formation rate as a function of simulation time $t$ for each isolated disc galaxy. Following the initial vertical collapse of the disc and the subsequent star formation `burst' from $t \sim 30$~Myr to $t \sim 250$~Myr, the SFR settles down to a rate of $\sim 2$-$4$~M$_\odot$~yr$^{-1}$. We make absolutely sure to consider each isolated disc in its equilibrium state by examining the cloud population during a later time interval, between $t = 600$~Myr and $t = 1$~Gyr (grey-shaded region). Over this period, the SFR declines only gradually, by a total of around $0.5$~M$_\odot$~yr$^{-1}$. These values are consistent with the current observed SFR in the Milky Way~\citep{Murray&Rahman10,Robitaille&Whitney10,Chomiuk&Povich,Licquia&Newman15}. We may also consider the resolved star-forming behaviour on scales of $750$~pc, as studied in nearby galaxies by~\cite{Bigiel08}. In Figure~\ref{Fig::KS-rln}, we display the star formation rate surface density as a function of gas surface density for each of our simulated galaxies, at a simulation time of $600$~Myr. The top row shows the 2D projection maps of the CO-bright molecular gas column density $\Sigma_{{\rm H}_2}$. These are computed via the total gas column density in Equation~(\ref{Eqn::LCO-to-SigmaH2}) and degraded using a Gaussian filter of ${\rm FWHM}=750$~pc. The corresponding projections of the star formation rate surface densities $\Sigma_{\rm SFR}$ are displayed in the central row. Details for the production of all maps are given in Appendix~\ref{App::analysis}. In the bottom row, the values of $\Sigma_{{\rm H}_2}$ and $\Sigma_{\rm SFR}$ in each pixel of the spatially-degraded projection maps are compiled to produce a single histogram. The loci of our simulation data fall close to the observed star formation relations obtained by~\cite{Bigiel08}, denoted by the orange contours, though with a population of points at lower densities and star formation rates than are reached by the observations. These points arise because we consider all CO-emitting gas down to a molecular hydrogen surface density of $\Sigma_{{\rm H}_2} = 10^{-3.5} \: {\rm M}_\odot {\rm pc}^{-2}$ (see Figure~\ref{Fig::DESPOTIC-threshold}). This avoids taking an arbitrary cut on $\Sigma_{\rm H_2}$, but also captures much lower levels of CO emission than could be detected by current observatories.

\subsection{Resolved disc stability} \label{Sec::MW-comparison-disc-stab}
Our simulated discs have approximately-uniform values of the line-of-sight velocity dispersion $\sigma_{\rm los, g}$ and the surface density $\Sigma_{\rm g}$ for atomic and molecular gas. The radial profiles for $\sigma_{\rm los, g}$ and $\Sigma_{\rm g}$ in each galaxy are displayed in the upper two panels Figure~\ref{Fig::radial-profiles}, computed in $50$ overlapping bins of width $1$~kpc. Only the gas component with temperature $\leq 10^4$~K is considered, and the line-of-sight turbulent velocity dispersion is calculated according to
\begin{equation} \label{Eqn::los-veldisp}
\sigma_{{\rm los}, g}^2 = \frac{\langle |\mathbf{v}_i-\langle \mathbf{v}_i \rangle|^2 \rangle}{3},
\end{equation}
where $\{\mathbf{v}_i\}$ are the velocity vectors of the gas cells in each radial bin, and angled brackets denote mass-weighted averages over these cells. The gas sound speed $c_{\rm s}=\sqrt{k_{\rm B}T/{\mu m_{\rm H}}} \sim 6$~kms$^{-1}$ for our galaxies (dashed lines in the upper panel) is consistent with the observed temperature of the neutral ISM phases in the Milky Way \citep[e.g.][]{Kalberla09}. Similarly, our combined turbulent/thermal velocity dispersion $\sqrt{c_{\rm s}^2+\sigma_{{\rm los}, \rm g}}$ (solid lines in the upper panel) also falls squarely within the observed range of $\sim 10 \pm 2$~kms$^{-1}$~\citep{Tamburro09}. Our total gas surface densities (solid lines in the central panel) are inside the observed range of $7$ to $11$~M$_\odot$~${\rm pc}^{-2}$ for galactocentric radii between $3$ and $16$~kpc in the Milky Way~\citep{Yin09}. Our molecular hydrogen surface densities (dashed lines in the central panel) fall between values of $1$ and $4$~M$_\odot$~${\rm pc}^{-2}$ for galactocentric radii from $2$ to $10$~kpc, in agreement with the Milky Way values from Figure 1 of~\cite{Wolfire03} and Figure 7 of~\cite{Kennicutt+Evans12}. Finally, the lower panel of Figure~\ref{Fig::radial-profiles} demonstrates that our values for the Toomre $Q$ parameter agree with the observed values in the discs of spiral galaxies, which are seen to vary across the range of $Q \sim [1,10]$~\citep{Leroy08}.

\subsection{ISM phase structure} \label{Sec::MW-comparison-phases}
The top panel of Figure~\ref{Fig::ISM-phases} displays the mass-weighted distribution of the gas temperature as a function of the gas volume density (phase diagram) for the FLAT simulation. The distribution is peaked along the `thermal equilibrium curve': the state of thermal equilibrium in which the rate of cooling (dominated by line emission from ${\rm C}^+$, ${\rm O}$ and ${\rm Si}^+$) balances the heating rate due to photoelectric emission from dust grains and PAHs. The position of our thermal equilibrium curve matches the analysis of~\cite{Wolfire03}, who studied the thermal structure of the ISM in the Milky Way. It also agrees with the thermal evolution expected from \textsc{SGChem} live chemistry, according to~\cite{GloverMacLow07a,GloverMacLow07b}. Along the curve, the gas can be divided into four major components: the warm neutral medium (WNM), the thermally-unstable phase (Unstable), the cold neutral medium (CNM) and the star-forming gas (SF). The non-equilibrium components of the gas, heated by stellar feedback from supernovae (SN) and HII regions (HII), fall above the thermal equilibrium curve. In particular, the horizontal line at $T \sim 7000$~K is formed from gas cells heated by HII region feedback, as described in Section~\ref{Sec::HII-heating}. The phase structure of the gas in our simulations agrees closely with the isolated galaxy simulations of~\cite{Goldbaum16}, as well as those of~\cite{Hopkins12,Agertz13,Keller14,Fujimoto18}, all of which use similar feedback models to ours. In the lower panel of Figure~\ref{Fig::ISM-phases}, we display the partitioning of gas mass between the phase components in the upper panel. The mass of CO-traced molecular gas ($M_{{\rm H}_2}$) is calculated using the {\sc DESPOTIC} model described in Section~\ref{Sec::chem-postproc}, and approximately equals the mass of star-forming gas (SF) in the upper panel. For all simulated discs, around half of the total gas mass is partitioned approximately-equally between the molecular, CNM and WNM components, such that $M_{{\rm H}_2}/(M_{\rm CNM} + M_{\rm WNM}) \in \{0.3, 0.4, 0.3\}$ for the FLAT, SLOPED and CORED simulations, in agreement with~\cite{Saintonge+11}.

Although the mass reservoirs for each of the phases described above are relatively static over the simulation times from $600$ to $1000$~Myr examined here, there exists an ISM baryon cycle that continually shifts gas between star-forming and non-star-forming phases, as explored in~\cite{Semenov17,Semenov18,Semenov19,ChevanceReview20}. The time spent in each of these reservoirs sets the global star-forming properties of the ISM (e.g.~gas depletion times and SFEs). In this work, we examine the time-independent properties of GMCs and their relation to the large-scale galactic-dynamical environment. In a follow-up paper~\citep{Jeffreson20}, we explore the influence of galactic dynamics on the time-evolving GMC lifecycle, and relate these findings to the environmental variation in the ISM baryon cycle.

\section{Theory} \label{Sec::Theory}
In~\cite{Jeffreson+Kruijssen18}, we introduced an analytic theory that quantifies the influence of galactic dynamics on molecular cloud evolution. Here we provide an overview of the theory, and explain how the environmental parameter space spanned by its variables is used to reveal the presence of galactic-dynamical trends in the physical properties of our simulated GMCs.

\begin{table*}
\begin{center}
  \caption{The dynamical time-scales used in the cloud lifetime theory of~\protect\cite{Jeffreson+Kruijssen18} and their physical interpretations.}
  \begin{tabular}{ m{1.3cm} m{8cm} m{2.5cm} m{2.5cm} }
  \hline
   Time-scale & Physical meaning & Analytic form & Physical variables \\
  \hline
    $\tau_\kappa$ &
    Time-scale for a molecular cloud to make a `maximal' orbital excursion along the galactic radial direction (defined as one quarter of the average orbital radius). & 
    $\frac{4\pi}{\Omega \sqrt{2(1+\beta)}} \frac{1}{\sqrt{3+\beta}}$ & $\Omega$, $\beta$ \\
    $\tau_{\rm ff,g}$ &
    Time-scale for the gravitational collapse of the ISM on sub-Toomre length scales, as in~\cite{Krumholz12}. & 
    $\sqrt{\frac{3\pi^2}{32\phi_P(1+\beta)}} \frac{Q}{\Omega}$ & $Q$, $\Omega$, $\beta$, $\phi_P$ \\
    $\tau_{\rm cc}$ &
    Average time-scale between cloud collisions~\citep{Tan00}. & 
    $\frac{2\pi Q}{9.4 f_G \Omega(1+0.3\beta)(1-\beta)}$ & $Q$, $\Omega$, $\beta$ \\
    $\tau_\beta$ &
    Time-scale for a spherical GMC to become ellipsoidal under the influence of galactic differential rotation (shear-induced azimuthal offset across the cloud becomes equal to its radial extent). &
    $\frac{2}{\Omega(1-\beta)}$ & $\Omega$, $\beta$ \\
  \hline
  \end{tabular}
\end{center}
\label{Tab::time-scales}
\vspace{-0.25cm}
\end{table*}

\subsection{Dynamical time-scales for GMC evolution}
In~\cite{Jeffreson+Kruijssen18}, the time-averaged influence of galactic dynamics on the evolution and destruction of GMCs is determined by the five dynamical time-scales for gravitational free-fall in the galactic mid-plane ($\tau_{\rm ff,g}$), galactic shear ($\tau_\beta$), spiral-arm interactions ($\tau_{\Omega_{\rm P}}$), cloud-cloud collisions ($\tau_{\rm cc}$) and orbital epicyclic perturbations ($\tau_\kappa$). With the exception of the time-scale for spiral arm perturbations, which is not relevant for the flocculent discs simulated in this work, each time-scale is defined in terms of its physical variables in Table~\ref{Tab::time-scales}. Here, $\Omega$ is the angular velocity of the mid-plane ISM around the galactic centre, and the galactic shear parameter is defined by
\begin{equation}
\label{Eqn::beta}
\beta = \frac{\dd \ln{v_c}}{\dd \ln{R}},
\end{equation}
for a circular velocity $v_{\rm c}(R)$ at galactocentric radius $R$. The \citet{Toomre64} $Q$ parameter for the gravitational stability of the mid-plane gas is given by
\begin{equation}
\label{Eqn::Toomre_Q}
Q = \frac{\kappa \sqrt{\sigma_{\rm g}^2+c_s^2}}{\pi G \Sigma_{\rm g}},
\end{equation}
with an epicyclic frequency $\kappa$, a mid-plane gas velocity dispersion $\sigma_{\rm g}$, a mid-plane sound speed $c_{\rm s}$, and a mid-plane gas surface density $\Sigma_{\rm g}$. The variable $\phi_P$ quantifies the relative gas and stellar contributions to the mid-plane hydrostatic pressure $P_{\rm mp} = \pi G \phi_{\rm P} \Sigma_{\rm g}^2/2$, defined in~\cite{Elmegreen89} as
\begin{equation}
\begin{split}
\label{Eqn::phi_P}
\phi_P &= 1+\frac{\Sigma_{\rm s}}{\Sigma_{\rm g}} \frac{\sigma_{\rm g}}{\sigma_{\rm s}} \\
&\approx 1 + \frac{\sigma_{\rm g}}{\Sigma_{\rm g}} \sqrt{\frac{2\rho_{\rm s}}{\pi G}}.
\end{split}
\end{equation}
Here, $\sigma_{\rm s}$, $\Sigma_{\rm s}$ and $\rho_{\rm s}$ denote the stellar velocity dispersion, surface density, and volume density, respectively. The second approximate equality is obtained by assuming that the scale-height of the stellar disc is much larger than the gas disc scale-height, so that the stellar disc maintains its own state of collisionless equilibrium, $\Sigma_{\rm s} = \sigma_{\rm s} \sqrt{2\rho_{\rm s}/\pi G}$~\citep[c.f.][]{BlitzRosolowsky2004}. The cloud-cloud `collision probability' parameter $f_{\rm G}$ is defined and constrained by comparison to observations in~\cite{Tan00}.

\begin{figure}
  \label{Fig::paramspace-og}
    \includegraphics[width=\linewidth]{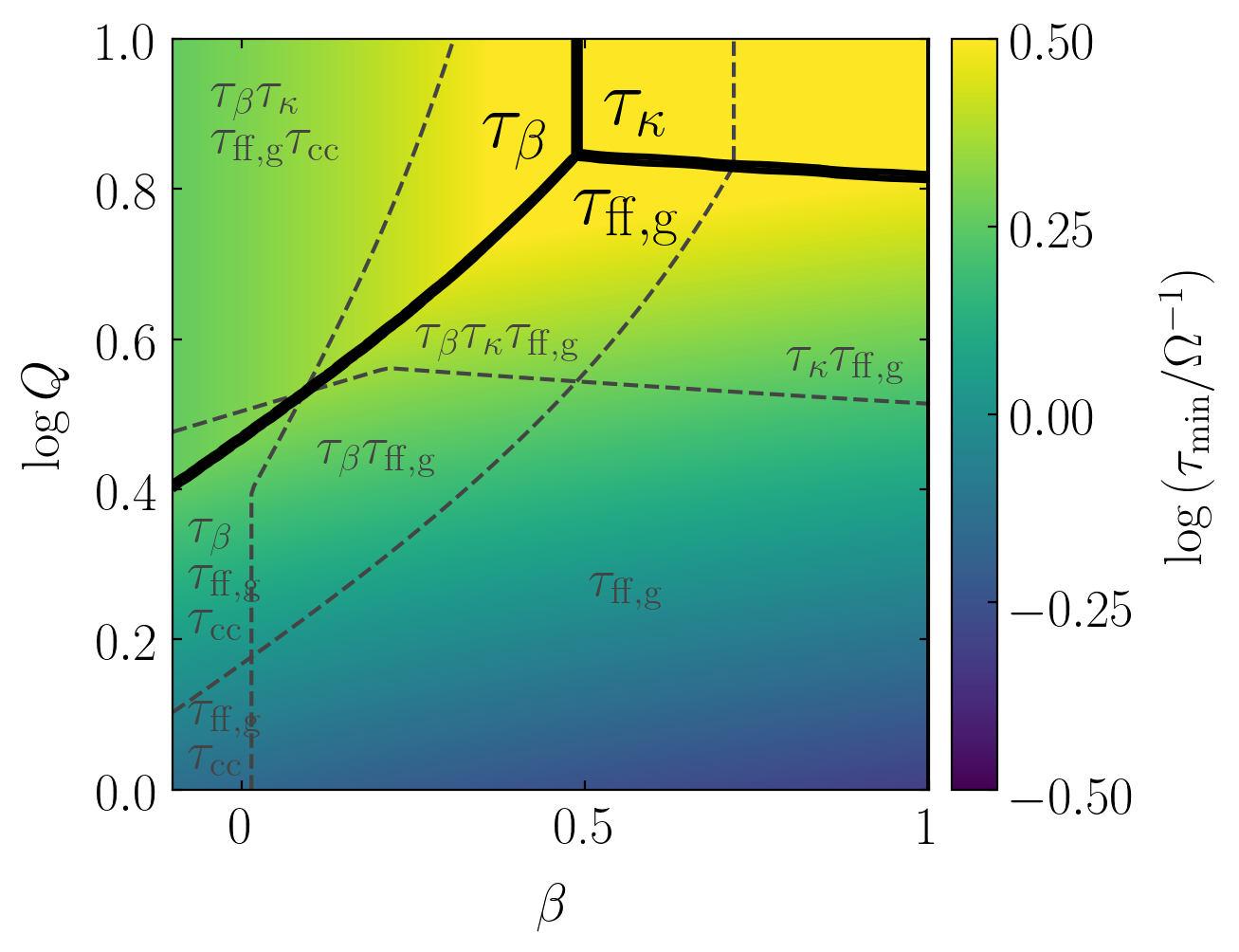}
    \caption{The minimum of the four galactic-dynamical time-scales for gravitational free-fall ($\tau_{\rm ff,g}$), galactic shear ($\tau_\beta$), cloud-cloud collisions ($\tau_{\rm cc}$), and orbital epicyclic perturbations ($\tau_\kappa$), normalised by the orbital period of the galaxy $\Omega^{-1}$, as a function of the galactic-dynamical parameters $\beta$ and $Q$. The solid black contours delineate the regions of parameter space in which each of the time-scales is shortest, and so has the greatest potential to influence molecular cloud evolution. The dashed grey contours delineate the regions for which each time-scale is shorter than twice the minimum dynamical time-scale ($\tau > 2\tau_{\rm min}$), or else shorter than twice the cloud lifetime ($\tau > 2\tau_{\rm life}$), if $\tau_{\rm life}>\tau_{\rm min}$. They provide a rough indication of the parameter space regions in which multiple dynamical mechanisms are likely to influence molecular cloud evolution. The positions of all contours are determined for the fiducial value of $\phi_{\rm P} = 3$.}
\end{figure}

\begin{figure*}
  \label{Fig::betaQ_explore}
    \includegraphics[width=\linewidth]{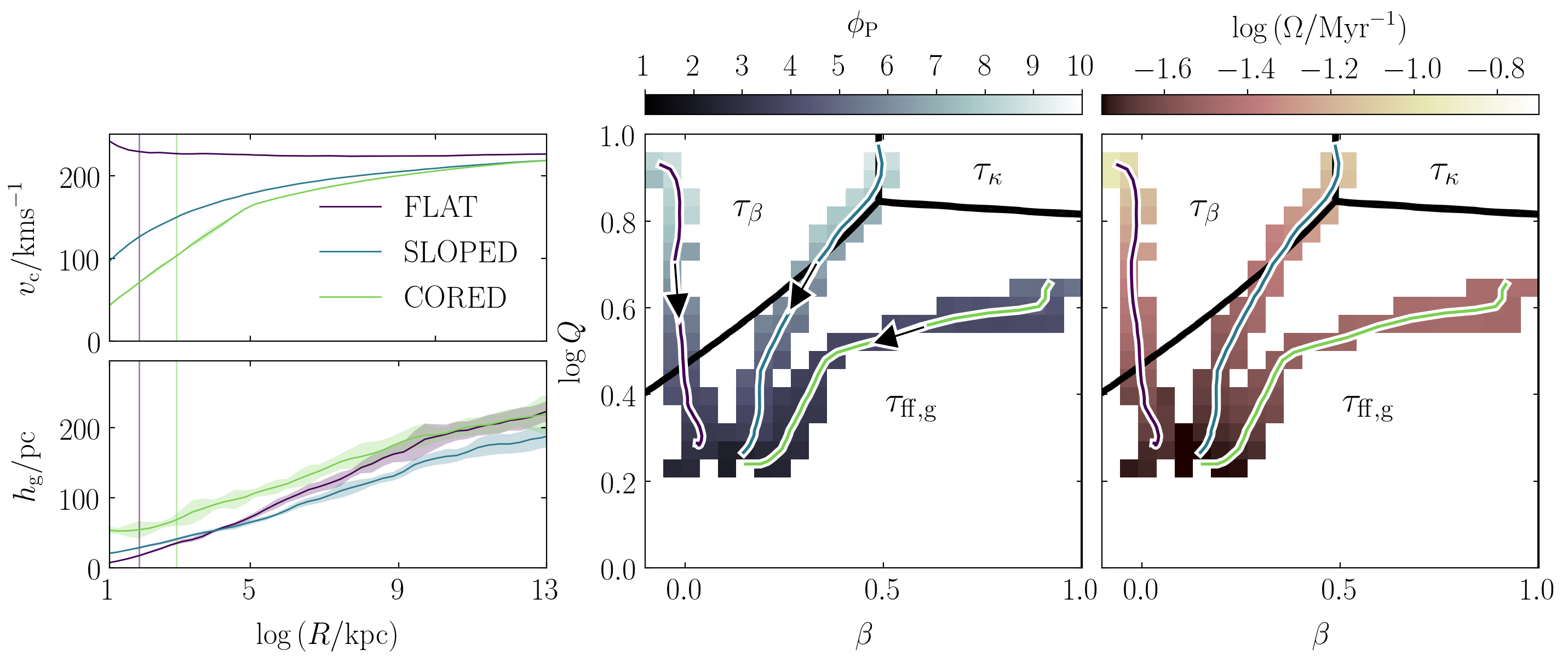}
    \caption{{\it Left:} Circular velocity $v_c$ (upper panel) and cold-gas ($\leq 10^4$~K) scale-height $h_{\rm g}$ (lower panel) for each simulated galaxy as a function of the galactocentric radius, averaged over the time interval from $600$~Myr to $1$~Gyr (solid lines). The temporal standard deviation at each radius is given by the translucent shaded regions, and the vertical lines denote the radial cuts for cloud identification (see Section~\ref{Sec::dynamical-span}). {\it Centre:} Median value of the $\phi_{\rm P}$ parameter in each simulation as a function of the shear $\beta$ and Toomre $Q$ parameters. The black arrows mark the direction in which the galactocentric radius increases. {\it Right:} Median value of the galactic angular velocity $\Omega$ in each simulation as a function of $\beta$ and $Q$. The temporally-averaged radial trajectories of each disc through the parameter space are denoted by the solid lines, and all values are compiled on the time interval from $600$~Myr to $1$~Gyr. The black solid lines enclose the regions of parameter space for which the minimum dynamical time-scale is $\tau_\kappa$ (orbital epicyclic perturbations), $\tau_\beta$ (galactic shear) and $\tau_{\rm ff,g}$ (gravitational free-fall).}
\end{figure*}

All of the time-scales in Table~\ref{Tab::time-scales} depend inversely on the angular velocity $\Omega$. As such, they can be compared within a parameter space spanned by the four physical variables $\beta$, $Q$, $\phi_{\rm P}$ and $f_G$. Of these, we fix $f_G=0.5$ to its fiducial value. \cite{Jeffreson+Kruijssen18} therefore describes the influence of galactic dynamics on GMC evolution within a fundamental parameter space spanned by $\beta$, $Q$ and $\phi_{\rm P}$. This parameter space is displayed in Figure~\ref{Fig::paramspace-og} for Milky Way-like environments, with $\beta \in [0,1]$, $\log{Q} \in [0, 10]$, and $\phi_{\rm P} = 3$. The minimum galactic-dynamical time-scale $\tau_{\rm min} = \min{(\tau_{\rm ff,g}, \tau_\beta, \tau_\kappa, \tau_{\rm cc})}$ is shown in colour, and we indicate the regions of parameter space in which each time-scale is shorter than all others, such that its corresponding dynamical process has the dominant influence on cloud evolution (solid black lines). With the grey dashed contours, we also indicate the regions of parameter space over which the galactic-dynamical time-scales have comparable values, to within a factor of two. Formally, these contours appear for a time-scale $\tau$ where $\tau = 2\tau_{\rm min}$, or where $\tau = 2\tau_{\rm life}$, if the cloud lifetime $\tau_{\rm life}$ is shorter than the minimum dynamical time-scale $\tau_{\rm min}$, where the cloud lifetime is defined in~\cite{Jeffreson+Kruijssen18} as the linear combination of dynamical Poisson rates, $\tau_{\rm life} = |\tau_{\rm ff,g}^{-1} - \tau_\beta^{-1} + \tau_{\rm cc}^{-1} + \tau_\kappa^{-1}|^{-1}$. In this expression, we assume a simple form for the support provided by galactic shear against gravitational collapse, so that $\tau_\beta^{-1}$ is subtracted from $\tau_{\rm ff,g}^{-1}$.

\begin{figure*}
  \label{Fig::betaQ_timescales}
    \includegraphics[width=\linewidth]{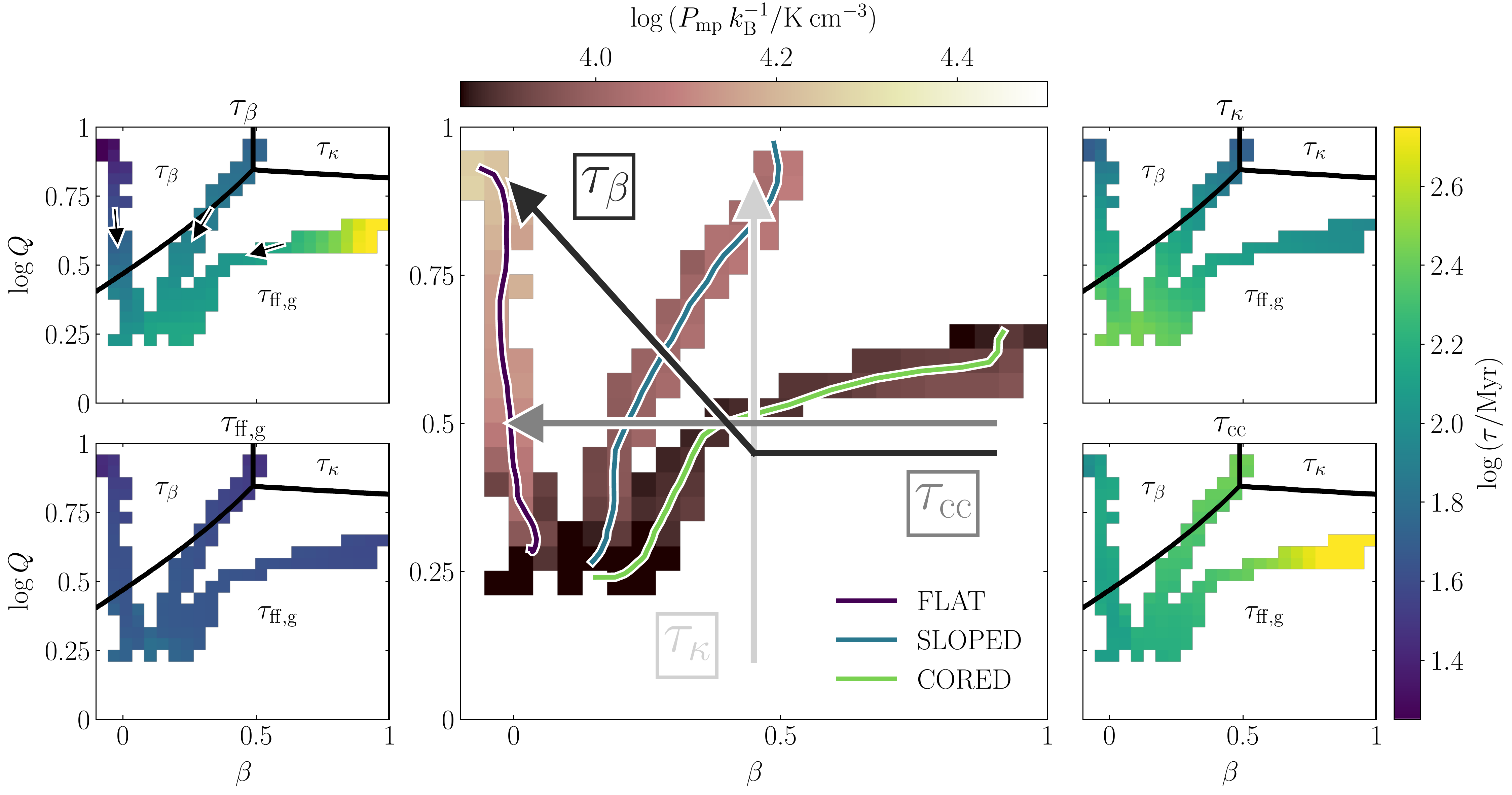}
    \caption{Mid-plane pressure ($P_{\rm mp}$, centre) and dynamical time-scales for galactic shear ($\tau_\beta$, upper left), gravitational free-fall ($\tau_{\rm ff,g}$, lower left), orbital excursions ($\tau_\kappa$, upper right) and cloud-cloud collisions ($\tau_{\rm cc}$, lower right), as a function of the shear parameter $\beta$ and the Toomre $Q$ stability parameter, i.e.~across the galactic-dynamical parameter space of~\protect\cite{Jeffreson+Kruijssen18}. The black arrows mark the direction in which the galactocentric radius increases. All quantities are compiled over the time interval from $600$~Myr to $1$~Gyr, and the time-scales are defined quantitatively in Table~\ref{Tab::time-scales}. The central panel additionally shows a schematic for the direction in which each dynamical rate {\it increases} (time-scale {\it decreases}), so becomes {\it more likely} to influence GMC evolution, along with solid lines denoting the positions of the FLAT, SLOPED and CORED galaxies.}
\end{figure*}

\subsection{Choice of Toomre $Q$ parameter} \label{Sec::theory-Toomre-Q}
In this work, we use the Toomre $Q$ stability parameter associated with the gaseous component of the interstellar medium, and do not include the influence of the stellar component as described in~\cite{Elmegreen95}. We do this for two reasons. Practically, we use an external background potential in our simulations to model the gravitational force due to the stellar disc and stellar bulge (see Section~\ref{Sec::ext-ptnl}). The live stellar particles formed during the simulation make up only 6~per~cent of the stellar mass, at maximum. We are therefore unable to obtain an accurate estimate of the stellar velocity dispersion from our simulations. Physically, the use of a background potential also means that the stellar component cannot respond dynamically to the growth of gravitational instabilities, such that the gas-only Toomre Q parameter is the best quantification of dynamical instability in our simulations. Both of the above considerations make the gas-only Toomre $Q$ the natural choice to compare the analytic theory presented here to the numerical results presented in Section~\ref{Sec::results-properties}.

\subsection{Simulated galaxies in galactic-dynamical parameter space} \label{Sec::dynamical-span}
Our galaxies resemble the Milky Way in their total masses ($\sim 10^{12}$~M$_\odot$), gas masses ($\sim 6 \times 10^9$~M$_\odot$), and total gas disc scale-lengths ($\sim 7$~kpc), but cover a range of different galactic-dynamical environments. Galactic-dynamical variation is induced via differences in the external gravitational potential (see Table~\ref{Tab::params}), leading to variations in the circular velocities and scale-heights of the gas discs. Figure~\ref{Fig::betaQ_explore} presents the span of our simulations within the galactic-dynamical parameter space of~\cite{Jeffreson+Kruijssen18}. The two left-hand panels show the azimuthally- and temporally-averaged values of the circular velocity (top) and of the cold-gas scale-height (bottom) for each simulation, as a function of the galactocentric radius $R$. These profiles are computed in $50$ overlapping environmental bins of width $1$~kpc, between galactocentric radii of $R = 1$~kpc and $R = 13$~kpc. The temporal standard deviation is indicated by the translucent shaded regions. In the two right-hand panels, we show the azimuthally- and temporally-averaged values of the parameters $\beta$ and $Q$ for each galaxy as solid lines of matching colour. We achieve the mapping $R \rightarrow (\beta, Q)$ by constructing the azimuthally-averaged profiles of $\beta$ and $Q$ for each galaxy, using the same $50$ overlapping bins. We then plot $Q$ as a function of $\beta$, where the black arrows in the central panel denote the direction in which $R$ increases. Further details on the computation of each radial profile are given in Appendix~\ref{App::analysis}.

The locus of the simulated GMC population on the time interval from $600$~Myr to $1$~Gyr is indicated by the spread of $(\beta, Q)$ bins behind each solid line. We sample both the cloud population and the radial profiles of $\beta$, $Q$, $\Omega$ and $\phi_{\rm P}$ at time intervals of $50$~Myr, treating each snapshot as a distinct population of clouds.\footnote{In a companion paper~\cite{Jeffreson20}, we study the time evolution of GMCs and find that their lifetimes have a maximum value of $\sim 40$-$50$~Myr. As such, we can be confident that there are very few duplicate clouds in our sample.} We calculate the position of each cloud in the parameter space by interpolating the smooth radial profiles of the four environmental variables $\beta$, $Q$, $\Omega$ and $\phi_{\rm P}$ at the galactocentric radius of the cloud centre of mass. Via this method, we retrieve a cumulative total of $\sim 80,000$ molecular clouds and $\sim 55,000$ HI clouds across the three simulations. The spread of the environmental bins behind each temporally-averaged solid line arises due to the small but significant temporal variation in the Toomre $Q$ profile (recall Figure~\ref{Fig::radial-profiles}). On the scale of $1$~kpc used to compute the environmental variables, the azimuthal variation in the value of the Toomre $Q$ parameter is negligible relative to its temporal variation, owing to the approximate axisymmetry of our simulated discs. In Figure~\ref{Fig::betaQ_explore} and in every following appearance of the parameter space, we show only those bins that contain $\ge 100$ GMCs, ensuring that a sufficiently-large distribution of clouds is present in each galactic-dynamical environment to reliably compute a mean and a standard deviation for each cloud property. We note that for the FLAT and CORED simulations, we take stricter minimum radii of $R=2$~kpc and $R=3$~kpc for cloud identification, respectively, indicated by the vertical lines that cut through the radial profiles in Figures~\ref{Fig::radial-profiles} and~\ref{Fig::betaQ_explore}. We do this to exclude the ring of zero star-formation at $R \sim 1$~kpc in the FLAT simulation, and to exclude the very low inner surface densities for the CORED simulation.

The colours of the pixels in the two right-hand panels of Figure~\ref{Fig::betaQ_explore} correspond to the mean values of the $\phi_{\rm P}$ parameter (blue pixels) and the orbital angular velocity $\Omega$ (pink pixels). Together, we see that the simulated galaxies span approximately an order of magnitude in the Toomre $Q$ stability parameter ($\log{Q} \in [0.2, 1]$), the angular velocity ($\log{(\Omega/{\rm Myr})} \in [-1.75, -0.75]$), and the parameter $\phi_{\rm P} \in [1, 9]$, and cover the full range of galactic shear parameters from $\beta \sim 0$ (flat rotation curve) to $\beta = 1$ (solid-body rotation). We note that the parameter $\phi_{\rm P}$ appears only in the time-scale for gravitational free-fall, to the power $1/2$. In determining the regions of parameter space for which each dynamical time-scale is minimum (enclosed by the solid black lines), we therefore set $\phi_{\rm P} \sim 3$, corresponding to its environment-averaged mean value.

Figure~\ref{Fig::betaQ_explore} shows that both $\Omega$ and $\phi_{\rm P}$ increase monotonically with the Toomre $Q$ parameter. The relation between $Q$ and $\Omega$ is almost linear, because the kpc-scale velocity dispersion and surface density of the gas disc are roughly constant with galactocentric radius (see Section~\ref{Sec::MW-comparison-disc-stab}), leaving the gravitational stability to vary with the degree of centrifugal support~\citep{Toomre64}, which is proportional to $\Omega$. This degeneracy between $\Omega$ and $Q$ has the important consequence that the ratio $Q/\Omega$ is roughly constant across all simulated environments, such that the time-scale $\tau_{\rm ff,g}$ varies only by a factor of two from $25$ up to $50$~Myr across galactic-dynamical parameter space.

\begin{figure*}
\label{Fig::correlation-summary}
  \includegraphics[width=\linewidth]{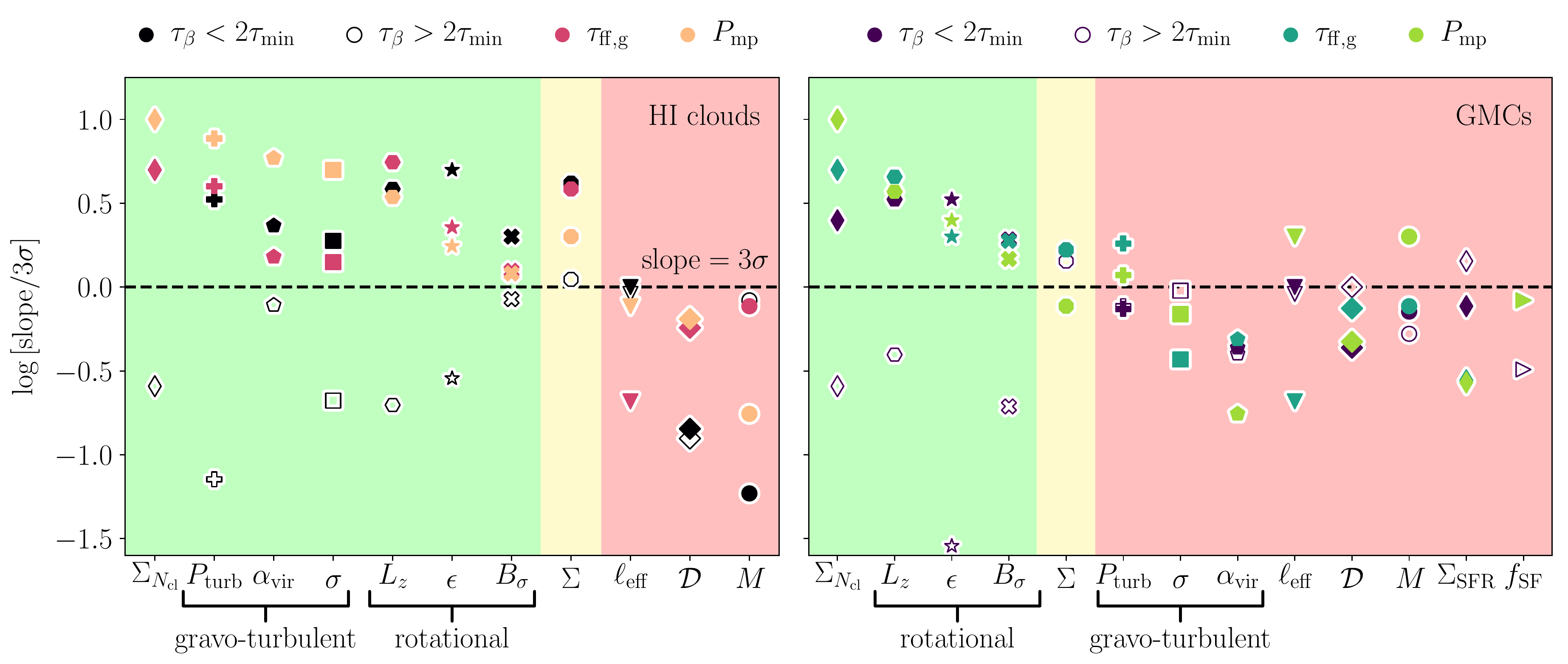}
  \caption{Summary of the correlations investigated in this work between galactic-dynamical quantities (top legend) and the properties ($x$-axis) of HI clouds (left) and GMCs (right). The $y$-axis gives the slope of each correlation, divided by its $3\sigma$ variation. The dashed line divides the trends that are significant on the $3\sigma$ level (above) from those that are not (below). The clouds are ordered so that the net significance of the dynamical correlations decreases towards the right. For both types of clouds, the green shaded region highlights the cloud properties that display galactic-dynamical correlations with $\tau_\beta$, $\tau_{\rm ff,g}$ and $P_{\rm mp}$ at the $3\sigma$ significance level. The red shaded region highlights the cloud properties that display significant correlations with at most one of these time-scales. The yellow-shaded region highlights cloud properties that display significant correlations, but over a very small dynamic range ($<0.05$~dex for the HI cloud and GMC surface densities). Correlations with the time-scale $\tau_\beta$ for galactic shearing are divided into two distinct regimes: one in which the shear time-scale is less than twice the minimum evolutionary time-scale (filled points), and one in which it is greater than twice the minimum evolutionary time-scale (unfilled points). We note that for cloud properties with significant galactic-dynamical trends in $\tau_\beta$, $\tau_{\rm ff,g}$ and $P_{\rm mp}$, there exists a clear break between the two regimes.}
\end{figure*}

\begin{table*}
\begin{center}
\label{Tab::results}
  \caption{Summary of the environmental trends in GMC properties that are discussed in Section~\ref{Sec::results-properties}. For each cloud property, we give the global mean value, the span (lowest-highest) of mean values across the $(\beta, Q, \Omega, \phi_{\rm P})$ parameter space, and the environmental trends followed by each property, when the correlation is significant at the $3\sigma$ confidence level. In the two right-most columns we give the Section in which each GMC property is discussed, along with the Figure showing its galactic-dynamical trend.}
  \begin{tabular}{@{}l|l||c|c||c|c||c|c||c|c @{}}
  \hline
   Cloud property/unit & Symbol & \multicolumn{2}{c||}{Mean} & \multicolumn{2}{c||}{Span of mean values} & \multicolumn{2}{c||}{$3 \sigma$ correlation} & Sec. & Fig. \\

  \hline
    & & ${\rm H}_2$ & HI & ${\rm H}_2$ & HI & ${\rm H}_2$ & HI & \\
  \hline
    Mass/$10^5 {\rm M}_\odot$ & $M$ & $1.3$ & $0.65$ & $1.0 \rightarrow 1.7$ & $0.49 \rightarrow 1.3$ & \cellcolor{red!25}None & \cellcolor{red!25}None & \ref{Sec::cloud-mass-and-size} & $-$ \\

    Size/pc & $\ell_{\rm eff}$ & $24$ & $65$ & $22 \rightarrow 26$ & $51 \rightarrow 80$ & \cellcolor{red!25}None & \cellcolor{red!25}None & \ref{Sec::cloud-mass-and-size} & $-$ \\

    Surface density/${\rm M}_\odot {\rm pc}^{-2}$ & $\Sigma$ & $226$ & $16$ & $202 \rightarrow 254$ & $14 \rightarrow 18$ & \multicolumn{2}{c||}{\cellcolor{yellow!25}Marginal} & \ref{Sec::gravity-turbulence} & \ref{Fig::turbulent-properties} \\

    Velocity dispersion/${\rm kms}^{-1}$ & $\sigma$ & $3.3$ & $6.7$ & $2.8 \rightarrow 3.8$ & $5.0 \rightarrow 8.1$ & \cellcolor{red!25}None & \cellcolor{green!25}$\tau_\beta$, $\tau_{\rm ff,g}$, $P_{\rm mp}$ & \ref{Sec::gravity-turbulence} & \ref{Fig::turbulent-properties} \\

    Virial parameter & $\alpha_{\rm vir}$ & $1.2$ & $22$ & $0.9 \rightarrow 1.5$ & $12 \rightarrow 32$ &\cellcolor{red!25}None & \cellcolor{green!25}$\tau_\beta$, $\tau_{\rm ff,g}$, $P_{\rm mp}$ & \ref{Sec::gravity-turbulence} & \ref{Fig::turbulent-properties} \\

    Turbulent pressure/$10^5 k_{\rm B} \: {\rm K} {\rm cm}^{-3}$ & $P_{\rm turb}$ & $2.3$ & $0.21$ & $1.9 \rightarrow 3.6$ & $0.14 \rightarrow 0.36$ & \cellcolor{red!25}None & \cellcolor{green!25}$\tau_\beta$, $\tau_{\rm ff,g}$, $P_{\rm mp}$ & \ref{Sec::gravity-turbulence} & \ref{Fig::turbulent-properties} \\

    Velocity divergence/${\rm kms}^{-1}$ & $\mathcal{D}$ & $-1.0$ & $-0.2$ & $-1.3 \rightarrow -0.6$ & $-1.6 \rightarrow 0.6$ & \cellcolor{red!25}None & \cellcolor{green!25}$\sigma_{\rm mp}$ & \ref{Sec::cloud-divergence} & $-$ \\

    Aspect ratio & $\epsilon$ & $2.3$ & $2.3$ & $1.9 \rightarrow 3.2$ & $1.7 \rightarrow 3.1$ & \multicolumn{2}{c||}{\cellcolor{green!25}$\tau_\beta$, $\tau_{\rm ff,g}$, $P_{\rm mp}$} & \ref{Sec::shear-properties} & \ref{Fig::betaQ_ellipticity} \\

    Angular momentum/${\rm pc}\:{\rm kms}^{-1}$ & $L_z$ & $4.2$ & $34$ & $3 \rightarrow 9$ & $17 \rightarrow 101$ & \multicolumn{2}{c||}{\cellcolor{green!25}$\tau_\beta$, $\tau_{\rm ff,g}$, $P_{\rm mp}$} & \ref{Sec::shear-properties} & \ref{Fig::betaQ_angmom} \\

    Velocity anisotropy & $B_\sigma$ & $-0.5$ & $-0.2$ & $-0.9 \rightarrow -0.3$ & $-0.4 \rightarrow 0.1$ & \multicolumn{2}{c||}{\cellcolor{green!25}$\tau_\beta$, $\tau_{\rm ff,g}$, $P_{\rm mp}$} & \ref{Sec::shear-properties} & \ref{Fig::betaQ_anisotropy} \\

    No.~clouds per unit area/${\rm kpc}^{-2}$ & $\Sigma_{N_{\rm cl}}$ & $7.4$ & $5$ & $2.9 \rightarrow 19$ & $2.4 \rightarrow 18$ & \multicolumn{2}{c||}{\cellcolor{green!25}$\tau_\beta$, $\tau_{\rm ff,g}$, $P_{\rm mp}$} & \ref{Sec::cloud-star-formation} & \ref{Fig::SFR_vs_P-mp} \\

    SFR surface density/${\rm M}_\odot {\rm kpc}^{-2} {\rm yr}^{-1}$ & $\Sigma_{{\rm SFR}, {\rm cl}}$ & $0.63$ & $-$ & $0.36 \rightarrow 0.89$ & $-$ & \cellcolor{red!25}None & $-$ & \ref{Sec::cloud-star-formation} & \ref{Fig::Pturb_vs_SFR} \\

    Fraction star-forming clouds & $f_{\rm SF}$ & $32$\% & $-$ & $24$\% $\rightarrow$ $43$\% & $-$ & \cellcolor{red!25}None & $-$ & \ref{Sec::cloud-star-formation} & \ref{Fig::Pturb_vs_SFR} \\
\end{tabular}
\end{center}
\end{table*}

\subsection{Simulated galaxies and their galactic-dynamical time-scales} \label{Sec::timescale-span}
In Figure~\ref{Fig::betaQ_timescales} we show the variation in the galactic-dynamical time-scales for galactic shear ($\tau_\beta$, top left panel), gravitational free-fall ($\tau_{\rm ff,g}$, bottom left panel), epicyclic perturbations ($\tau_\kappa$, top right panel) and cloud-cloud collisions ($\tau_{\rm cc}$, bottom right panel) across the environments spanned by our three simulations. The direction in which each time-scale decreases in value (and so the rate of the associated dynamical process increases) is indicated by the arrows in the central panel. We can make the following key observations.
\begin{enumerate}
  \item The galactic-dynamical environments spanned by our simulations are partitioned between two regimes: a `gravity-dominated regime' for which the gravitational free-fall time-scale $\tau_{\rm ff,g}$ is the shortest dynamical time-scale, and a `shear-dominated regime' for which the time-scale $\tau_\beta$ for galactic shearing is the shortest.
  \item $\tau_{\rm ff,g}$ is always close in value to the shortest dynamical time-scale, even in environments for which $\tau_\beta$ is the shortest time-scale.
  \item $\tau_{\rm ff,g}$ varies over a small dynamic range of just $\sim 0.4$~dex across our Milky Way-pressured environments.
  \item $\tau_\kappa$ and $\tau_{\rm cc}$ are around an order of magnitude longer than the free-fall time-scale across all simulated environments, and so epicyclic perturbations and cloud-cloud collisions are not likely to be significant drivers of cloud properties, relative to gravitational free-fall and galactic shearing.
\end{enumerate}
The time-scale for gravitational free-fall is short across our simulated environments, but has a small dynamic range. In the following, we also consider the mid-plane hydrostatic pressure $P_{\rm mp} = \pi G/2 \: \phi_{\rm P} \Sigma_{\rm g}^2$~\citep[see][]{Elmegreen89} and~\cite{BlitzRosolowsky2004}, which is closely related to the free-fall time, as
\begin{equation} \label{Eqn::freefall-time-midplane-pressure}
\tau_{\rm ff,g} = \sigma_{\rm g} \sqrt{\frac{3\pi}{32G P_{\rm mp}}},
\end{equation}
but has a greater dynamic range across our sample of galactic environments. The variation in $P_{\rm mp}$ across the galactic-dynamical parameter space of~\cite{Jeffreson+Kruijssen18} is shown in the central panel of Figure~\ref{Fig::betaQ_timescales}.

\section{Galactic-dynamical trends in cloud properties} \label{Sec::results-properties}
In this section, we analyse the physical properties of the GMC and HI cloud populations across the FLAT, SLOPED and CORED galaxies at simulation times from $600$~Myr to $1$~Gyr. We consider the variation of the mean for each property as a function of its galactic-dynamical environment in the parameters $\beta$, $Q$, $\Omega$ and $\phi_{\rm P}$, demonstrating the interplay between stabilising dynamical influences (galactic rotation and pressure) and de-stabilising dynamical influences (gravity) in driving the evolution of clouds. We find that the variation of GMC and HI cloud properties across this parameter space indicates the presence of statistically-significant correlations between these properties and the key galactic-dynamical time-scales $\tau_\beta$ and $\tau_{\rm ff,g}$, as well as with the mid-plane pressure $P_{\rm mp}$.

\subsection{Overview of galactic-dynamical correlations} \label{Sec::results-summary}
In Figure~\ref{Fig::correlation-summary} and Table~\ref{Tab::results}, we summarise the galactic-dynamical correlations between all physical cloud properties analysed in this work and the dynamical variables $\tau_\beta$, $\tau_{\rm ff,g}$ and $P_{\rm mp}$. The correlations themselves are presented in Appendix~\ref{App::stat-sig}, where we also describe in detail the procedures used for constraining the slope of each relationship. We find that at the $3\sigma$ confidence level, statistically-significant correlations are present in all three variables for $7/11$ HI cloud properties, and for $4/13$ GMC properties, shaded green in Figure~\ref{Fig::correlation-summary}. The $y$-axis in both panels gives the slope of the best-fit relationship between variables, divided by the $3\sigma$ variation on this slope. The statistically-significant correlations are therefore given by the points that fall above the dashed line. The red-shaded regions highlight the cloud properties that display statistically-significant correlations with fewer than three of the dynamical variables (none, in most cases). The yellow-shaded region highlights the cloud surface density $\Sigma$, which displays a significant correlation in all three variables for HI clouds, but over a very small dynamic range ($< 0.05$~dex). Figure~\ref{Fig::correlation-summary} demonstrates the following three general results for our simulated cloud population.
\begin{enumerate}
  \item HI clouds display galactic-dynamical correlations in their gravo-turbulent cloud properties (internal turbulent pressures $P_{\rm turb}$, virial parameters $\alpha_{\rm vir}$ and velocity dispersions $\sigma$), but GMCs do not.
  \item Both HI clouds and GMCs display statistically-significant galactic-dynamical correlations in their rotational cloud properties (aspect ratios $\epsilon$, specific angular momenta $L_z$ and velocity anisotropies $B_{\sigma}$).
  \item There exist two distinct regimes for the time-scale $\tau_\beta$ for galactic shearing. The unfilled points in both panels represent the statistical significance of correlations between each physical cloud property and the shear time-scale, in galactic environments for which the shear time-scale is long (more than twice the length of the minimum dynamical time-scale, $\tau_\beta > 2\tau_{\rm min}$). For the dynamically-correlated cloud properties (green shaded regions in Figure~\ref{Fig::correlation-summary}) the break in the slope of the correlation between the two regimes manifests itself as a division of filled points ($\tau_\beta < 2\tau_{\rm min}$) and unfilled points ($\tau_\beta > 2\tau_{\rm min}$) across the dashed line. By contrast, the time-scale for gravitational free-fall $\tau_{\rm ff,g}$ has a very small dynamic range and so always remains close in value to the minimum dynamical time-scale. No such break is seen for dynamical correlations with $\tau_{\rm ff,g}$.
\end{enumerate}
In the following sections we examine each cloud property in detail, and so shed light on the galactic-dynamical trends in the gravo-turbulent and rotational properties of GMCs and HI clouds.

\begin{figure}
\label{Fig::cloud-spectra}
  \includegraphics[width=\linewidth]{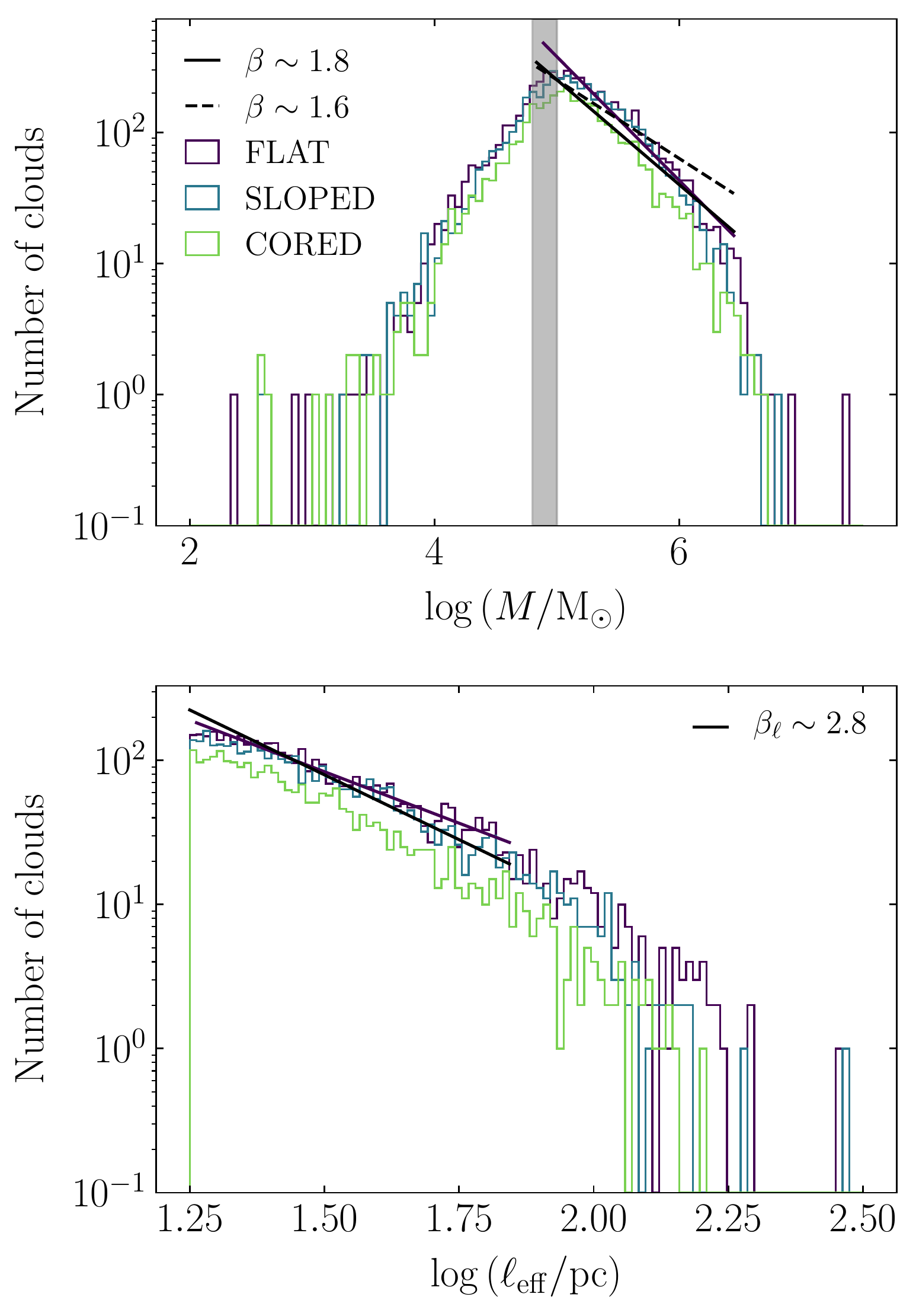}
  \vspace{-0.5cm}
  \caption{\textit{Upper panel:} Mass distribution of the GMCs identified in each simulated galaxy at $t = 600$~Myr. The solid black and dashed black lines denote the range of power-law slopes for the observed cloud mass distribution in the Milky Way, given by $\dd N/\dd M \propto M^{-\beta}$ with $\beta \in [1.6,1.8]$. The purple line gives the power-law fit to mass spectrum of the FLAT simulation, corresponding to $\beta = 1.94 \pm 0.05$. The grey shaded region indicates the range of observed values for the turnover of the mass spectrum at $M \sim 10^{4.8}$-$10^5$~M$_\odot$. \textit{Lower panel:} Size distribution of GMCs at $t = 600$~Myr. The black line denotes the power-law slope of the observed cloud size distribution in the Milky Way, given by $\dd N/\dd M \propto M^{-\beta_\ell}$ with $\beta_\ell \sim 2.8$. The purple line gives the power-law fit to the size spectrum of the FLAT simulation, corresponding to $\beta_\ell = 2.43 \pm 0.06$.}
\end{figure}

\begin{figure*}
\label{Fig::turbulent-properties}
  \includegraphics[width=\linewidth]{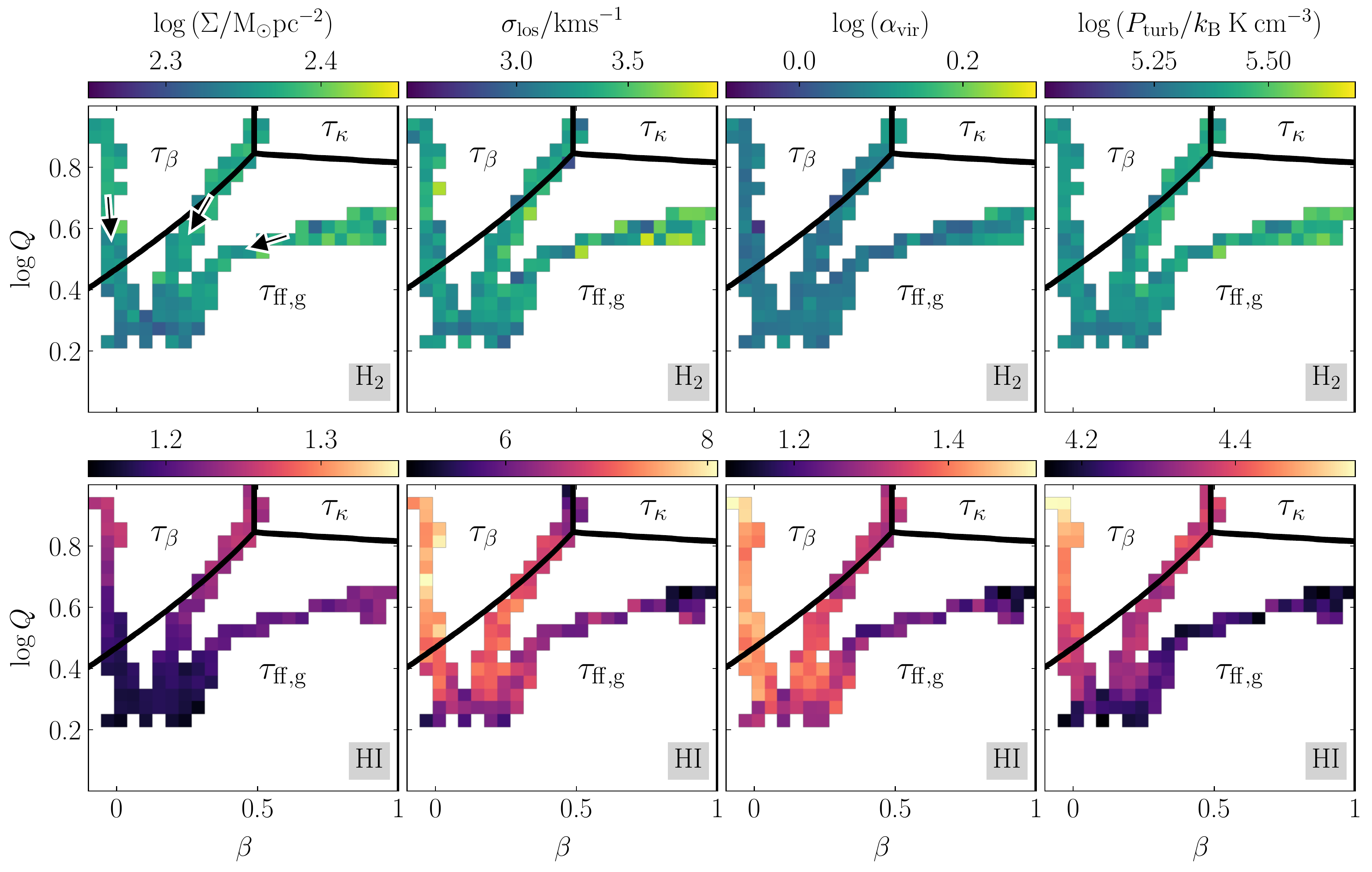}
  \caption{Mean surface density $\Sigma$ (left), line-of-sight velocity dispersion $\sigma_{\rm los}$ (centre-left), virial parameter $\alpha_{\rm vir}$ (centre-right), and turbulent pressure $P_{\rm turb}$ (right) of the GMCs (upper panels) and HI clouds (lower panels) in our simulations, as a function of the shear parameter $\beta$ and the Toomre $Q$ stability parameter. Each distinct set of connected pixels corresponds to the total cloud population of one isolated disc galaxy, compiled across the simulation time interval from $600$~Myr to $1$~Gyr, at a sampling interval of $50$~Myr. The color of each pixel represents the mean value of the relevant cloud property at the indicated value of $(\beta, Q)$. The black arrows in the upper left panel mark the direction in which the galactocentric radius increases. The black solid lines enclose the regions of parameter space for which the minimum dynamical time-scale is $\tau_\kappa$ (orbital epicyclic perturbations), $\tau_\beta$ (galactic shear) and $\tau_{\rm ff,g}$ (gravitational free-fall).}
\end{figure*}

\subsection{Mass and size} \label{Sec::cloud-mass-and-size}
The masses and sizes of GMCs and HI clouds in our simulations are not significantly correlated with the galactic dynamical environment. Their mean values are approximately invariant under changes in the galactic-dynamical time-scales $\tau_\beta$ and $\tau_{\rm ff,g}$, and in the mid-plane pressure $P_{\rm mp}$. On average, the molecular clouds are around three times smaller but two times more massive than the HI clouds in each environment, with a global mean diameter of $24$~pc (relative to $65$~pc for HI clouds) and a global mean mass of $1.3 \times 10^5$~M$_\odot$ (relative to $0.65 \times 10^5$~M$_\odot$ for HI clouds). In fact, it is plausible that the mean GMC diameter would appear even smaller at a higher simulation resolution, given that its average value is close to the minimum value of $18$~pc enforced by our GMC identification criterion (see Section~\ref{Sec::cloud-ID}). The small mean size and high mean volume density of the identified GMCs relative to the HI clouds is consistent with the idea that in Milky Way-pressure galaxies like ours, the GMCs are high-density `iceberg tips' poking up above the CO emissivity threshold.

In the upper panel of Figure~\ref{Fig::cloud-spectra}, we demonstrate that the mass distribution of GMCs in each galaxy reproduces the upper limit of $\sim 3$ to $8 \times 10^6$~M$_\odot$ observed by~\cite{Rosolowsky03} in M33, by~\cite{Freeman17} in M83 and by~\cite{Miville-Deschenes17} and~\cite{Colombo+19} in the Milky Way. This upper limit has been predicted to arise due to a combination of centrifugal forces and stellar feedback~\citep{Reina-Campos17}. We also find a turnover in the mass spectrum between $10^{4.8}$ and $10^5$~M$_\odot$, consistent with the behaviour of the GMC mass distribution in the Milky Way~\citep{Miville-Deschenes17}, although we cannot rule out the possibility that the turnover we see in the simulations is influenced by their limited mass resolution. Above the turnover, the GMC mass function has a power-law form with $\beta \sim 1.9$, close to the observed range of $\beta \in [1.6, 1.8]$ for clouds in the Milky Way~\citep{Solomon87,Williams&McKee97,Heyer+09,Roman-Duval+10,Miville-Deschenes17,Colombo+19} over the same mass range ($\log{M} \in [4.8, 6.5]$). In the lower panel of Figure~\ref{Fig::cloud-spectra}, we display the spectrum of GMC sizes for each simulated galactic disc, given by the effective cloud radius $\ell_{\rm eff}$, such that
\begin{equation}
\ell_{\rm eff} = 1.91 \sqrt{\Delta \ell_{\rm maj}^2 + \Delta \ell_{\rm min}^2},
\end{equation}
where $\Delta \ell_{\rm maj}$ and $\Delta \ell_{\rm min}$ are the second moments of an ellipse fitted to the footprint of each cloud in the galactic mid-plane, using {\sc Astrodendro}. We adopt this definition of the cloud size in order to make a direct comparison to works in the existing observational literature~\citep[e.g.][]{Solomon87,BertoldiMcKee1992,RosolowskyLeroy06,Colombo+19}. The factor of $1.91$ is the correction first defined by~\cite{Solomon87} for converting the RMS cloud extent to an estimate of the spherical cloud size. The smallest resolved cloud has a diameter of $18$~pc, so we do not capture the observed turnover of the distribution at $\sim 30$~pc~\citep{Miville-Deschenes17}. Likewise, our largest clouds slightly exceed the truncation size of $70$~pc observed by~\cite{Colombo+19}, with a maximum diameter of $\sim 200$~pc. Importantly, we do approximately reproduce the observed power-law slope of $\dd N/\dd R \sim R^{-\beta_\ell}$ with $\beta_\ell \sim 2.8$~\citep{Colombo+19}. This is given by the black line in Figure~\ref{Fig::cloud-spectra}, while our best fit to the simulation data over the observed range of cloud sizes $\ell_{\rm eff} \in [18,70]$~pc is given by the purple line, with a slightly shallower slope of $\beta_\ell = 2.43 \pm 0.06$.

\begin{figure}
\label{Fig::scaling-rln}
  \includegraphics[width=\linewidth]{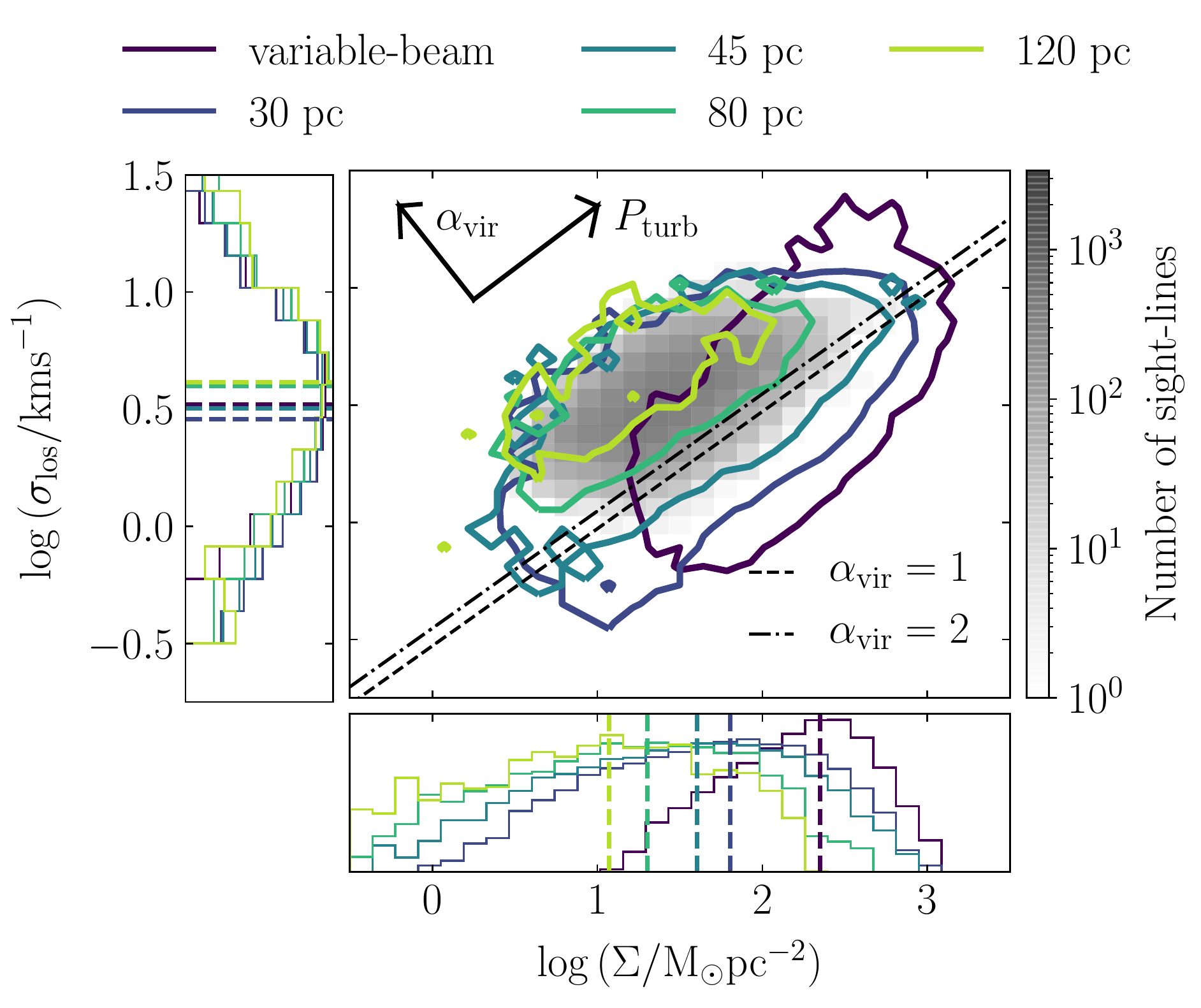}
  \vspace{-0.5cm} \caption{Relationship between the molecular gas surface density $\Sigma$ and the line-of-sight velocity dispersion $\sigma_{\rm los}$ for the combined GMC populations of our three simulations between simulation times of $600$~Myr and $1$~Gyr, sampled at $50$~Myr intervals. The dark purple contour encloses 90~per~cent of data points for GMCs identified using {\sc Astrodendro}, as described in Section~\ref{Sec::cloud-ID}. The remaining contours enclose 90~per~cent of the data points for molecular gas `sight-lines' at fixed spatial scales of $30$, $45$, $80$ and $120$~pc, following the cloud-identification method described in~\protect\cite{Sun18}. The grey-shaded histogram in the central panel contains the observational data at $120$ pc-scale from~\protect\cite{Sun18}, combining the results from the flocculent Milky Way-mass galaxies NGC 628, NGC 2835, NGC 5068, NGC 4254 and NGC 6744. In the lower- and side-panels, the distributions of $\Sigma$ and $\sigma_{\rm los}$ are shown on a logarithmic $y$-axis scale, where the median values at each beam resolution are given by the dashed lines.}
\end{figure}

\subsection{Cloud self-gravity and turbulence} \label{Sec::gravity-turbulence}
Observationally, giant molecular clouds living in similar galactic environments exhibit a tight correlation between their surface densities $\Sigma$ and their line-of-sight turbulent velocity dispersions $\sigma_{\rm los}$~\citep[e.g.][]{Larson1981,Heyer+09,Longmore+13,Leroy17b,Sun18,Colombo+19}. For each GMC or HI cloud in our simulations, we define these quantities as
\begin{equation} \label{Eqn::surfdens}
\Sigma_x = \frac{\sum_i^N{m_{i,x}}}{A_x},
\end{equation}
and
\begin{equation}
\sigma_{{\rm los}, x} = \frac{\sigma_x}{\sqrt{3}} = \sqrt{\frac{\langle |\mathbf{v}_i-\langle \mathbf{v}_i \rangle_{x}|^2 \rangle_{i,x}}{3}},
\end{equation}
where $x = \{{\rm H}_2, {\rm HI}\}$. That is, $\{m_{i, x}\}$ are the masses of ${\rm H}_2$ or ${\rm HI}$ in the gas cells $i=1...N$ of each cloud, $\{\mathbf{v}_i\}$ are the velocities of the gas cell centroids, $A_x$ is the pixel-by-pixel area of the cloud's footprint on the galactic mid-plane, and $\langle ... \rangle_{x}$ denotes a mass-weighted average. The exact position of each cloud in the $\Sigma$-$\sigma_{\rm los}$ plane probes its virial parameter
\begin{equation} \label{Eqn::virial-param}
\alpha_{\rm vir} = \frac{5 \sigma_{\rm los}^2}{G\sqrt{\pi M \Sigma}},
\end{equation}
and turbulent pressure
\begin{equation} \label{Eqn::P-turb}
P_{\rm turb} \approx \rho \sigma_{\rm los}^2 \approx \frac{\Sigma \sigma_{\rm los}^2}{L},
\end{equation}
which provide insight into the cloud's state of self-gravitational and external confinement, respectively~\citep[see also][]{Sun18}. We have adopted the standard definition of the virial parameter from~\cite{MacLaren1988,BertoldiMcKee1992}, and have assumed a spherical cloud\footnote{We use the standard definition of the virial parameter to allow for direct comparison to the existing literature, but our GMCs and HI clouds are ellipsoidal rather than spherical, with an average aspect ratio of $\epsilon \sim 2.3$ (see Table~\ref{Tab::results}). According to Figure 2 and Equation (21) of~\cite{BertoldiMcKee1992}, the virial parameter is reduced by a factor of $0.8$ for an aspect ratio of $\epsilon \sim 2.3$. Each cloud is slightly more tightly-bound than the presented values would indicate} of radius $L$ to arrive at the right-hand side of Equation~(\ref{Eqn::P-turb}). A collapsing molecular cloud that is over-pressured relative to its ambient environment must be gravitationally-bound, with $\alpha_{\rm vir} \la 2$, though we emphasise that this threshold is approximate, as the virial parameter is an approximate measure of gravitational boundedness~\citep[e.g.][]{Mao19}. Conversely, a super-virial cloud with $\alpha_{\rm vir} \ga 2$ may still be confined if its internal turbulent pressure $P_{\rm turb}$ is exceeded by that of the surrounding medium. To discern whether or not a cloud is confined, measurements both of $\alpha_{\rm vir}$ and of $P_{\rm turb}$ are required.

\subsubsection{The dynamical decoupling of GMCs relative to HI clouds}
Figure~\ref{Fig::turbulent-properties} demonstrates that the molecular clouds (top row) in our simulations display a much lesser degree of environmental variation in their gravo-turbulent properties than do the HI clouds (bottom row). On average, they have surface densities that are ten times smaller than those of the GMCs ($\Sigma_{\rm HI} \sim 16$~M$_\odot$~${\rm pc}^{-2}$ vs. $\Sigma_{\rm H_2} \sim 226$~M$_\odot$~${\rm pc}^{-2}$), and velocity dispersions that are two times higher ($\sigma_{\rm los, HI} \sim 6.7 \: {\rm km} \: {\rm s}^{-1}$ vs. $\sigma_{\rm los, H_2} \sim 3.7 \: {\rm km} \: {\rm s}^{-1}$). This means that the turbulent pressures of the GMCs in our simulations are ten times larger than those of the HI clouds ($P_{\rm turb, H_2} \sim 2.3 \times 10^5 \: k_{\rm B} \: {\rm K} \: {\rm cm}^{-3}$ vs. $P_{\rm turb, HI} \sim 2.1 \times 10^4 \: k_{\rm B} \: {\rm K} \: {\rm cm}^{-3}$), and while the GMCs are gravitationally-bound across all galactic-dynamical environments ($\alpha_{\rm vir, H_2} \sim 1.2$), the HI clouds are far from gravitational equilibrium, with an average virial parameter of $\alpha_{\rm vir, HI} \sim 22$. The large difference in the mean pressures and densities of the GMCs relative to the HI clouds points to an explanation for the dichotomy in their degree of environmental variation: the molecular clouds in our Milky Way-pressured galaxies are too over-dense and over-pressured to be turbulently-coupled to the ambient medium.

The dynamical decoupling of our GMCs is best examined in the $\Sigma$-$\sigma_{\rm los}$ plane, and in the plane comparing the mid-plane hydrostatic pressure to the cloud turbulent pressure $P_{\rm mp}$-$P_{\rm turb}$, as studied in observations by~\cite{Sun2020}. In Figure~\ref{Fig::scaling-rln}, we compare the combined population of GMCs from all three simulations (coloured contours) to the observed molecular $\Sigma$-$\sigma_{\rm los}$ distribution of~\cite{Sun18} for five flocculent, bar-less Milky Way-mass galaxies, at $120$~pc-resolution (grey-shaded histogram). The GMCs identified by these authors are independent of any clump-finding algorithm; instead they are assumed to fill a `sight-line' of size equal to the observational resolution. Only those sight-lines with signal-to-noise ratios of ${\rm S}/{\rm N} \geq 5$ in two consecutive channels of CO emission are considered. We therefore produce ray-tracing maps of $\Sigma_{\rm H_2}$ at resolutions of $\Delta x = 120$~pc, $80$~pc, $45$~pc and $30$~pc, and identify all pixels with $\log{(\Sigma_{\rm H_2}/{\rm M}_\odot) {\rm pc}^{-2}} > -3.5$ as `GMC sight-lines'. This removes the low-level CO emission as depicted in Figure~\ref{Fig::DESPOTIC-threshold}, and we calculate the properties of each sight-line using the gas cells masked by each pixel, in the same way as for the `beam-filling' clouds of variable scale (i.e.~the clouds that we identify as isodensity contours in {\sc Astrodendro}). Figure~\ref{Fig::scaling-rln} demonstrates good agreement between simulations and observations for GMC sight-lines at a resolution of $120$~pc (light-green contour), but shows a systematic decrease in the median virial parameter as the resolution is increased, combined with a systematic increase in the turbulent pressure. The black arrows denote the direction of increasing value for each of these quantities in the $\Sigma$-$\sigma_{\rm los}$ plane. The virial parameter is lowest, and the turbulent pressure highest, for the beam-filling clouds (purple contour). The reason for this is revealed in the lower- and left-hand side panels: the molecular gas velocity dispersion is approximately independent of the resolution $\Delta x$, but the molecular gas surface density depends on the beam filling-factor, which scales as $(\Delta x)^{-2}$. This leads to an increase in the virial parameter and a decrease in the turbulent pressure, in proportion to $(\Delta x)^2$. While the sight-lines at $120$~pc have an average overdensity of just $\sim 10 \times$ relative to the ambient medium, the beam-filling GMCs have an overdensity of $\sim 100 \times$.

\begin{figure}
\label{Fig::P-DE_vs_Pturb}
  \includegraphics[width=\linewidth]{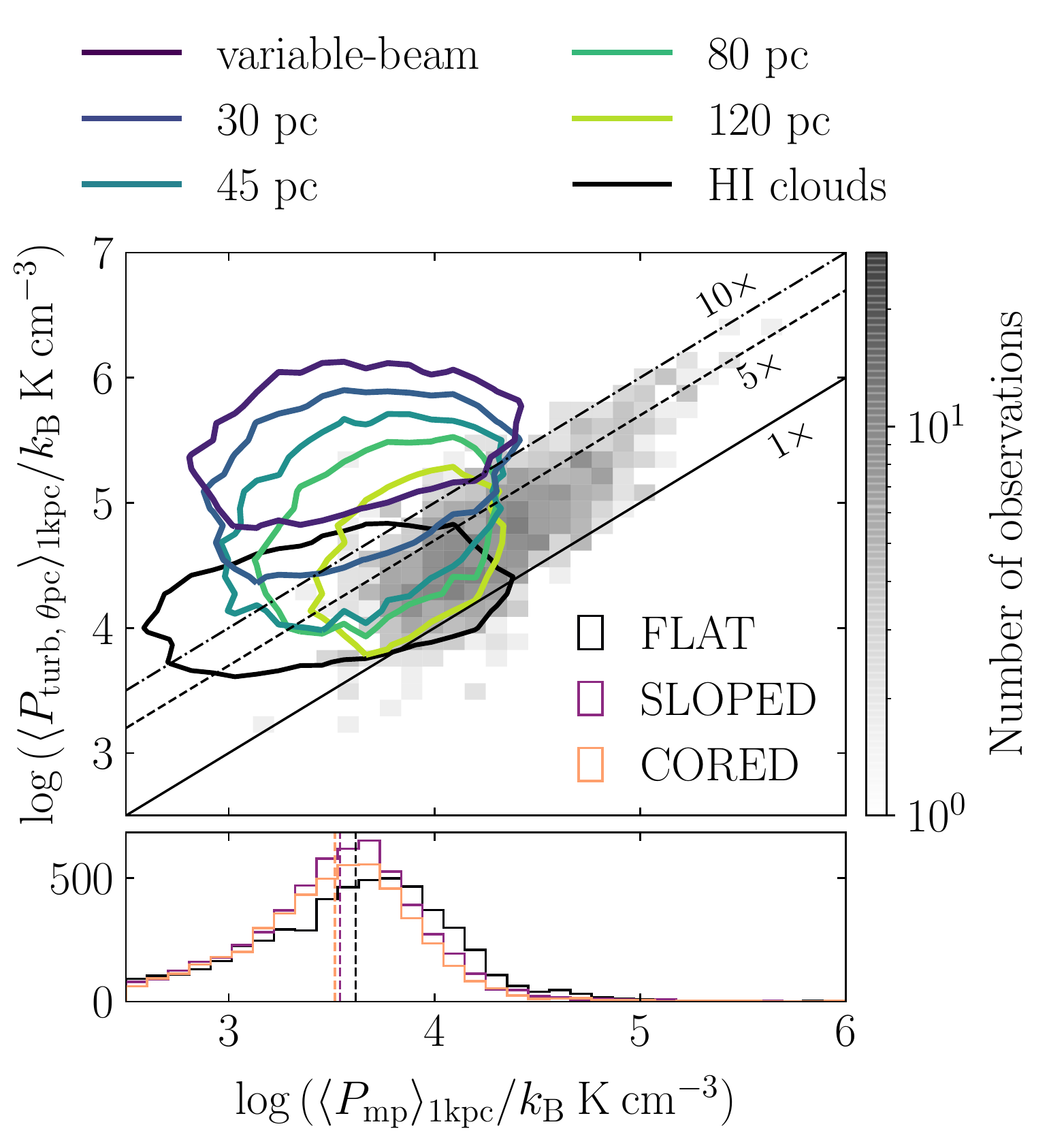}
  \vspace{-0.5cm} \caption{The kpc-scale averaged turbulent pressure $\langle P_{{\rm turb}, \: \theta {\rm pc}} \rangle_{1 {\rm kpc}}$ at varied spatial resolution $\theta$ for the combined GMC population across our three simulations at times between $600$~Myr and $1$~Gyr, as a function of the kpc-averaged mid-plane hydrostatic pressure, $\langle P_{\rm mp} \rangle_{1 {\rm kpc}}$. The dark purple contour encloses 90~per~cent of data points for molecular clouds identified using {\sc Astrodendro}, as described in Section~\ref{Sec::cloud-ID}. The remaining contours enclose 90~per~cent of data points for molecular gas `sight-lines' at fixed spatial scales of $30$, $45$, $80$ and $120$~pc, following the cloud-identification method described in~\protect\cite{Sun18}. The grey-shaded histogram in the upper panel contains the observational data at $\theta = 120$~pc from~\protect\cite{Sun2020}, across their sample of $28$ nearby star-forming galaxies. The solid, dashed and dot-dashed lines denote overpressures $\langle P_{\rm turb, \theta pc} \rangle_{1{\rm kpc}}/\langle P_{\rm mp} \rangle_{1{\rm kpc}}$ of $1$, $5$ and $10$ times, respectively. In the lower panel, the cloud populations of the FLAT, SLOPED and CORED runs are separated to show the similarity in their mid-plane turbulent pressure distributions, on kpc scales.}
\end{figure}

In Figure~\ref{Fig::P-DE_vs_Pturb}, we demonstrate the impact of this over-density on the over-pressure of beam-filling GMCs. The horizontal axis gives the kpc-scale mid-plane hydrostatic pressure $P_{\rm mp} = \pi G/2 \: \phi_{\rm P} \Sigma_{\rm g}^2$, combining the ISM and stellar-gravity contributions. We calculate this using 2D ray-tracing projections at 120 pc-resolution, as described in Appendix~\ref{App::analysis}, and perform the kpc-average by resampling the array into larger bins. The vertical axis of Figure~\ref{Fig::P-DE_vs_Pturb} gives the cloud-scale turbulent pressure according to Equation~(\ref{Eqn::P-turb}). Again, the coloured contours enclose 90~per~cent of the combined simulation data at resolutions of $\Delta x = 120$~pc (light-green contour), $80$~pc, $45$~pc and $30$~pc, along with the beam-filling clouds (purple contour). The grey-shaded histogram shows the combined data from the sample of $28$ nearby galaxies studied by~\cite{Sun2020}. As seen in the $\Sigma$-$\sigma_{\rm los}$ distribution, the overlap between observations and simulations is greatest for molecular gas sight-lines at 120~pc-resolution. We find an average over-pressure of $\sim 4 \times$ for these sight-lines: very close to the observed value of $\sim 2.8 \times$. The GMC turbulent pressure increases with resolution, reaching a peak over-pressure of $\sim 25 \times$ for the beam-filling GMCs. By contrast, 90~per~cent of the data for the HI clouds is enclosed by the thin black line. We see that their internal turbulent pressures are much closer to the mid-plane hydrostatic pressure than are the internal turbulent pressures of the beam-filling GMCs.

Given the information contained in Figures~\ref{Fig::scaling-rln} and~\ref{Fig::P-DE_vs_Pturb}, the dichotomy in the degree of dynamical variation for GMC and HI cloud gravo-turbulent properties is likely caused by the difference in their internal turbulent pressures. Comparing the purple and black contours in Figure~\ref{Fig::P-DE_vs_Pturb}, this offset spans around two orders of magnitude. The HI clouds are close to the pressure of the galactic mid-plane, and so are dynamically-coupled to it. The GMCs sit high above the pressure of the galactic mid-plane, and so are dynamically-decoupled. At mid-plane pressures higher than those in Milky Way-like galaxies (right-hand side of the grey-shaded histogram in Figure~\ref{Fig::P-DE_vs_Pturb}), we hypothesise that the dynamical-coupling of GMCs will increase substantially. A higher fraction of the mid-plane gas will be molecular, and so the beam filling-factor at $120$~pc resolution will be much higher. That is, the purple contour (beam-filling clouds) and the green contour (clouds identified at a fixed resolution of $120$~pc) will overlap at high $\langle P_{\rm mp} \rangle_{1{\rm kpc}}$, leading to galactic-dynamical trends in $\sigma_{\rm los, H_2}$, $\alpha_{\rm vir, H_2}$ and $P_{\rm turb, H_2}$ for beam-filling GMCs. In galaxies with significantly-higher mid-plane pressures than the Milky Way, we expect to find substantial variation in these properties as a function of $(\beta, Q, \Omega, \phi_{\rm P})$. A first observational indication for a this trend has been identified by~\cite{Chevance20}, who find that GMC lifetimes in galaxies with kpc-scale molecular gas surface densities $\Sigma_{{\rm H}_2}>8~\msun~\pc^{-2}$ correlate with galactic dynamical time-scales rather than internal dynamical ones, contrary to those at lower surface densities.

\subsubsection{Self-gravity and turbulence in HI clouds} \label{Sec::grav-turb-HIclouds}
The velocity dispersions $\sigma_{\rm HI}$, virial parameters $\alpha_{\rm vir, HI}$ and turbulent pressures $P_{\rm turb, HI}$ of the HI clouds in our simulations display a clear environmental variation across the galactic-dynamical parameter space $(\beta, Q, \Omega, \phi_{\rm P})$. In the bottom right-hand panel of Figure~\ref{Fig::turbulent-properties}, we show that the HI cloud turbulent pressure increases towards high-shear, high gravitational-stability environments with $\beta \rightarrow 0$ and $Q \rightarrow 10$. The trend comes predominantly from an increase in the cloud velocity dispersion, while the cloud surface density varies over a much smaller dynamic range (lower-left and centre-left panels of Figure~\ref{Fig::turbulent-properties}). As such, the HI cloud virial parameter (centre-right panel of Figure~\ref{Fig::turbulent-properties}) also increases with the degree of shearing and gravitational stability. This environmental variation in each HI cloud gravo-turbulent property corresponds to a statistically-significant correlation with the two shortest galactic-dynamical time-scales for shearing ($\tau_\beta$) and gravitational free-fall ($\tau_{\rm ff,g}$), as well as with the mid-plane pressure $P_{\rm mp}$, as summarised in Figure~\ref{Fig::correlation-summary}. The HI cloud pressures therefore trace the ambient pressure, density and dynamical state of the galactic mid-plane, while the GMCs are dynamically-decoupled and gravitationally-collapsing (discussed further in Section~\ref{Sec::cloud-divergence}).

\begin{figure}
\label{Fig::betaQ_ellipticity}
  \includegraphics[width=\linewidth]{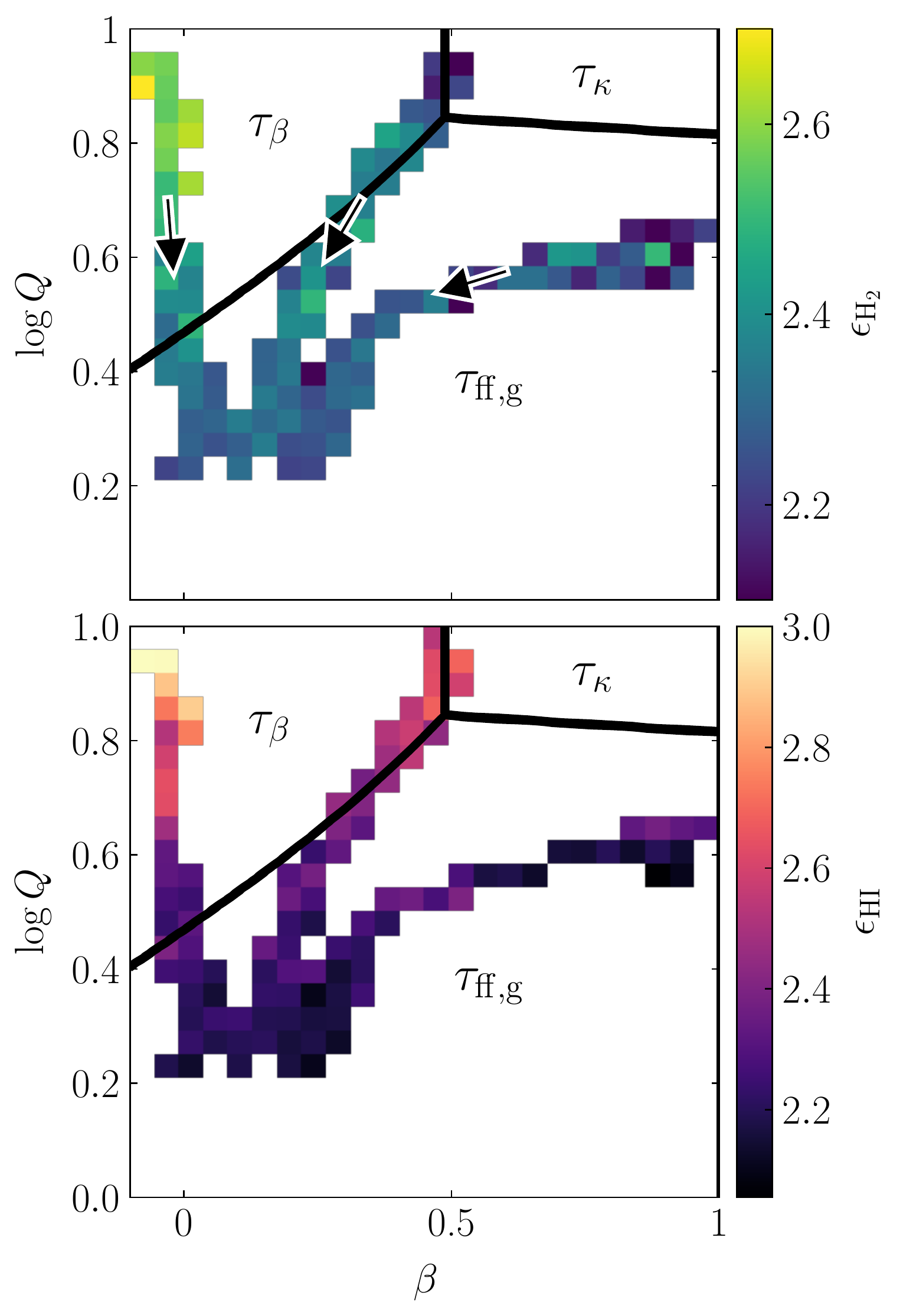}
  \caption{The mean aspect ratio $\epsilon$ of the GMCs (upper panel) and HI clouds (lower panel) in our simulations, as a function of the shear parameter $\beta$ and the Toomre $Q$ stability parameter. Each distinct set of connected pixels corresponds to the total cloud population of one isolated disc galaxy, compiled across the simulation time interval from $600$~Myr to $1$~Gyr, at a sampling interval of $50$~Myr. The color of each pixel represents the mean value of $\epsilon$ at the indicated value of $(\beta, Q)$. The black arrows in the top panel mark the direction in which the galactocentric radius increases. The black solid lines enclose the regions of parameter space for which the minimum dynamical time-scale is $\tau_\kappa$ (orbital epicyclic perturbations), $\tau_\beta$ (galactic shear) and $\tau_{\rm ff,g}$ (gravitational free-fall).}
\end{figure}

\begin{figure*}
\label{Fig::betaQ_angmom}
  \includegraphics[width=\linewidth]{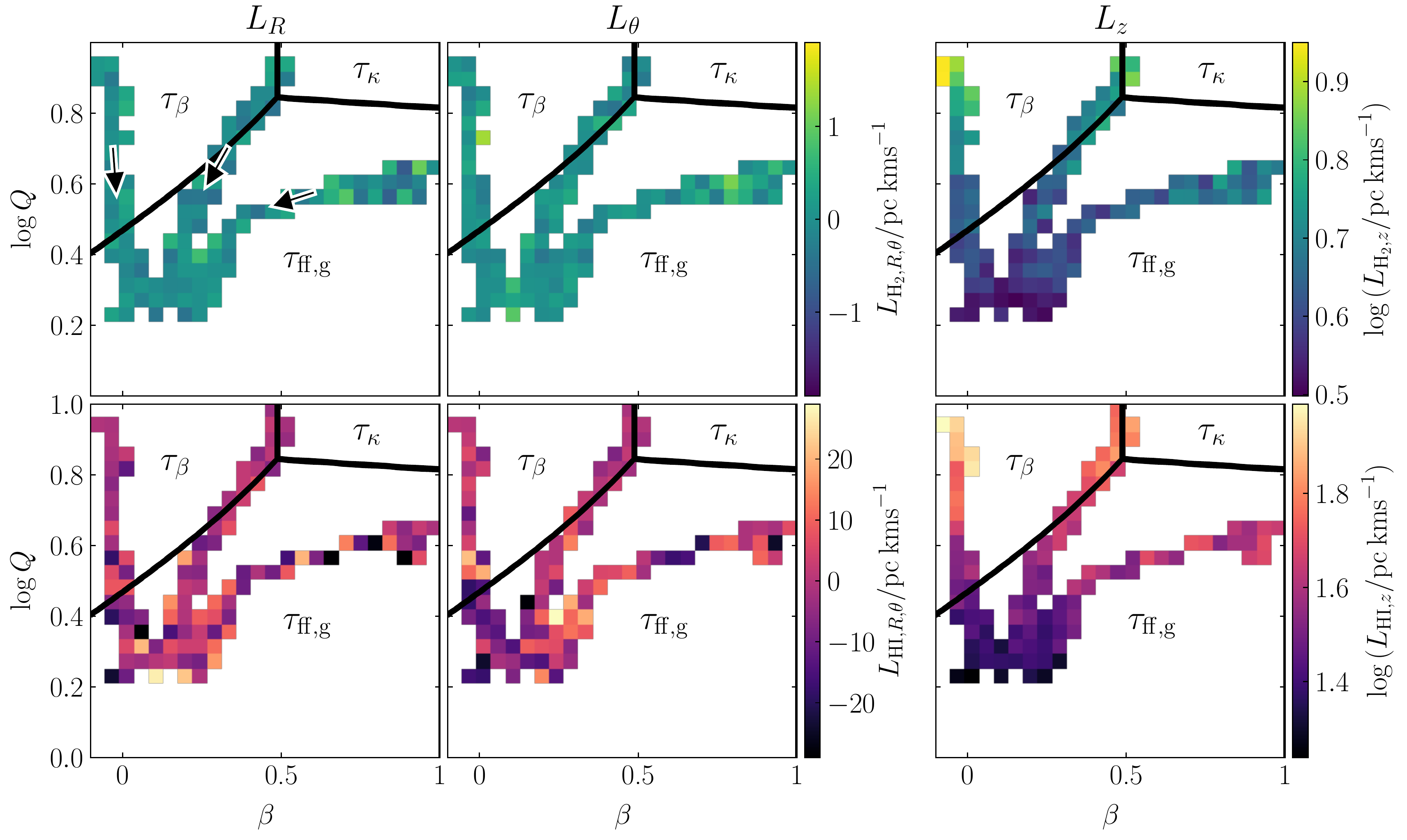}
  \caption{The mean angular velocity $\mathbf{L}$ of the GMCs (upper panel) and HI clouds (lower panel) in our simulations, as a function of the shear parameter $\beta$ and the Toomre $Q$ stability parameter. The three columns display the three orthogonal components of the angular momentum vector $\mathbf{L}=(L_R, L_\theta, L_z)$. Each distinct set of connected pixels corresponds to the total cloud population of one isolated disc galaxy, compiled across the simulation time interval from $600$~Myr to $1$~Gyr, at a sampling interval of $50$~Myr. The color of each pixel represents the mean value of the $L_R$, $L_\theta$ or $L_R$ at the indicated value of $(\beta, Q)$. The black arrows in the upper left panel mark the direction in which the galactocentric radius increases. The black solid lines enclose the regions of parameter space for which the minimum dynamical time-scale is $\tau_\kappa$ (orbital epicyclic perturbations), $\tau_\beta$ (galactic shear) and $\tau_{\rm ff,g}$ (gravitational free-fall).}
\end{figure*}

\subsection{Cloud collapse} \label{Sec::cloud-divergence}
The degree to which a cloud is collapsing globally towards its centre of mass can be quantified by the magnitude of its internal radial velocity streaming, as
\begin{equation}
\begin{split}
\mathcal{D}_x &= \langle v_{r,i} \rangle_{x}, \; \; \; x \in \{{\rm H}_2, {\rm HI}\}
\end{split}
\end{equation}
where $\{v_{r,i}\}$ are the radial velocities of the gas cells in the cloud, with respect to the velocity of its centre of mass. We find that the GMC velocity divergence is negative and invariant across all galactic-dynamical environments, with a mean value of $\mathcal{D}_{\rm H_2} \sim (-0.99 \pm 0.08)$~km s$^{-1}$, where the error indicates the standard deviation of the measurements. That is, averaged across dynamical environments, the simulated molecular clouds are collapsing at a uniform rate towards their centres of mass, consistent with the high degree of gravitational boundedness indicated by their virial parameters, $\alpha_{\rm vir, H_2} \sim 1.2$. The parent HI clouds display a much weaker level of collapse (mean value $\mathcal{D}_{\rm HI} \sim (-0.2 \pm 0.4)$~kms$^{-1}$), which follows a statistically-significant correlation with the mid-plane velocity dispersion of the galaxy, shown in Appendix~\ref{App::stat-sig}. At the highest mid-plane velocity dispersions of $\sigma_{\rm mp} \sim 10$~kms$^{-1}$, HI clouds are not collapsing, reaching near-zero velocity divergences. At the lowest velocity dispersions ($\sigma_{\rm mp} \sim 6$~kms$^{-1}$), HI clouds are collapsing at their maximum average rate of $\mathcal{D}_{\rm HI} \sim -1$~kms$^{-1}$. This trend is consistent with HI clouds that are coupled to the galactic mid-plane, so are more-effectively supported against collapse at higher mid-plane velocity dispersions.

\subsection{Cloud rotational properties} \label{Sec::shear-properties}
The GMC population in our simulations is highly over-dense and over-pressured relative to the ambient ISM, leading to its turbulent and gravitational decoupling from the properties of the galactic mid-plane. Nevertheless, we find statistically-significant galactic-dynamical trends in the rotational properties of our clouds (aspect ratios, angular momenta and velocity anisotropies). As shown in Figure~\ref{Fig::correlation-summary}, each of these properties correlates with the time-scales $\tau_\beta$ and $\tau_{\rm ff,g}$ for galactic shear and gravitational free-fall, as well as with the mid-plane pressure $P_{\rm mp}$. In the environmental parameter space $(\beta, Q, \Omega, \phi_{\rm P})$, this manifests itself as a monotonic gradient towards the top left-hand corner of the $(\beta, Q)$ plane, as seen in the top left-hand panel of Figure~\ref{Fig::betaQ_timescales}. The rotational trends observed for GMCs are mirrored by the parent HI clouds, indicating either that (1) both GMCs and HI clouds are affected in the same way by the interplay between galactic rotation and gravitational free-fall, or (2) that GMCs retain the rotational properties of their parent HI clouds as they collapse.

\subsubsection{Aspect ratio} \label{Sec::cloud-ellipticity}
The elongation of each GMC and HI cloud within the galactic mid-plane can be quantified by the aspect ratio, $\epsilon_x$, as
\begin{equation}
\begin{split}
\epsilon_x &= \frac{\Delta \ell_{{\rm maj}, x}}{\Delta \ell_{{\rm min}, x}}, \; \; \; x \in \{{\rm H}_2, {\rm HI}\}
\end{split}
\end{equation}
where $\Delta \ell_{\rm maj}$ and $\Delta \ell_{\rm min}$ are the major and minor axes, respectively, of an ellipse fitted to the footprint of each cloud in the galactic mid-plane. Both the GMC and HI cloud populations in our galaxies have the same mean aspect ratio ($\epsilon_{\rm H_2}, \epsilon_{\rm HI} \sim 2.3$), and both increase monotonically with the degree of differential rotation in the galactic mid-plane ($\beta \rightarrow 0$) and with the degree of gravitational stability ($Q \rightarrow 10$). This qualitative trend is shown in Figure~\ref{Fig::betaQ_ellipticity}. The particular role of galactic shearing in stretching HI clouds and molecular clouds is indicated by the following two results.
\begin{enumerate}
  \item The direction of elongation points along the direction of galactic rotation in every dynamical environment, with mean position angles of $-1.7^{\circ} \pm 9^{\circ}$ and $-2.3^{\circ} \pm 9^{\circ}$ for GMCs and HI clouds, respectively.
  \item In environments for which clouds are controlled primarily by gravity (region $\tau_{\rm ff,g}$ of the dynamical parameter space), the HI clouds and GMCs have approximately-constant and equal aspect ratios of $\epsilon_{\rm H_2}, \epsilon_{\rm HI} \sim 2.2$, but in environments for which galactic shear dominates (region $\tau_\beta$), the GMC aspect ratio anticorrelates with the shear time-scale as
\begin{equation}
\epsilon_{\rm H_2} \propto (-0.39 \pm 0.03) \log{\tau_\beta},
\end{equation}
and the HI cloud aspect ratio anticorrelates as
\begin{equation}
\epsilon_{\rm HI} \propto (-0.81 \pm 0.04) \log{\tau_\beta},
\end{equation}
as demonstrated in Appendix~\ref{App::stat-sig}.
\end{enumerate}
The HI clouds in our simulations are significantly more suceptible to elongation than are the GMCs, which is not surprising given their lower over-densities and over-pressures, and their correspondingly larger degree of dynamical coupling.

\subsubsection{Specific angular momentum} \label{Sec::cloud-angmom}
The specific angular momentum of GMCs and HI clouds provides a measure of their internal rotation about the cloud centre of mass. The specific angular momentum vector for a given cloud is defined as
\begin{equation}
\begin{split}
\mathbf{L}_x &= L_{z,x} \hat{\mathbf{z}} + L_{\theta,x} \hat{\mathbf{\theta}} + L_{R,x} \hat{\mathbf{R}} \\
&= \langle \mathbf{r}_i \times \mathbf{v}_i \rangle_{x}, \; \; \; x \in \{{\rm H}_2, {\rm HI}\}
\end{split}
\end{equation}
where $\{\mathbf{r}_i\}$ are the positions of the gas cell centroids relative to the cloud centre of mass, and $\{\hat{\mathbf{R}}, \hat{\mathbf{\theta}}, \hat{\mathbf{z}}\}$ are the galactic unit vectors in cylindrical polar coordinates. The component $L_z$ is therefore the magnitude of the angular momentum within the galactic mid-plane and about an axis perpendicular to it, while $L_\theta \hat{\mathbf{\theta}}$ and $L_R \hat{\mathbf{R}}$ are the components of the vector within the galactic mid-plane.

In the parameter space of Figure~\ref{Fig::betaQ_angmom}, the cloud angular momentum displays a clear variation with the galactic-dynamical environment. This environmental variation leads to clear, statistically-significant correlations with the galactic-dynamical time-scales $\tau_\beta$ and $\tau_{\rm ff,g}$, presented in Figure~\ref{Fig::correlation-summary} and in Table~\ref{Tab::results}. However, given the preferentially prograde sense of rotation for our simulated cloud population, these trends are likely to be driven by the variation of $L_z$ with the orbital angular velocity $\Omega$, which dictates the strength of the fictitious Coriolis force in the galactic rest-frame, rather than the strength of shearing. For our simulated galaxies, all of which have shear parameters in the range $0 < \beta < 1$, we would expect shear-driven rotation to be exclusively retrograde. Only for a rising rotation curve with $\beta \geq 1$ would we expect clouds to be spun up in a prograde direction by galactic shear. Therefore, while the degree of internal rotation for our GMCs and HI clouds is determined by the galactic rotation curve, it is not due to differential rotation across a cloud. We also find that the GMCs in our sample display a factor of ten reduction in angular momentum relative to the HI clouds across all galactic-dynamical environments, which is a factor of three greater than can be accounted for by their smaller sizes. It is therefore unlikely that the GMCs themselves are torqued by galactic rotation, but rather inherit their angular momentum from their parent HI clouds. On average, some of the inherited angular momentum is then partially converted into turbulent motion of the surrounding gas during gravitational collapse.

\subsubsection{Velocity anisotropy} \label{Sec::cloud-vel-anisotropy}
In addition to the line-of-sight velocity dispersion presented in Section~\ref{Sec::gravity-turbulence}, we can compute the radial $\sigma_r$ and tangential $\sigma_t$ components of the cloud velocity dispersion with respect to the cloud centre of mass, such that
\begin{equation}
\sigma_r^2 = \langle |\mathbf{v}_i \cdot \hat{\mathbf{r}}_i - \langle \mathbf{v}_i \cdot \hat{\mathbf{r}}_i \rangle_{i, x}|^2 \rangle_{i, x}
\end{equation}
and
\begin{equation}
\begin{split}
\sigma_t^2 &= \sigma_\theta^2 + \sigma_\phi^2 \\
\sigma_\theta^2 &= \langle |\mathbf{v}_i \cdot \hat{\mathbf{\theta}}_i - \langle \mathbf{v}_i \cdot \hat{\mathbf{\theta}}_i \rangle_{i, x}|^2 \rangle_{i, x} \\
\sigma_\phi^2 &= \langle |\mathbf{v}_i \cdot \hat{\mathbf{\phi}}_i - \langle \mathbf{v}_i \cdot \hat{\mathbf{\phi}}_i \rangle_{i, x}|^2 \rangle_{i, x}.
\end{split}
\end{equation}
providing the velocity anisotropy as
\begin{equation}
\begin{split}
B_{\sigma, x} &= 1 - \frac{\sigma_{t, x}}{2\sigma_{r, x}},
\end{split}
\end{equation}
with $x \in \{{\rm H}_2, {\rm HI}\}$. In the above equations, $\{\hat{\mathbf{\theta}}_i\}$ are the longitudinal unit vectors of the gas cells in each cloud with respect to the cloud centre of mass, and $\{\hat{\mathbf{\phi}}_i\}$ are the corresponding azimuthal unit vectors. If all motions within a cloud are radial, then $\sigma_t=0$ and $B_\sigma = 1$. Conversely, if all motions are circular, then $\sigma_r=0$ and $B_\sigma \rightarrow -\infty$~\citep{BinneyTremaine1987}. The velocity anisotropy of the internal turbulent motion in a GMC or HI cloud therefore tells us about the origin of turbulence in this cloud ~\citep[e.g.][]{Ossenkopf2002,Lazarian02,Esquivel&Lazarian11,Burkhart14,Otto17}. Significant radial bias ($B_\sigma>0$) indicates that turbulence is primarily driven by spherically-symmetric processes such as global gravitational collapse. Significant tangential bias ($B_\sigma<0$) indicates that turbulence is driven either by non-spherically symmetric processes in the cloud's environment (e.g.~galactic shearing or compression driven by external feedback) or by processes that occur inside the cloud but that may be far from its centre of mass (e.g.~hierarchical collapse and internal feedback). We note that magnetic fields in particular (not included in our simulations) may significantly alter the velocity anisotropy by directing the bulk gas flow within the ambient ISM~\citep[e.g.][]{KimCG&Ostriker17,Seifried17,Seifried2020}. We discuss this further in Section~\ref{Sec::caveats}.

\begin{figure}
\label{Fig::betaQ_anisotropy}
  \includegraphics[width=\linewidth]{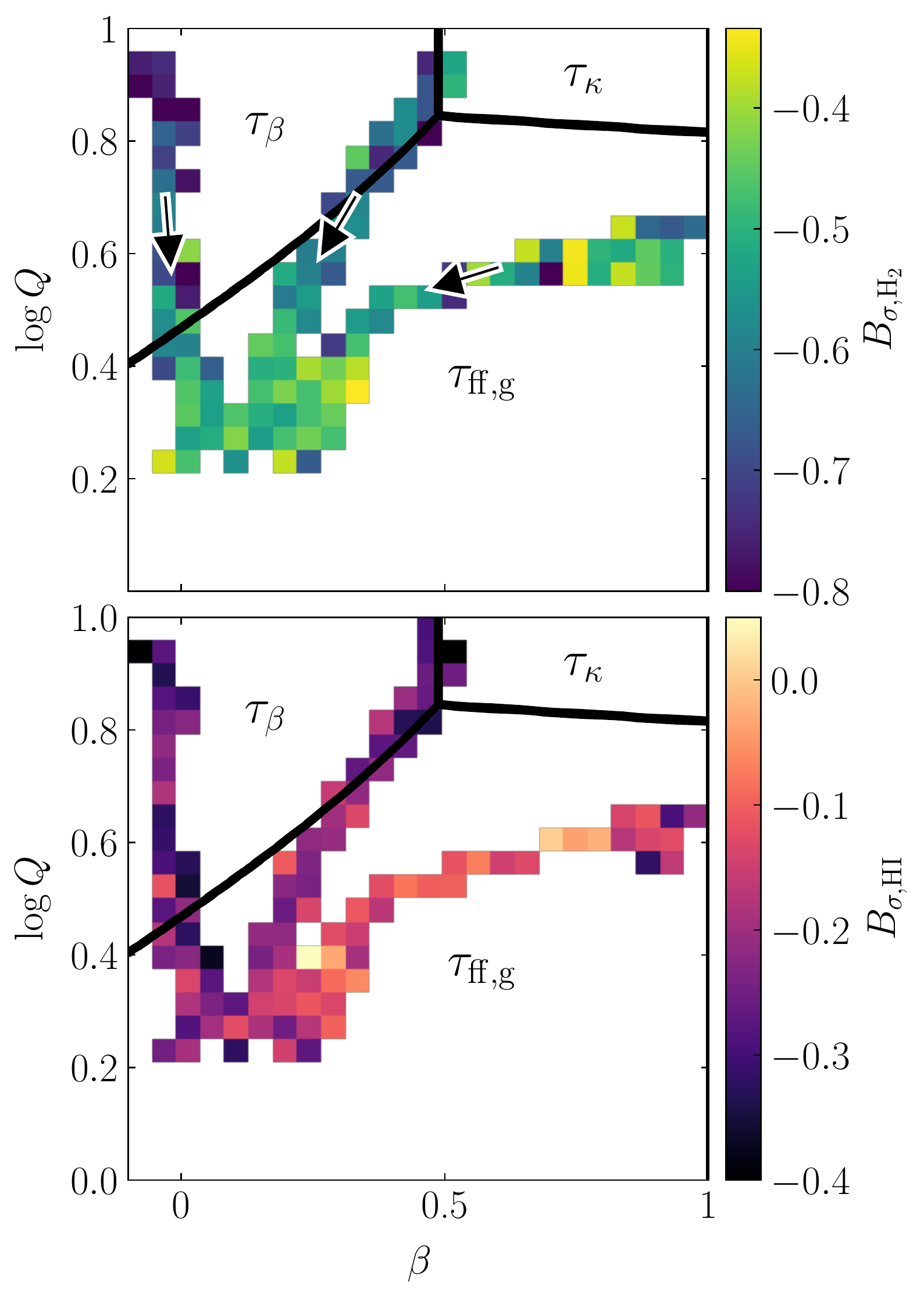}
  \caption{The mean velocity anisotropy $B_\sigma$ of the GMCs (upper panel) and HI clouds (lower panel) in our simulations, as a function of the shear parameter $\beta$ and the Toomre $Q$ stability parameter. Each distinct set of connected pixels corresponds to the total cloud population of one isolated disc galaxy, compiled across the simulation time interval from $600$~Myr to $1$~Gyr, at a sampling interval of $50$~Myr. The color of each pixel represents the mean value of the anisotropy at the indicated value of $(\beta, Q)$. The black arrows mark the direction in which the galactocentric radius increases. The black solid lines enclose the regions of parameter space for which the minimum dynamical time-scale is $\tau_\kappa$ (orbital epicyclic perturbations), $\tau_\beta$ (galactic shear) and $\tau_{\rm ff,g}$ (gravitational free-fall).}
\end{figure}

Figure~\ref{Fig::betaQ_anisotropy} shows that in all galactic-dynamical environments, our simulated GMCs and HI clouds are tangentially-biased, with mean anisotropy parameters ranging from $-0.9$ to $-0.3$ for $B_{\sigma, {\rm H_2}}$ and from $-0.4$ to $0.1$ for $B_{\sigma, {\rm HI}}$. For the GMCs, this means that the tangential velocity dispersions are 30 to 90~per~cent higher than the radial velocity dispersions, displaying a monotonic increase in the degree of tangential bias with the rate of galactic shearing, as
\begin{equation}
B_{\sigma, {\rm H_2}} \propto (0.35 \pm 0.04)\log{\tau_\beta}.
\end{equation}
This monotonic increase is consistent with the presence of shear-induced internal circular motions, although given the prograde orientation of the specific angular momentum, we cannot rule out the contribution of the fictitious Coriolis force in the rotating galactic frame. The slope is shallower for the HI clouds in our sample, with a value of
\begin{equation}
B_{\sigma, {\rm HI}} \propto (0.18 \pm 0.03)\log{\tau_\beta}.
\end{equation}
That is, both the degree of anisotropy, and its dependence on the shear rate, is reduced. In fact, in the gravity-dominated regime $\tau_{\rm ff,g}$ of the galactic-dynamical parameter space, the tangential velocity dispersion of the HI clouds approaches equality with the radial component ($B_\sigma \rightarrow 0$). The disparity between the shear-correlated degree of anisotropy in GMCs and HI clouds is likely to arise from the conservation of angular momentum during the collapse of the GMCs, which enhances the shear-driven tangential velocity dispersion present in their parent HI clouds~\citep[see also][]{Kruijssen2019b}.

\subsection{Molecular cloud star formation rate} \label{Sec::cloud-star-formation}
\begin{figure*}
\label{Fig::SFR_vs_P-mp}
  \includegraphics[width=\linewidth]{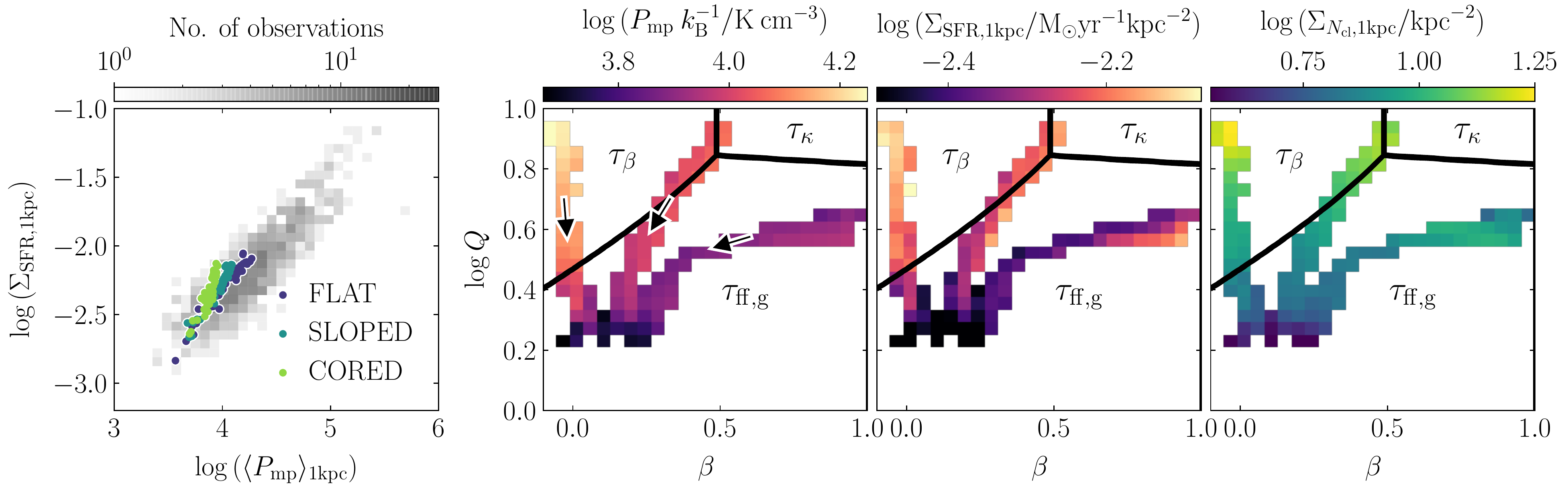}
  \caption{{\it Left:} The kpc-scale star formation rate surface density $\langle \Sigma_{\rm SFR} \rangle_{1{\rm kpc}}$ as a function of the kpc-scale mid-plane pressure $\langle P_{\rm mp} \rangle_{1{\rm kpc}}$. The grey-shaded histogram shows the observational data of~\protect\cite{Sun2020}, aggregated across 28 nearby galaxies. The three sets of scattered points correspond to the azimuthally-averaged values for our FLAT, SLOPED and CORED simulations at intervals of $50$~Myr between simulation times of $600$~Myr and $1$~Gyr. The simulation data fall at the lowest mid-plane pressures present in the observed sample. {\it Centre-left:} The azimuthally-averaged value of $\langle \Sigma_{\rm SFR} \rangle_{1{\rm kpc}}$ as a function of the galactic-dynamical variables $(\beta, Q)$. The black arrows mark the direction in which the galactocentric radius increases. {\it Centre-right:} The azimuthally-averaged value of $\langle P_{\rm mp} \rangle_{1{\rm kpc}}$, as a function of the galactic-dynamical variables $(\beta, Q)$. This obeys a very similar environmental trends to $\langle \Sigma_{\rm SFR} \rangle_{1{\rm kpc}}$. {\it Right:} The number of GMCs per unit area of the galactic mid-plane, as a function of the galactic-dynamical variables $(\beta, Q)$. This obeys a very similar trend to the kpc-scale star formation rate, indicating that the global star formation rate correlates with the number of star-forming regions. The black solid lines enclose the regions of parameter space for which the minimum dynamical time-scale is $\tau_\kappa$ (orbital epicyclic perturbations), $\tau_\beta$ (galactic shear) and $\tau_{\rm ff,g}$ (gravitational free-fall).}
\end{figure*}

Across the galactic-dynamical environments $(\beta, Q, \Omega, \phi_{\rm P})$ spanned by our simulations, we find a quantitative correlation between the kpc-scale star formation rate $\langle \Sigma_{\rm SFR} \rangle_{1{\rm kpc}}$ and the kpc-scale mid-plane hydrostatic pressure $\langle P_{\rm mp} \rangle_{1{\rm kpc}}$. The relationship between the azimuthally-averaged values of $\langle P_{\rm mp} \rangle_{1{\rm kpc}}$ and $\langle \Sigma_{\rm SFR} \rangle_{1{\rm kpc}}$ for our simulations shows a large degree of overlap with the observed relationship presented by~\cite{Sun2020} (also originally by~\cite{BlitzRosolowsky2004}) for 28 nearby galaxies, as shown in the left-hand panel of Figure~\ref{Fig::SFR_vs_P-mp}. In turn, this is consistent with a state of approximate dynamical equilibrium between feedback and gravitational collapse~\citep[see][]{Ostriker+10,OstrikerShetty2011,Benincasa16}. As for the relationship between the cloud-scale turbulent pressure and the mid-plane hydrostatic pressure (Section~\ref{Sec::gravity-turbulence}) our simulated discs fall at the low-pressure end of the observed relationship, corresponding to the discs of Milky Way-pressure, flocculent galaxies.

\begin{figure*}
\label{Fig::Pturb_vs_SFR}
  \includegraphics[width=\linewidth]{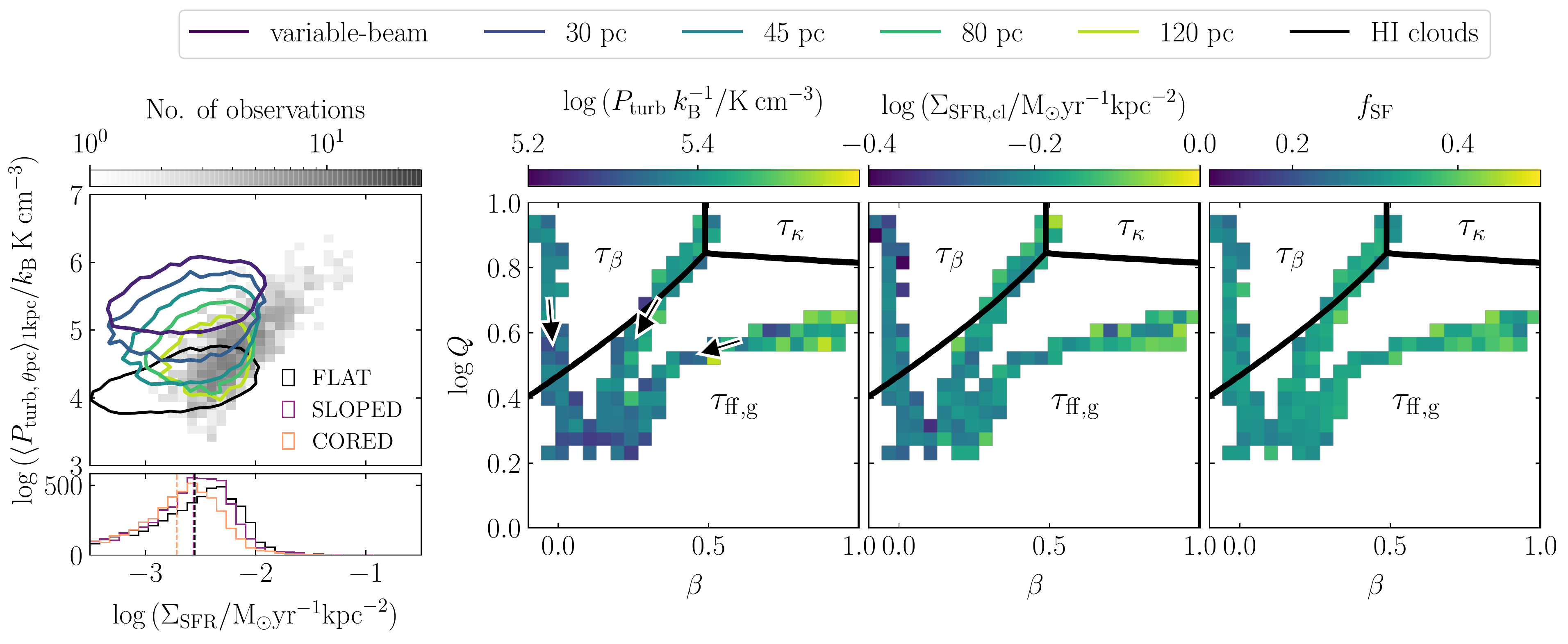}
  \caption{{\it Left:} The kpc-scale averaged turbulent pressure $\langle P_{{\rm turb}, \: \theta {\rm pc}} \rangle_{1 {\rm kpc}}$ at varied spatial resolution $\theta$ for the combined GMC population across our three simulations at times between $600$~Myr and $1$~Gyr, as a function of the kpc-averaged star formation rate surface density $\langle \Sigma_{\rm SFR} \rangle_{1{\rm kpc}}$. The dark purple contour encloses 90~per~cent of beam-filling molecular clouds identified using {\sc Astrodendro}. The remaining contours represent molecular gas `sight-lines' at fixed spatial scales of $30$, $45$, $80$ and $120$~pc, following the method of cloud identification described in~\protect\cite{Sun18}. The grey-shaded histogram in the upper panel contains the observational data at $\theta = 120$~pc from~\protect\cite{Sun2020}, across a sample of $28$ nearby star-forming galaxies. {\it Centre-left: } Turbulent pressure of the beam-filling GMCs in our simulations, as a function of the galactic-dynamical variables $(\beta, Q)$. The black arrows mark the direction in which the galactocentric radius increases. {\it Centre-right/Right:} Star-forming properties of the simulated GMCs: the mean turbulent pressure $P_{\rm turb}$, star formation rate surface density per molecular cloud $\Sigma_{\rm SFR, cl}$, and the fraction of clouds in each environment that are star-forming $f_{\rm SF}$. The scale used for each colorbar is expanded to 0.4 dex, to match the span of global star formation rate surface densities across our simulations. The dynamic ranges of these cloud-scale properties are much smaller than that of the global galactic star forming properties, and of the number of GMCs per unit area of the galactic mid-plane. The black solid lines enclose the regions of parameter space for which the minimum dynamical time-scale is $\tau_\kappa$ (orbital epicyclic perturbations), $\tau_\beta$ (galactic shear) and $\tau_{\rm ff,g}$ (gravitational free-fall).}
\end{figure*}

Within our sample, we find that the rise in star formation rate surface density with mid-plane pressure is primarily due to a rise in the number of GMCs per unit area of the galactic mid-plane, and not due to environmental variations in the properties of the GMCs themselves. In the central and right-hand panels of Figure~\ref{Fig::SFR_vs_P-mp}, we show the environmental trends in $\langle \Sigma_{\rm SFR} \rangle_{1{\rm kpc}}$ and $\langle P_{\rm mp} \rangle_{1{\rm kpc}}$, and compare these to the environmental trend in the number of GMCs per unit area of the galactic mid-plane on kpc-scales, $\langle \Sigma_{N_{\rm cl}} \rangle_{1{\rm kpc}}$. Qualitatively, the trends match across all galactic environments. In Figure~\ref{Fig::correlation-summary} and in Appendix~\ref{App::stat-sig}, we show that the correlation between $\langle \Sigma_{N_{\rm cl}} \rangle_{1{\rm kpc}}$ and $\langle P_{\rm mp} \rangle_{1{\rm kpc}}$ is statistically-significant at the $3\sigma$ confidence level.

By comparison to the dynamic ranges of the environmental trends in $\langle \Sigma_{\rm SFR} \rangle_{1{\rm kpc}}$, $\langle P_{\rm mp} \rangle_{1{\rm kpc}}$ and $\langle \Sigma_{N_{\rm cl}} \rangle_{1{\rm kpc}}$, the qualitative variation in the internal star-forming properties of GMCs is much smaller. On the right-hand side of Figure~\ref{Fig::Pturb_vs_SFR}, we show the mean turbulent pressure $P_{\rm turb}$ of GMCs in our simulations as a function of the galactic-dynamical parameters $(\beta, Q)$, as well as their mean internal star formation rate surface densities $\Sigma_{\rm SFR, cl}$ and the fraction $f_{\rm SF}$ of identified GMCs that are currently star-forming. We have expanded the colorbars of each plot to match the dynamic range of the global kpc-scale star formation rate of the host galaxy, emphasising the small contribution to the overall trend in star formation that is made by each cloud-scale property. On the left-hand side of Figure~\ref{Fig::Pturb_vs_SFR}, we demonstrate the reason for the environmental independence of the star-forming properties of our GMCs: as discussed in Section~\ref{Sec::gravity-turbulence}, they are decoupled from the galactic-midplane, with much higher pressures ($\sim 25 \times$), densities ($\sim 100 \times$) and star formation rates ($\sim 100 \times$) than the surrounding medium. The dark purple contour encloses 90~per~cent of the beam-filling GMCs analysed in this work, while the lighter-coloured contours represent molecular gas sight-lines corresponding to those analysed in~\cite{Sun2020} across 28 nearby galaxies. These observational data are given by the grey-shaded histogram. At the observational $120$~pc-resolution (light-green contour), a good level of overlap between simulations and observations is recovered. At higher resolutions, the measured cloud-scale turbulent pressures increase relative to the galactic mid-plane pressure, pushing their values above the SFR-pressure correlation that is observed on larger scales. This decoupling of GMCs from the mid-plane, which we also discussed in Section~\ref{Sec::gravity-turbulence}, is responsible for the lack of a correlation between the internal star-forming properties of GMCs and the properties of the mid-plane. In order to observe GMCs for which the star-forming properties are significantly influenced by the ambient environment and by its galactic dynamics, we would need to additionally simulate high-pressure, high-star formation rate environments towards the upper end of the observed pressure-SFR relation. These environments can be found in the spiral arms and bars of massive galaxies, and in galaxy discs with higher masses and gas fractions.

\begin{figure*}
\label{Fig::rotation-discussion}
  \includegraphics[width=\linewidth]{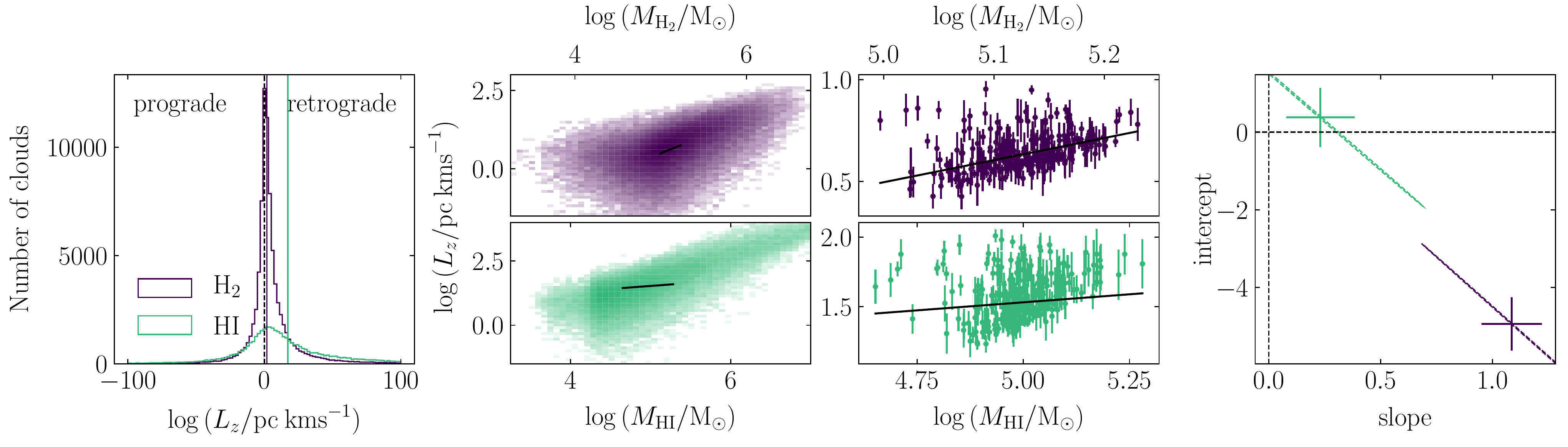}
  \caption{Analysis of GMC and HI cloud rotation, compiled across the FLAT, SLOPED and CORED discs at simulation times between $600$~Myr and $1$~Gyr. {\it Left:} Distribution of molecular cloud angular momenta (purple line) and HI cloud angular momenta (green line). The vertical lines give the means of the distributions. The black dashed vertical line denotes the state of zero angular momentum. We note that both distributions are skewed towards retrograde rotation (positive values of $L_z$. {\it Centre left:} Distribution of GMC (purple) and HI cloud (green) angular momenta, as functions of their respective masses. We note a clear correlation between the cloud mass and angular momentum in both cases. {\it Centre right:} Mean angular momentum of GMCs (purple) and of HI clouds (green) as a function of mean cloud mass, when the clouds are binned according to their galactic-dynamical environments, $(\beta, Q, \Omega, \phi_{\rm P})$. The error bars denote the error on the mean in each environmental bin. The black lines (in all panels) show the non-linear least-squares fit to the environmentally-binned data. {\it Right:} The $1\sigma$ (filled) and $3\sigma$ (dashed lines) confidence ellipses for each fit. The error-bars correspond to the projection of the $1\sigma$ confidence ellipse onto each of the parameter axes. Neither trend is significant at the $3\sigma$ level.}
\end{figure*}

\section{Discussion} \label{Sec::discussion}
\subsection{Comparison to previous work} \label{Sec::comp-to-prev-work}
This work addresses the statistical variation in GMC properties as a function of the galactic-dynamical environment. To this end, we have produced a large sample of $\sim 80,000$ molecular clouds that spans the dynamical parameter space of~\cite{Jeffreson+Kruijssen18}, encompassing the full range of shear parameters $\beta$, and spanning around an order of magnitude in the Toomre $Q$ parameter, the angular velocity $\Omega$ and the stellar contribution $\phi_{\rm P}$ to the mid-plane pressure. Although the analysis and coverage of dynamical environments is unique, a number of other works have simulated and studied large samples of GMCs in isolated disc galaxies. The single-isothermal-phase simulations of~\cite{Dobbs06}, and the two-isothermal-phase simulations of~\cite{Dobbs07,Dobbs08} studied the role of spiral shocks, agglomeration and self-gravity in forming populations of GMCs in spiral galaxies. \cite{Dobbs11} additionally include magnetic fields and thermal SNe feedback, finding that the low degree of gravitational binding in GMCs can be attributed to a combination of feedback and cloud-cloud collisions, producing virial parameters in the range $0.1 \la \alpha_{\rm vir} \la 10$. In the current work, we find the same range of GMC virial parameters, despite the fact that our feedback prescription is physically-motivated and significantly more sophisticated (we include momentum and thermal energy injection due to SNe and HII-region feedback, and mass injection from stellar winds). At the typical hydrogen number densities of molecular clouds ($n_H \ga 30$~${\rm cm}^{-3}$), this suggests that the self-regulation of turbulence does not depend on the mode of energy injection. The GMC virial parameter also appears to be independent of the method used for identifying clouds (we use contours in projected CO luminosity, rather than clump-finding in total 3D gas density). In this work, we find that the mean virial parameter of GMCs is independent of the galactic-dynamical environment, with a global mean value of $\alpha_{\rm vir} \sim 1.6$. Our GMCs are therefore marginally gravitationally-bound on average. Our virial parameters are independent of the time-scale for cloud-cloud collisions, which is up to ten times longer than the time-scale for gravitational free-fall.

More similar to our discs are the spiral galaxies of~\cite{Dobbs11b,Dobbs12,DobbsPringle13}, and the simulated M33-analogues of~\cite{Dobbs18,Dobbs19}, which include thermal feedback from supernovae, ISM heating and cooling with ${\rm H}_2$ and CO formation, and live stellar and gas particles at a mass resolution of 400 ${\rm M}_\odot$. As in our simulations, the M33-like galaxies reproduce the observed cloud virial relation and the observed range of cloud velocity dispersions. They also reproduce the observed distribution of cloud angular momenta in M33, spanning a range of $-100 {\rm pc} \: {\rm kms}^{-1} \la L_z \la 100 {\rm pc} \: {\rm kms}^{-1}$. The authors find that the GMC angular momentum correlates with the GMC mass, which we confirm for our simulated GMCs and HI clouds in the central-left panel of Figure~\ref{Fig::rotation-discussion}. This correlation persists when the clouds are binned by galactic-dynamical environment. The significance of the trend between these two observables is demonstrated in the central-right and right-hand panels of Figure~\ref{Fig::rotation-discussion}, where we find a non-zero slope at the $3\sigma$ signficance level. Finally, in the left-hand panel of Figure~\ref{Fig::rotation-discussion}, we show the spectrum of angular momenta for our sample of GMCs and HI clouds. We find that the span and skew in angular momentum space is comparable to that shown in~\cite{Dobbs19}~\citep[see also][]{Braine2020}.

The angular momenta of our simulated GMCs and HI clouds can also be compared to the recent study of cloud-scale circulating velocities in isolated disc simulations, conducted by~\cite{Utreras20}. These authors quantify the degree of coupling between cloud-scale circulation and large-scale galactic rotation by modelling the galactic velocity field as a linear superposition of two components: the contribution from rotation plus a Gaussian random velocity field. Applying this analytic theory to three simulated discs with flat rotation curves, the authors find that cloud-scale spin is coupled to large-scale rotation only in the very centres of galaxies ($\la 3$-$5$~kpc). By contrast, we have found that, in galaxies with flat rotation curves, the radial gradient of the galactic rotation curve (i.e.~the rate of galactic shearing) correlates significantly with the cloud angular momentum out to galactocentric radii of $\ga 10$~kpc. A possible explanation for the additional rotational decoupling seen in~\cite{Utreras20} is evident in the surface density profiles of their galaxies. Their disc morphologies are highly-filamentary relative to ours, with surface densities reaching up to ten times higher in the densest clumps, despite their lower spatial resolution of $\sim 30$~pc, and their comparable gas, stellar and halo masses. This may be due to ineffective stellar feedback in their simulations, which is unable to disrupt the densest gas. The only source of momentum injection in their simulations is from photo-ionisation and stellar winds, both of which inject hundreds of times less energy than SNe explosions, which they treat thermally. The importance of momentum injection from unresolved SNe blast-waves has been confirmed by~\cite{KimmCen14,Slyz2005,Smith2018}, among others. In the absence of effective disruption, the cloud-scale gas in the simulations of~\cite{Utreras20} may be dominated to a greater extent by self-gravity than the GMCs and HI clouds in our simulations, explaining the lower degree of coupling between cloud rotation and galactic rotation. This effect may also be enhanced by the additional numerical viscosity in Eulerian codes such as {\sc Enzo}, relative to the moving-mesh of {\sc Arepo}~\citep[see][]{Benincasa13}.

The Milky Way-mass galaxies with different rotation profiles studied by~\cite{Nguyen18} have similarly-high densities and filamentary morphologies to the simulations of~\citep{Utreras20}, due to the total absence of stellar feedback. They therefore contain no hot ISM phase, and the resulting GMCs are systematically larger and more massive than ours. The authors focus on the formation of GMCs via dynamics and cooling, but also conclude that the properties of GMCs (including the virial parameter, angular momentum, mass and size) are largely unaffected by the galactic rotation curve. The difference in stellar feedback prescription makes little difference to the distribution of virial parameters and turbulent velocity dispersions, which we find to be similar between this work,~\cite{Nguyen18} and~\cite{Dobbs11}, as mentioned earlier. It therefore appears that stellar feedback does not play a major role in the self-regulation of gravity and turbulence at molecular cloud scales and densities, in Milky Way-pressured galaxies: a finding that is consistent with the results of the smaller-scale simulations by~\cite{Seifried18} and~\cite{Ibanez-Mejia16}. We also concur that GMC masses and sizes are unaffected by galactic rotation, and in fact are largely unaffected by any dynamical process in Milky Way-pressured galaxies. However, we again note that for our simulations, a statistically-significant coupling between galactic shearing and cloud rotation is retrieved, in contrast to the conclusion of~\cite{Nguyen18} that cloud rotation is largely invariant of galactic rotation, in all but the largest clouds. This may be attributed to a smaller number of highly self-gravitating, high-density clumps in our simulations, owing to the efficacy of stellar feedback in disrupting such dense regions, consistent with observations (see below).

In terms of physics and resolution, the simulations most similar to ours are presented by~\cite{Fujimoto18,Fujimoto19}, which include thermal energy and momentum from SNe, thermal energy from HII regions, mass injection from stellar winds, stochastic stellar population synthesis, and ISM heating and cooling. For a sample of GMCs in an isolated Milky Way-like disc galaxy, reaching $\sim 7$~pc resolution in the densest gas, these authors are able to reproduce all of the galaxy-scale, kpc-scale and cloud-scale observables that we show in this work, with the exception of the observed decorrelation between stellar and gas peaks on small spatial scales~\citep{Kruijssen18a,Kruijssen2019,Chevance20}. Given that the major difference between the two sets of simulations is the momentum-injection model for unresolved and marginally-resolved HII regions presented in Section~\ref{Sec::numerical-methods}, we tentatively attribute the adequate (rapid) disruption of gas peaks in our galaxies to this component of our models. This finding will be discussed further in an upcoming paper,~\cite{Jeffreson20b}.

Moving beyond isolated disc galaxies to galactic encounters and mergers,~\cite{Tress20} have compared the ISM on cloud scales in an isolated M51-analogue to the ISM in a simulation of the interaction between M51 and its companion. The authors find that although the interaction produces spiral structure and generates galaxy-scale gas flows, it does not have a significant influence on the global star formation rate or on the fraction of gas that is dense and molecular. Although the influence of galaxy interactions is beyond the scope of this work, the finding of~\cite{Tress20} is broadly consistent with star-forming dense gas that is over-pressured and over-dense, therefore decoupled from the ambient medium in its gravo-turbulent and star-forming properties, as seen for our GMCs. Such gas would be moved around by large-scale gas flows, but its internal properties would not be significantly altered.

Finally, on cosmological scales and at lower resolutions,~\cite{Benincasa2019} have studied GMC populations in Milky Way-like zoom-in simulations selected from the FIRE-2 sample~\citep{Hopkins18}. These authors find a variation of the molecular cloud lifetime with galactocentric radius, confirming observational~\citep[e.g.][]{Kruijssen2019,Chevance20}, numerical~\citep[e.g.][]{DobbsPringle13}, and theoretical~\citep[e.g.][]{Jeffreson+Kruijssen18} findings that the cloud lifetime is environmentally-dependent. They use a complete suite of stellar feedback mechanisms, including radiation pressure, photo-ionisation, stellar winds and supernovae. In a follow-up paper,~\cite{Jeffreson20}, we study the explicit influence of the galactic-dynamical environment on the time-dependent properties of molecular clouds (i.e.~their lifetimes, merger histories and evolution) and will connect these to the instantaneous properties studied in this work. We will then be able to make a more direct comparison to the cosmological-scale work of~\cite{Benincasa2019}.

\subsection{Caveats of our simulations} \label{Sec::caveats}
In order to produce a realistic population of GMCs in our simulations, we require a realistic model of the interstellar medium. In Section~\ref{Sec::galaxy-properties}, we have therefore compared the kpc-scale and galaxy-scale properties of our simulations to the corresponding sets of observables, and have found a high level of agreement. However, our simulations do not include every physical mechanism known to have a relevant influence on the baryon cycle and on the structure of the ISM. Local, higher-resolution studies such as the TIGRESS simulations~\citep{KimCG&Ostriker17,KimCG&Ostriker18} and the SILCC simulations~\citep{Walch15}, demonstrate the relevance of magnetic fields, binary stars, cosmic rays, resolved SN bubbles and the self-consistent injection of momentum from stellar winds and outflows in setting the properties and structure of the three-phase ISM.

Firstly, we do not include the effect of magnetic fields in our simulations. Although GMCs themselves are generally observed to be magnetically super-critical~\citep[see the review by][]{Crutcher12}, the lower-density gas that forms the bulk of the ISM volume is threaded by field-lines and evolves according to the equations of magnetohydrodynamics. There is no clear observational consensus as to the role of these field lines in the formation of molecular clouds~\citep[see review by][]{Hennebelle&Inutsuka19}. The scales on which magnetic field lines are ordered and regular may be similar to the galactic dynamical scales considered in our work~\citep{Beck01}, and may be coupled to galactic-dynamical processes such as differential rotation and the evolution of spiral arm structure~\citep{Beck96,KimCG&Ostriker15b}. As such, large-scale magnetic fields may affect the rates of GMC formation and destruction by galactic dynamics, as well as the environmental trends in the GMC properties presented here. The local velocity anisotropy of the ISM (Section~\ref{Sec::cloud-vel-anisotropy}) is expected to vary substantially with the magnetic field direction~\citep[e.g.][]{Ossenkopf2002,Otto17}, so the introduction of magnetic fields may substantially disrupt the shear-correlated trend in velocity anisotropy shown in Figure~\ref{Fig::betaQ_anisotropy}. Simulations by~\cite{Girichidis18} find that magnetic fields may also control the shapes and fragmentation of the sub-critical overdensities from which GMCs form, with the potential to influence their aspect ratios (Section~\ref{Sec::cloud-ellipticity}).

Although we include mass injection from stellar winds, we do not include the momentum that they impart to the ISM. At solar metallicities, simulations of stellar winds within molecular clouds have demonstrated a profound influence on the masses of the stellar clusters they host~\citep[e.g.][]{Gatto17,Peters17}, and therefore on GMC morphologies and the duration of GMC lifetimes. The evolution of the HII regions within each cloud may also be altered by stellar winds, changing the time-scale over which pre-SN feedback is deposited into the ISM, along with the geometry of the subsequent SN explosions~\citep[e.g.][]{Shull80}. At our resolutions, we expect that the major effect of neglecting stellar wind energy is to shorten the duration of early-time stellar feedback and thus to reduce its overall efficacy in destroying GMCs.

Lastly, the resolution of our simulations is lower than in studies such as TIGRESS and SILCC, such that we do not explicitly resolve the Sedov-Taylor phase of SN blast-wave expansion. Instead, we use the mechanical feedback algorithm outlined in Section~\ref{Sec::SN-and-winds} to compute the final momentum of the blast-wave.~\cite{KimCG&Ostriker17} and~\cite{KimCG&Ostriker18} have shown that the hot gas generated during the resolved Sedov-Taylor phase of expansion is required to correctly reproduce the phase structure of the superbubbles driven by clustered supernovae, and so to correctly model the mass-loading and the phase structure of the galactic outflows associated with these bubbles. However, given that the assembly of our discs is not dependent on the high-redshift phase of galactic evolution, during which strong outflows may strongly affect the structure of the ISM, it is unlikely that the properties of our simulated GMCs and HI clouds are strongly affected by the exact multiphase structure of our winds.~\citep{Muratov2015} in particular have shown that the behaviour of such unresolved winds is roughly-correct in the low-redshift case, and consistent with these results, our model produces outflows out to several kpc above the galactic mid-plane (see Figure~\ref{Fig::disc-morphology}).

In addition to the missing physics outlined above, we note the possibility that our procedure for cloud identification has an influence on the GMC and HI cloud properties presented here, as explored by~\cite{Grisdale18}. These authors model a time-evolving GMC population in a Milky Way-like disc at a spatial resolution of $4.5$~pc, with thermal and kinetic stellar feedback from supernovae, stellar winds and HII regions. In agreement with our results, they find a close match between the properties of their simulated cloud population (specifically scaling relations, masses, sizes, densities and virial parameters) and observations of GMCs in the Milky Way. However, they also note that the cloud scaling relations are significantly steepened when GMCs are identified in 3D rather than 2D. In this work we have aimed to provide a direct comparison to observations by identifying clouds in 2D only, but in the future it will also be possible to examine the ways in which each GMC property is influenced by the number of spatial dimensions included during cloud identification, along with the algorithm used to define the boundaries of each GMC.

\subsection{Future work} \label{Sec::future-work}
In the simulations presented here, we can distinguish clear galactic-dynamical correlations with the time-scales for gravitational free-fall and galactic shearing, as well as with the mid-plane hydrostatic pressure. However, we note that the dynamic range of the free-fall time-scale is small, and that the time-scales for epicyclic perturbations and for cloud-cloud collisions are ten times longer than the minimum dynamical time-scale, such that they play a negligible role in determining the physical properties of GMCs and HI clouds in Milky Way-pressured galaxies. In the future, we will extend the range of environments studied to include early-type galaxies (shorter epicyclic time-scales), spiral galaxies (shorter cloud-collision time-scales), and higher-pressure discs such as those observed at higher redshifts $z \sim 2$ (greater dynamic range in the free-fall time-scale). In the latter case, we expect GMCs to have lower over-pressures and over-densities relative to the galactic mid-plane, and so to be more strongly-coupled to the galactic-dynamical environment.

\section{Summary of results} \label{Sec::conclusion}
In this work, we have investigated the large-scale galactic-dynamical trends in the properties of GMCs and HI clouds within Milky Way-pressured galaxies. We have found that:
\begin{enumerate}
  \item The shortest galactic-dynamical time-scale across Milky Way-pressured environments is either the time-scale for galactic shearing $\tau_\beta$ or the time-scale for gravitational free-fall, $\tau_{\rm ff,g}$. Such environments span $\sim 1$~dex in the Toomre $Q$ parameter, the orbital angular velocity $\Omega$ and the stellar contribution $\phi_{\rm P}$ to the mid-plane hydrostatic pressure, along with all values of the shear parameter from $\beta=0$ (flat rotation curve) to $\beta=1$ (solid-body rotation). Clouds that evolve under the influence of galactic dynamics in such galaxies are therefore most likely to have properties that correlate with $\tau_\beta$ and $\tau_{\rm ff,g}$.
  \item The simulated GMCs in Milky Way-like environments are small, gravitationally-bound and gravitationally-collapsing, as well as highly over-dense and over-pressured relative to the ambient ISM. As such, they are decoupled from the gravitational and turbulent properties of the galactic mid-plane, and follow no galactic-dynamical trends in these properties. That is, the GMC velocity dispersions $\sigma_{\rm H_2}$, virial parameters $\alpha_{\rm vir, H_2}$ and turbulent pressures $P_{\rm turb, H_2}$ are decoupled from the galactic-dynamical environment.
  \item The simulated HI clouds are gravitationally-unbound across all Milky Way-like environments, with much lower pressures and densities than the GMCs. Their velocity dispersions $\sigma_{\rm HI}$, virial parameters $\alpha_{\rm vir, HI}$ and turbulent pressures $P_{\rm turb, HI}$ all display statistically-significant (at the $3\sigma$ confidence level) trends with the mid-plane hydrostatic pressure $P_{\rm mp}$, with the time-scale $\tau_{\rm ff,g}$ for gravitational collapse, and with the time-scale $\tau_\beta$ for galactic shearing.
  \item By contrast with the turbulent/gravitational properties of the GMCs and HI clouds, the rotational properties of both types of clouds display statistically-significant correlations with the galactic-dynamical variables $P_{\rm mp}$, $\tau_{\rm ff,g}$ and $\tau_\beta$. The aspect ratio $\epsilon$ (along the galactic azimuthal direction) increases monotonically with the shear and free-fall rates. The plane-perpendicular angular momentum $L_z$ also increases with the degree of shearing and gravitational stability, but its preferentially-prograde orientation implies that its magnitude is controlled primarily by the strength of the fictitious Coriolis force in the galaxy frame. Steeper trends and higher values of $\epsilon$ and $L_z$ are observed for the HI clouds than for the GMCs. This is likely due to two effects: (1) GMCs have smaller sizes, higher densities and higher pressures than HI clouds, so are less susceptible to galactic-dynamical influences, and (2) GMCs inherit their rotational properties from their parent HI clouds, and the dynamical imprints in these properties are partially erased during their collapse.
  \item Like the cloud aspect ratios and angular momenta, the cloud velocity anisotropy $B_\sigma$ is correlated with the rates of galactic shearing and gravitational free-fall, both for GMCs and for HI clouds. The clouds become more tangentially-biased at higher levels of galactic differential rotation and gravitational stability, indicating an increase in the prevalence of circulating gas about the cloud centre of mass, relative to radially-streaming gas. Both the absolute degree of anisotropy and its galactic-dynamical correlation is increased for GMCs, possibly due to the conservation of angular momentum during their collapse.
  \item The number of GMCs per unit area of the galactic mid-plane sets the kpc-scale star formation rate across Milky Way-pressured galactic environments, which in turn is correlated with the mid-plane pressure, in agreement with observations.
  \item The internal star-forming properties of GMCs are uncorrelated with the global SFR surface density, and are approximately-constant across all galactic-dynamic environments. That is, like the GMC turbulent pressures and velocity dispersions, the star-forming properties of GMCs are decorrelated from the galactic-dynamical environment.
  \item The masses, sizes and surface densities of GMCs and HI clouds are uncorrelated with the galactic-dynamical environment.
  \item All statistically significant, best-fitting relations between cloud properties and galactic-dynamical quantities are tabulated in Appendix~\ref{App::stat-sig}. We expect these predicted correlations to be testable using cloud-scale observations of the atomic and molecular ISM in nearby galaxies, such as those obtained by the PHANGS collaboration~\citep[e.g.][Leroy et al.~in prep.; Rosolowsky et al.~in prep.]{Sun18}.
\end{enumerate}
We also make a comparison between the measured properties of beam-filling GMCs, identified via clump-finding, and molecular gas sight-lines at fixed spatial scale, independent of any clump-finding algorithm. We compare the latter to observations of molecular gas velocity dispersion, surface density, virial parameter and turbulent pressure by~\cite{Sun18,Sun2020}, and find the following:
\begin{enumerate}
  \item The molecular gas sight-lines at $120$-pc resolution have an average over-pressure of $\sim 4 \times$, relative to the galactic mid-plane pressure on kpc-scales: approximately consistent with the average over-pressure of $2.8 \times$ observed at the same resolution by~\cite{Sun2020}. As we increase the fixed resolution through $80$, $45$ and $30$~pc, the over-pressure increases monotonically. We attribute the dependence of over-pressure on resolution to the dependence of the measured cloud surface density on the beam/sight-line filling factor.
  \item Our beam-filling GMCs have an average over-pressure of $\sim 25 \times$ and an over-density of $\sim 100 \times$, relative to the galactic mid-plane. It is therefore unsurprising that the turbulent, gravitational and star-forming properties of GMCs are decoupled from the galactic-dynamical environment.
  \item By contrast, our HI clouds have an average over-pressure of only $\sim 2 \times$, relative to the galactic mid-plane. The HI clouds show clear galactic-dynamical trends in their turbulent and gravitational properties.
  \item Given the above observations, we predict that galactic-dynamical trends {\it can} be obtained for the turbulent, gravitational and star-forming properties of GMCs. Such trends should be visible in higher-pressure galaxies (i.e.~galaxies with $\log{(P/k_{\rm B}~{\rm K}~{\rm cm}^{-3})}>4.5$), where a larger fraction of the total mid-plane gas is molecular, and so the cloud turbulent pressures are closer to the hydrostatic pressures of the galactic mid-plane. This is consistent with the observations of~\cite{Chevance20}, who find that the decoupling of GMC evolution from galactic dynamical processes occurs for kpc-scale molecular gas surface densities of $\Sigma_{{\rm H}_2}<8~{\rm M}_\odot~{\rm pc}^{-2}$.
\end{enumerate}
In closing, we expect the galactic-dynamical trends in GMC and HI cloud properties identified here to produce corresponding trends in their lifecycles. In a follow-up paper, we will discuss the interplay between galactic dynamics and the cloud lifecycle in these simulations.

\section*{Acknowledgements}
We thank the anonymous referee for an attentive and insightful report that improved the presentation of the results in our manuscript. We thank Volker Springel for allowing us access to {\sc Arepo}. We thank Mark Krumholz for calculating and providing the {\sc Despotic} table of CO luminosities for interpolation. The authors acknowledge support by the High Performance and Cloud Computing Group at the Zentrum f\"{u}r Datenverarbeitung of the University of T\"{u}bingen, the state of Baden-W\"{u}rttemberg through bwHPC and the German Research Foundation (DFG) through grant no INST 37/935-1 FUGG. SMRJ, JMDK and MC gratefully acknowledge funding from the Deutsche Forschungsgemeinschaft (DFG, German Research Foundation) through an Emmy Noether Research Group (grant number KR4801/1-1) and the DFG Sachbeihilfe (grant number KR4801/2-1). SMRJ, JMDK and BWK gratefully acknowledge funding from the European Research Council (ERC) under the European Union’s Horizon 2020 research and innovation programme via the ERC Starting Grant MUSTANG (grant agreement number 714907). BWK acknowledges funding in the form of a Postdoctoral Research Fellowship from the Alexander von Humboldt Stiftung. JMDK and SCOG acknowledge funding from the German Research Foundation (DFG) via the collaborative research center (SFB 881, Project-ID 138713538) `The Milky Way System' (subproject B2 for JMDK and B1, B2 and B8 for SCOG). SCOG acknowledges financial support through Germany’s Excellence Strategy project EXC-2181/1 - 390900948 (the Heidelberg STRUCTURES Cluster of Excellence). We are very grateful to Bruce Elmegreen, Mark Krumholz, Dominik Lentrodt, Adam Leroy, Eve Ostriker, Matthew Smith, Jiayi Sun, Robin Tre{\ss} and Jacob Ward for helpful discussions.

\bibliographystyle{mnras}
\bibliography{bibliography}

\bsp

\section*{Supporting information}
\begin{flushleft}
Supplementary data are available at MNRAS online.
\end{flushleft}

\begin{flushleft}
\textbf{Tables.} Catalogues of the physical properties and galactic-dynamical parameters of all GMCs and HI clouds presented in this paper.
\end{flushleft}

\begin{flushleft}
Please note: Oxford University Press is not responsible for the content or functionality of any supporting materials supplied bby the authors. Any queries (other than missing material) should be directed to the corresponding author for the article.
\end{flushleft}

\appendix
\section{Calculation of large-scale galaxy properties} \label{App::analysis}
The results presented in Section~\ref{Sec::results-properties} depend on the large-scale environmental properties of the ISM in each simulated galaxy. Here we describe the data analysis procedures used to calculate the variation of these properties on kpc-scales as 1D radial profiles and as 2D projection maps.

\subsection{1D radial profiles of $(\beta, Q, \Omega, \phi_{\rm P})$}
In order to examine the cloud properties presented in Section~\ref{Sec::results-properties} as a function of the galactic-dynamical environment, we must compute the radial profiles of $\beta$, $Q$, $\Omega$ and $\phi_{\rm P}$, as described in Section~\ref{Sec::dynamical-span}. Between galactocentric radii of $R = 1$~kpc and $R = 13$~kpc, we use $50$ overlapping bins with widths of $1$-kpc, collecting the gas cells $i=1...N$ and star particles $*=1...M$ in each bin by performing a tree-walk of radius $\Delta R = 0.5$~kpc about the bin centre, and discarding gas cells with temperatures $T>10^4$~K. We then calculate the gas line-of-sight velocity dispersion $\sigma_{\rm los, g}$, as
\begin{equation} \label{Eqn::analysis-sigma-los}
\sigma_{\rm los, g}(R) = \frac{\sigma_{\rm g}}{\sqrt{3}} = \sqrt{\frac{\langle |\mathbf{v}_i - \langle \mathbf{v}_i \rangle|^2\rangle}{3}},
\end{equation}
where the angled brackets denote the mass-weighted average over the gas cells in each radial bin, with velocity vectors $\{\mathbf{v}_i\}$. By using the above isotropic approximation, we assume that the turbulent length-scale is much shorter than the length-scale over which galactic rotation is noticeable, and also much shorter than the pressure scale-height of the galactic disc. In addition, we calculate the gas sound speed as
\begin{equation}
c_s = \Big\langle \Big(\frac{k_B T_i}{\mu m_p}\Big)^{\tfrac{1}{2}} \Big\rangle,
\end{equation}
the gas and live stellar surface densities $\Sigma_{\rm g}$ and $\Sigma_{\rm s_{\rm live}}$ as
\begin{equation}
\Sigma_{\rm g, s_{\rm live}}(R) = \frac{1}{\pi[(R+\Delta R)^2 - (R-\Delta R)^2]} \sum_{i,* = 1}^{N,M} m_{i,*},
\end{equation}
and the circular velocity $v_c$ at the mid-plane, as
\begin{equation}\label{Eqn::analysis-v-circ}
v_{\rm c}(R) = \Big\langle-\frac{y_i}{R} v_{x,i} + \frac{x_i}{R} v_{y,i}\Big\rangle.
\end{equation}
Using the circular velocity $v_c$, we obtain the angular velocity $\Omega(R) = v_{\rm c}/R$ and the shear parameter $\beta(R) = \partial \ln{v_{\rm c}}/\partial \ln{R}$, simultaneously smoothing the profile and performing the first-order derivative using a Savitzky-Golay kernel of window-length 35 and polynomial order 4. The parameters $\beta$ and $\Omega$ allow us to calculate the epicyclic frequency $\kappa = \Omega\sqrt{2(1+\beta)}$, and from this quantity we obtain the Toomre $Q$ parameter as
\begin{equation}
Q = \frac{\kappa \sqrt{\sigma^2+c_s^2}}{\pi G \Sigma_{\rm g}}.
\end{equation}
Finally, the parameter $\phi_{\rm P}$ is obtained via Equation~(\ref{Eqn::analysis-phiP}), using the 1D radial profiles of $\sigma_{\rm g}$, $\Sigma_{\rm g}$, $\rho_{\rm s}$ and $\Sigma_{\rm s, live}$.

\subsection{2D projection maps of $\Sigma_{\rm H_2}$, $\Sigma_{\rm HI}$ and $\Sigma_{\rm g}$} \label{Sec::analysis-gas-surfdens-maps}
We produce projection maps of gas surface densities using {\sc Arepo}'s native ray-tracing algorithm. Within each map, the value of the 2D field $F(x,y)$ at every pixel is computed by sending a ray from $1$~kpc below to $1$~kpc above the galactic mid-plane, then taking the density-weighted average of the relevant 3D field $f(x,y,z)$ over the set of intersected gas cells. Only those cells with temperatures $T \la 1 \times 10^4$~K are considered, to ensure that the 2D projection approximates the value of the field at the galactic mid-plane. Given that the scale height of the cold gas in each disc is at maximum $\sim$400~pc, this vertical range safely captures all gas cells of interest. The resulting density-weighted projection $F$ at the position $(x,y)$ may be written as
\begin{equation} \label{Eqn::analysis-2D-proj}
F(x,y) = \Sigma_{\rm g}(x,y)^{-1}\int_{z = -1 \: {\rm kpc}}^{z = +1 \: {\rm kpc}} dz \: \rho_{\rm g} f(x,y,z),
\end{equation}
where $\Sigma_{\rm g}(x,y) = \int^\infty_{-\infty}{\dd z \rho_{\rm g}(z)}$ is the total gas column density at the same position. The corresponding field pairs $[f(x,y,z),F(x,y)]$ for each gas surface density map are given by
\begin{equation} \label{Eqn::analysis-voldens-pairs}
\begin{split}
&f_{\rm H_2} = \frac{2.3 \times 10^{-29}{\rm M}_\odot ({\rm ergs}^{-1})^{-1}}{m_{\rm H}[{\rm M}_\odot]} L_{\rm CO}[{\rm ergs}^{-1}] \\
&f_{\rm HI} = x_{\rm HI} \\
&f_{\rm g} = 1, \\
\end{split}
\end{equation}
where $x_{\rm HI}$ is the abundance of atomic hydrogen per unit mass, and
\begin{equation} \label{Eqn::analysis-surfdens-pairs}
\begin{split}
&F_{\rm H_2} = \Sigma_{\rm H_2} \Sigma_{\rm g}^{-1} \\
&F_{\rm HI} = \Sigma_{\rm HI} \Sigma_{\rm g}^{-1} \\
&F_{\rm g} = 1.
\end{split}
\end{equation}
We then obtain $\Sigma_{\rm H_2}$ or $\Sigma_{\rm HI}$ by taking the product with $\Sigma_{\rm g}$.

\subsection{2D projection maps of $\Sigma_*$} \label{Sec::analysis-stellar-surfdens-maps}
The star particles in {\sc Arepo} are not contained in the Voronoi mesh but are treated as collisionless point particles. As such, we use the \texttt{pynbody.sph} analysis module~\citep{pynbody} to produce the 2D projection maps of all stellar properties. Within each simulation snapshot, an M4 spline SPH kernel is assigned to all star particles, with a variable smoothing length that encloses its $32$ nearest neighbours. The value of the 2D field at each pixel is then computed by taking the kernel-weighted average of the relevant 3D field $f(x,y,z)$ over the set of star particles whose smoothing lengths overlap with the coordinate $(x,y)$.

\subsection{2D projection maps of $\Sigma_{\rm SFR}$}
The central row of panels in Figure~\ref{Fig::KS-rln} displays 2D maps of the SFR surface density $\Sigma_{\rm SFR}$ for each simulated galaxy. These are produced by calculating the stellar surface density maps for young stars only, with ages $0$-$5$~Myr, approximately-matching the ages of stars traced by H-$\alpha$ emission. We simply divide the values in this map by the time interval ($5$~Myr) to obtain an estimate of $\Sigma_{\rm SFR}$.

\subsection{2D projection maps of $\phi_{\rm P}$} \label{Sec::analysis-phiP}
To quantify the relative influence of the stellar and gaseous components of the ISM in producing gravitational instability and setting the mid-plane pressure of the galactic disc, we have used the parameter $\phi_{\rm P}$ introduced in~\cite{Elmegreen89}, defined as
\begin{equation}
\phi_{\rm P} = 1 + \frac{\Sigma_{\rm s}}{\Sigma_{\rm g}}\frac{\sigma_{\rm g}}{\sigma_{\rm s}},
\end{equation}
where $\Sigma_{\rm g}$, $\Sigma_{\rm s}$ are the gas and stellar surface densities, and $\sigma_{\rm g}$, $\sigma_{\rm s}$ are the gas and stellar velocity dispersions. Given that we use an external potential to model the stellar disc and bulge components, we are not able to obtain an accurate estimate of $\sigma_{\rm s}$. We therefore additionally assume that the scale-height of the stellar disc is significantly higher than that of the gas disc, so that the stars maintain their own state of collisionless equilibrium, independent of the gas~\citep[c.f.][]{BlitzRosolowsky2004}. This leads to
\begin{equation} \label{Eqn::analysis-phiP}
\phi_{\rm P} = 1 + \frac{\sigma_{\rm g}}{\Sigma_{\rm g}} \sqrt{\frac{2\rho_{\rm s}}{\pi G}},
\end{equation}
where $\rho_{\rm s}$ is the mid-plane stellar volume density, and we compute this quantity by combining the contribution of the live stellar particles with that of the external potential, such that
\begin{equation}
\rho_{\rm s} = \frac{1}{4\pi G} \frac{1}{R} \frac{\partial}{\partial R} \Big[R \frac{\partial (\Phi_{\rm d} + \Phi_{\rm b})}{\partial R}\Big]\Big|_{z=0} + \frac{\Sigma_{\rm s_{\rm live}}}{4H},
\end{equation}
where $\Sigma_{\rm s_{\rm live}}$ is the stellar surface density of the live stellar particles (calculated per Section~\ref{Sec::analysis-stellar-surfdens-maps}), $H$ is the stellar disc scale-height (calculated per Section~\ref{Sec::analysis-scale-height}), and $\Phi_{\rm d}$ and $\Phi_{\rm b}$ are the disc and bulge components of the applied potential, given by Equations~(\ref{Eqn::d-ptnl}) and~(\ref{Eqn::b-ptnl}), respectively. The final two quantities in the 2D projection of Equation~(\ref{Eqn::analysis-phiP}) are then the total gas surface density $\Sigma_{\rm g}$ (Section~\ref{Sec::analysis-gas-surfdens-maps}) and the total gas velocity dispersion $\sigma_{\rm g}$ (Section~\ref{Sec::analysis-gas-veldisp}).

\subsection{Gas velocity dispersion, $\sigma_{\rm g}$} \label{Sec::analysis-gas-veldisp}
In order to compute the 2D projection of $\phi_{\rm P}$, we require a 2D projection of the gas velocity dispersion $\sigma_{\rm g}$. The three-dimensional turbulent velocity dispersion at the position of each gas cell is not tracked during our simulations. We calculate this quantity in post-processing using the particle positions, densities and velocity vectors in each simulation snapshot, according to~\cite{Gensior20}. For a given Voronoi cell, $\sigma_{\rm g}$ is given by
\begin{equation}
\sigma_{\rm g} = \langle |\mathbf{v}_i - \langle \mathbf{v}_i \rangle|^2 \rangle,
\end{equation}
where the angled brackets denote a (cubic spline) kernel-weighted average over the set of nearest-neighbour gas cells with velocity vectors $\{\mathbf{v}_i\}$, chosen according to the variable tree-walk radius defined in Equation (7) of~\cite{Gensior20}. Briefly, this radius is set to $|\langle\rho_{\rm g}\rangle/\langle \nabla{\rho_{\rm g}}\rangle|$, where $\langle\nabla{\rho_{\rm g}}\rangle$ is the kernel-weighted average of the volume density gradient with respect to the radial distance from the central cell. In every case, the smoothing length for the cubic spline kernel is chosen to enclose the $32$ nearest neighbour cells. We refer the reader to the cited work for a more detailed explanation.

Following the computation of the $\sigma_{\rm g}$ for each gas cell, we compute its 2D density-weighted projection along the $z$-axis using Equation~(\ref{Eqn::analysis-2D-proj}). The corresponding 2D projection of the line-of-sight velocity dispersion is then given by
\begin{equation}
\sigma_{\rm los} = \sigma_{\rm g}/\sqrt{3}.
\end{equation}

\subsection{Disc scale height and scale-length} \label{Sec::analysis-scale-height}
The gas and stellar disc scale-heights $z_{\rm d}$ or disc scale-lengths $R_{\rm d}$ for an exponential disc are uniquely determined by the ratio of the mass enclosed within $z_{\rm d}$ or $R_{\rm d}$, relative to the total mass of the galactic disc, such that
\begin{align}
\int^{z_{\rm d}}_0 \rho(R, z) \: \dd z &= (1-e^{-1})  \int^\infty_0 \rho(R, z) \: \dd z \\
\int^{R_{\rm d}}_0 \rho(R, z) \: \dd R &= (1-2e^{-1})  \int^\infty_0 \rho(R, z) \: \dd R.
\end{align}
Azimuthally-averaged values for each disc are given in Table~\ref{Tab::params} and in Figure~\ref{Fig::betaQ_explore}.

\section{Statistical significance of galactic-dynamical trends} \label{App::stat-sig}
In Table~\ref{Tab::results} and Figure~\ref{Fig::correlation-summary}, we present the relative statistical significance of the correlation between each physical cloud property and the two minimum galactic-dynamical time-scales ($\tau_\beta$ and $\tau_{\rm ff,g}$) across Milky Way-pressured environments, along with the pressure $P_{\rm mp}$ at the galactic mid-plane. We determine the statistical significance by performing a non-linear least squares optimisation to fit a straight line $y = ax + b$ to each data set, where $x$ corresponds to a set of $50$ non-overlapping bins in one of the independent variables $\tau_\beta$, $\tau_{\rm ff,g}$ or $P_{\rm mp}$. The dependent variable $y$ is mean value of a given cloud property in each bin, $a$ is the slope of the fit and $b$ is its $y$-axis intercept. We use the {\sc curvefit} function from {\sc SciPy}~\citep{SciPy-NMeth2020} to perform the optimisation and to calculate the covariance matrix of the dataset as
\begin{equation}
C(x,y) =
\begin{bmatrix}
    {\rm Var}(x) & {\rm Cov}(x,y) \\
    {\rm Cov}(x,y) & {\rm Var}(y) \\
\end{bmatrix}.
\end{equation}
We then compute the eigenvalues $(E_1, E_2)$ and eigenvectors $(\hat{\mathbf{e}}_1, \hat{\mathbf{e}}_2)$ of the covariance matrix to obtain the confidence ellipse of the parameters $a$ and $b$. The confidence ellipse is centred on the optimum values $(a=a_0, b=b_0)$ for the fit parameters, has major and minor axes lengths of $N\sqrt{E_1}=3\sqrt{E_1}$ and $N\sqrt{E_2}=3\sqrt{E_2}$ at the $3\sigma$ level, and has a position angle of $\theta = \arctan{(\hat{\mathbf{e}}_2 \cdot \hat{\mathbf{x}}/\hat{\mathbf{e}}_1 \cdot \hat{\mathbf{x}})}$. Crucially, the confidence ellipse shows both the correlation and the spread of the fit parameters. For each of our cloud properties, we find that the variance of the slope is typically small, however it displays a significant degree of correlation with the $y$-intercept, which has a much larger variance. The correlation can be seen in the elongation of any of the confidence ellipses presented in Figures~\ref{Fig::HI-veldisp-stat} to~\ref{Fig::H2-div-stat-veldisp}. Rather than using the variance $E_1$ on the fit parameter $a$ to calculate its error, we therefore consider the slope to be consistent with zero at the $3\sigma$ level if the ellipse crosses the line $a=0$ at any point. If the ellipse makes no such crossing, then we consider that the data are consistent with a non-zero correlation between independent and dependent variables at the $3\sigma$ level. Correspondingly, the error on the slope is quoted as the projection of the $1\sigma$ confidence ellipse onto the $a$-axis of the fit parameter space.

In the case of the two galactic dynamical time-scales $\tau_\beta$ and $\tau_{\rm ff,g}$, we consider two distinct dynamical regimes. In one, the time-scale $\tau_X$ is competitive with the minimum galactic-dynamical time-scale $\tau_{\rm min}$ to within a factor of two, such that $\tau_X \le \tau_{\rm min}$. In the other, it is longer and therefore less competitive, such that $\tau_X > \tau_{\rm min}$. We expect that in the former regime, the influence of a given dynamical mechanism on the properties of GMCs and HI clouds is greater. As discussed in Section~\ref{Sec::Theory}, a key result of this work is that the time-scale for gravitational free-fall $\tau_{\rm ff,g}$ is competitive across all Milky Way-pressured environments, and as such we need only to consider the first regime for $\tau_{\rm ff,g}$, represented by the filled data points in Figures~\ref{Fig::HI-veldisp-stat} to~\ref{Fig::H2-div-stat-veldisp}. Conversely, the time-scale $\tau_\beta$ for galactic shearing partitions the simulation data between the two regimes. Where a statistically-significant correlation with $\tau_\beta$ is present, we observe a clear break in the slope at $\tau_\beta = \tau_{\rm min}$, and so we fit the two regimes separately.

In Tables~\ref{Tab::P-mp-corr}, \ref{Tab::tau-ffg-corr} and \ref{Tab::tau-beta-corr}, we list those cloud properties that display statistically-significant correlations with the dynamical variables $P_{\rm mp}$, $\tau_{\rm ff,g}$ and $\tau_\beta$, respectively. We give the range of the correlation in each independent variable, its slope and its intercept, along with the associated errors.

\label{lastpage}
\clearpage
\begin{table*}
\begin{center}
\label{Tab::P-mp-corr}
  \caption{Statistically-significant correlations between each cloud property (dependent variable, $y$) and the logarithm of the hydrostatic mid-plane pressure, $P_{\rm mp}$ (independent variable, $x$).}
  \begin{tabular}{@{}l||c|c|c||c @{}}
  \hline
   $y$-property & \multicolumn{3}{c||}{$y \propto \log{(P_{\rm mp}/k_{\rm B} {\rm K} \: {\rm cm}^{-3})}$} & Figure \\
  \hline
    & $x$-range & Slope & $y$-intercept & \\
  \hline
    $\sigma_{\rm HI}/{\rm kms}^{-1}$ & $3.6 \rightarrow 4.3$ & $4.30 \pm 0.19$ & $-9.5 \pm 0.8$ & \ref{Fig::HI-veldisp-stat} \\

    $\log{\alpha_{\rm vir, HI}}$ & $3.6 \rightarrow 4.3$ & $0.42 \pm 0.02$ & $-0.29 \pm 0.06$ & \ref{Fig::HI-avir-stat} \\

    $\log{(P_{\rm turb, HI}/k_{\rm B} {\rm K} \: {\rm cm}^{-3})}$ & $3.6 \rightarrow 4.3$ & $0.62 \pm 0.02$ & $1.91 \pm 0.07$ & \ref{Fig::HI-Pturb-stat} \\
    
    $\epsilon_{\rm H_2}$ & $3.6 \rightarrow 4.3$ & $0.54 \pm 0.05$ & $0.18 \pm 0.19$ & \ref{Fig::H2-ellipticity-stat} \\

    $\epsilon_{\rm HI}$ & $3.6 \rightarrow 4.3$ & $0.74 \pm 0.09$ & $-0.6 \pm 0.4$ & \ref{Fig::HI-ellipticity-stat} \\

    $\log{(L_{z, {\rm H_2}}/{\rm pc} \: {\rm kms}^{-1})}$ & $3.6 \rightarrow 4.3$ & $0.51 \pm 0.03$ & $-1.40 \pm 0.13$ & \ref{Fig::H2-angmom-stat} \\

    $\log{(L_{z, {\rm HI}}/{\rm pc} \: {\rm kms}^{-1})}$ & $3.6 \rightarrow 4.3$ & $0.83 \pm 0.05$ & $-1.75 \pm 0.20$ & \ref{Fig::HI-angmom-stat} \\

    $B_{\sigma, {\rm H_2}}$ & $3.6 \rightarrow 4.3$ & $-0.49 \pm 0.06$ & $1.4 \pm 0.3$ & \ref{Fig::H2-anisotropy-stat} \\

    $\log{(\Sigma_{N_{\rm cl}, {\rm H_2}}/{\rm kpc}^{-2})}$ & $3.6 \rightarrow 4.3$ & $0.391 \pm 0.004$ & $-0.69 \pm 0.01$ & \ref{Fig::H2-no-density-stat} \\

    $\log{(\Sigma_{N_{\rm cl}, {\rm HI}}/{\rm kpc}^{-2})}$ & $3.6 \rightarrow 4.3$ & $0.376 \pm 0.003$ & $-0.643 \pm 0.013$ & \ref{Fig::HI-no-density-stat} \\
\end{tabular}
\end{center}
\end{table*}

\begin{table*}
\begin{center}
\label{Tab::tau-ffg-corr}
  \caption{Statistically-significant correlations between each cloud property (dependent variable, $y$) and the logarithm of the mid-plane free-fall time-scale $\tau_{\rm ff,g}$ (independent variable, $x$).}
  \begin{tabular}{@{}l||c|c|c||c @{}}
  \hline
   $y$-property & \multicolumn{3}{c||}{$y \propto \log{(\tau_{\rm ff,g}/{\rm Myr})}$} & Figure \\
  \hline
    & $x$-range & Slope & $y$-intercept & \\
  \hline
    $\sigma_{\rm HI}/{\rm kms}^{-1}$ & $1.4 \rightarrow 1.9$ & $-4.0 \pm 0.6$ & $14.1 \pm 1.1$ & \ref{Fig::HI-veldisp-stat} \\

    $\log{\alpha_{\rm vir, HI}}$ & $1.4 \rightarrow 1.9$ & $-0.28 \pm 0.06$ & $1.84 \pm 0.1$ & \ref{Fig::HI-avir-stat} \\

    $\log{(P_{\rm turb, HI}/k_{\rm B} {\rm K} \: {\rm cm}^{-3})}$ & $1.4 \rightarrow 1.9$ & $-0.84 \pm 0.05$ & $5.73 \pm 0.08$ & \ref{Fig::HI-Pturb-stat} \\
    
    $\epsilon_{\rm H_2}$ & $1.4 \rightarrow 1.9$ & $-0.83 \pm 0.08$ & $3.71 \pm 0.14$ & \ref{Fig::H2-ellipticity-stat} \\

    $\epsilon_{\rm HI}$ & $1.4 \rightarrow 1.9$ & $-1.8 \pm 0.2$ & $5.2 \pm 0.3$ & \ref{Fig::HI-ellipticity-stat} \\

    $\log{(L_{z, {\rm H_2}}/{\rm pc} \: {\rm kms}^{-1})}$ & $1.4 \rightarrow 1.9$ & $-1.26 \pm 0.05$ & $2.70 \pm 0.09$ & \ref{Fig::H2-angmom-stat} \\

    $\log{(L_{z, {\rm HI}}/{\rm pc} \: {\rm kms}^{-1})}$ & $1.4 \rightarrow 1.9$ & $-2.04 \pm 0.10$ & $4.91 \pm 0.17$ & \ref{Fig::HI-angmom-stat} \\

    $B_{\sigma, {\rm H_2}}$ & $1.4 \rightarrow 1.9$ & $1.05 \pm 0.12$ & $-2.3 \pm 0.2$ & \ref{Fig::H2-anisotropy-stat} \\

    $\log{(\Sigma_{N_{\rm cl}, {\rm H_2}}/{\rm kpc}^{-2})}$ & $1.4 \rightarrow 1.9$ & $-1.44 \pm 0.06$ & $3.24 \pm 0.10$ & \ref{Fig::H2-no-density-stat} \\

    $\log{(\Sigma_{N_{\rm cl}, {\rm HI}}/{\rm kpc}^{-2})}$ & $1.4 \rightarrow 1.9$ & $-0.75 \pm 0.05$ & $1.98 \pm 0.10$ & \ref{Fig::HI-no-density-stat} \\
\end{tabular}
\end{center}
\end{table*}

\begin{table*}
\begin{center}
\label{Tab::tau-beta-corr}
  \caption{Statistically-significant correlations between each cloud property (dependent variable, $y$) and the logarithm of the time-scale $\tau_\beta$ for galactic shearing (independent variable, $x$).}
  \begin{tabular}{@{}l||c|c|c||c @{}}
  \hline
   $y$-property & \multicolumn{3}{c||}{$y \propto \log{(\tau_\beta/{\rm Myr})}$} & Figure \\
  \hline
    & $x$-range & Slope & $y$-intercept & \\
  \hline
    $\sigma_{\rm HI}/{\rm kms}^{-1}$ & $1.2 \rightarrow 2.1$ & $-2.8 \pm 0.3$ & $13.0 \pm 0.6$ & \ref{Fig::HI-veldisp-stat} \\

    $\log{\alpha_{\rm vir, HI}}$ & $1.2 \rightarrow 2.1$ & $-0.22 \pm 0.02$ & $1.80 \pm 0.03$ & \ref{Fig::HI-avir-stat} \\

    $\log{(P_{\rm turb, HI}/k_{\rm B} {\rm K} \: {\rm cm}^{-3})}$ & $1.2 \rightarrow 2.1$ & $-0.36 \pm 0.02$ & $5.05 \pm 0.04$ & \ref{Fig::HI-Pturb-stat} \\
    
    $\epsilon_{\rm H_2}$ & $1.2 \rightarrow 2.1$ & $-0.39 \pm 0.03$ & $3.10 \pm 0.05$ & \ref{Fig::H2-ellipticity-stat} \\

    $\epsilon_{\rm HI}$ & $1.2 \rightarrow 2.1$ & $-0.81 \pm 0.04$ & $3.89 \pm 0.07$ & \ref{Fig::HI-ellipticity-stat} \\

    $\log{(L_{z, {\rm H_2}}/{\rm pc} \: {\rm kms}^{-1})}$ & $1.2 \rightarrow 2.1$ & $-0.41 \pm 0.02$ & $1.41 \pm 0.04$ & \ref{Fig::H2-angmom-stat} \\

    $\log{(L_{z, {\rm HI}}/{\rm pc} \: {\rm kms}^{-1})}$ & $1.2 \rightarrow 2.1$ & $-0.79 \pm 0.05$ & $3.04 \pm 0.10$ & \ref{Fig::HI-angmom-stat} \\

    $B_{\sigma, {\rm H_2}}$ & $1.2 \rightarrow 2.1$ & $0.35 \pm 0.05$ & $-1.23 \pm 0.07$ & \ref{Fig::H2-anisotropy-stat} \\

    $\log{(\Sigma_{N_{\rm cl}, {\rm H_2}}/{\rm kpc}^{-2})}$ & $1.2 \rightarrow 2.1$ & $-0.77 \pm 0.04$ & $2.29 \pm 0.06$ & \ref{Fig::H2-no-density-stat} \\

    $\log{(\Sigma_{N_{\rm cl}, {\rm HI}}/{\rm kpc}^{-2})}$ & $1.2 \rightarrow 2.1$ & $-0.72 \pm 0.06$ & $1.97 \pm 0.09$ & \ref{Fig::HI-no-density-stat} \\
\end{tabular}
\end{center}
\end{table*}

\begin{figure*}
  \label{Fig::H2-veldisp-stat}
  \includegraphics[width=.9\linewidth]{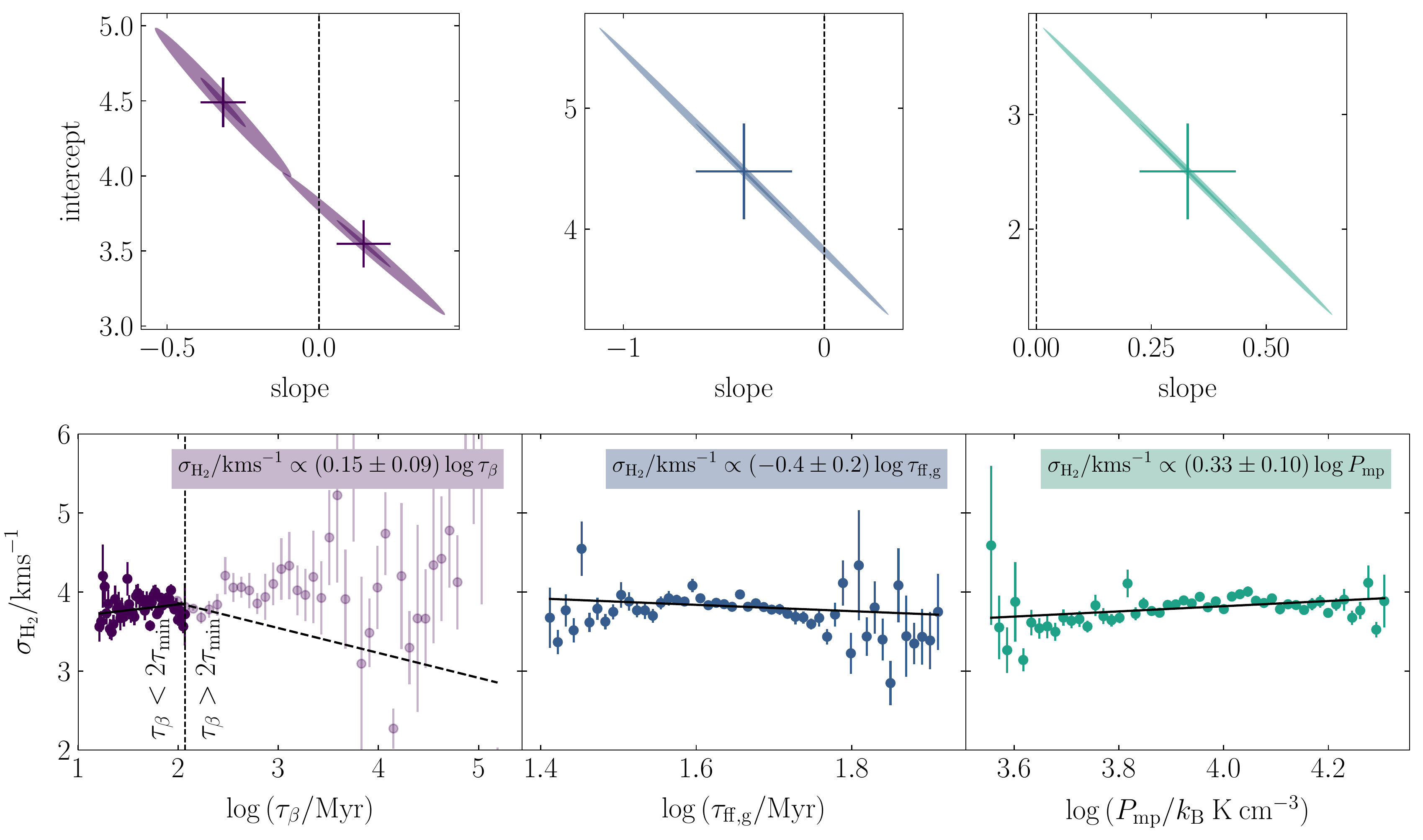}
  \caption{{\it Lower panels:} Mean line-of-sight GMC velocity dispersion $\sigma_{\rm H_2}$ as a function of the time-scale for galactic shearing ($\tau_\beta$, left), the time-scale for gravitational free-fall ($\tau_{\rm ff,g}$, centre) and the mid-plane pressure ($P_{\rm mp}$, right). The error-bars correspond to the standard deviation on the mean for the distribution of values in each bin. The equations in each panel give the non-linear least-squares fit to the data in each panel. In the case of the shear time-scale (left), the data are fitted in two distinct regimes: one in which the shear time-scale is shorter than twice the minimum galactic-dynamical time-scale $\tau_{\rm min}$ (dark purple points, left of the vertical dashed line) and one in which it is longer (transparent points, right of the vertical dashed line). {\it Upper panels:} The $1\sigma$ (dark-coloured) and $3\sigma$ (light-coloured) confidence ellipses for each fit. The error-bars correspond to the projection of the $1\sigma$ confidence ellipse onto each of the parameter axes.}
\end{figure*}

\begin{figure*}
  \label{Fig::HI-veldisp-stat}
  \includegraphics[width=.9\linewidth]{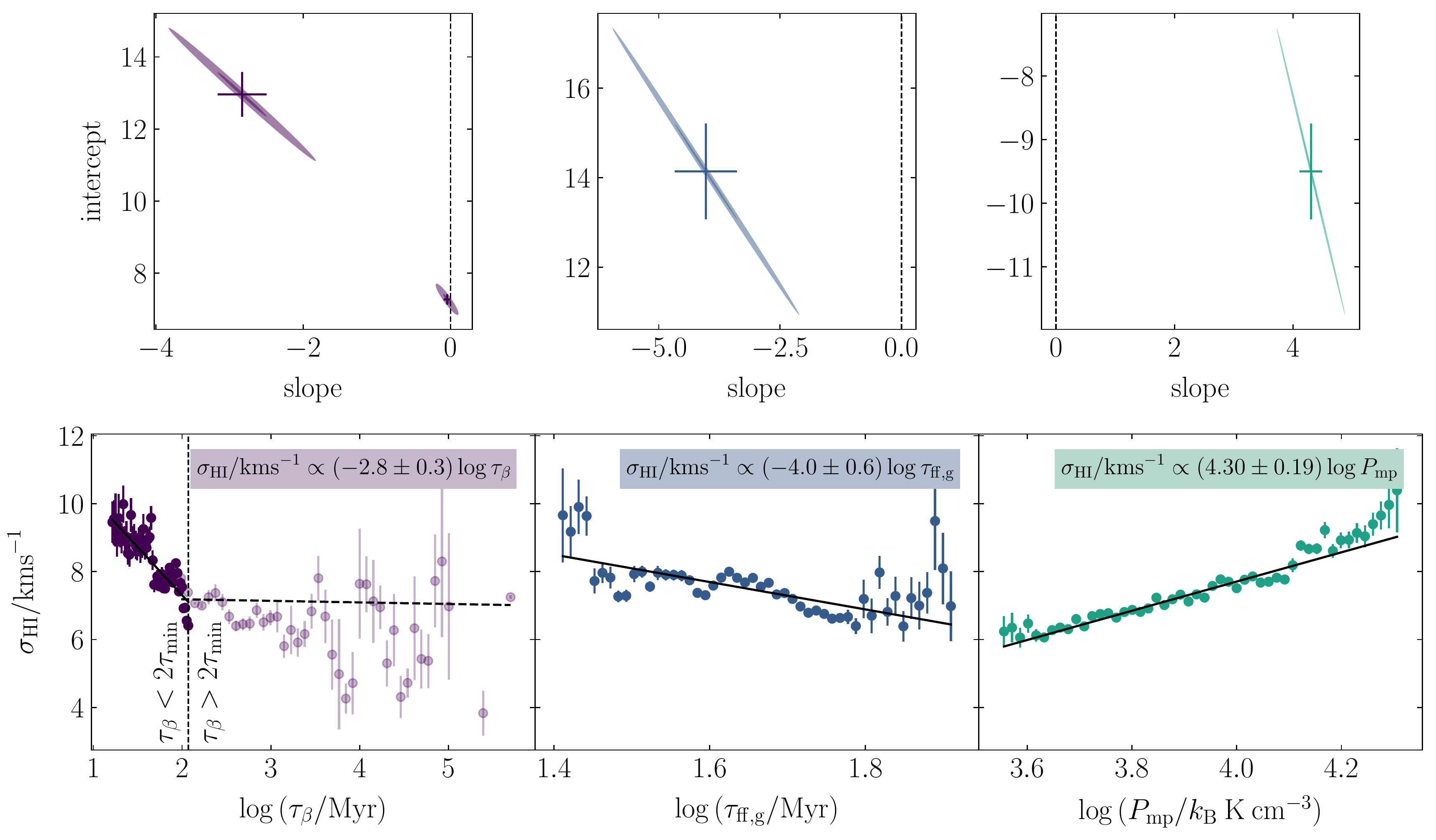}
  \caption{Same as Figure~\protect\ref{Fig::H2-veldisp-stat}, but for the HI cloud velocity dispersion, $\sigma_{\rm HI}$.}
\end{figure*}

\begin{figure*}
  \label{Fig::H2-avir-stat}
  \includegraphics[width=.9\linewidth]{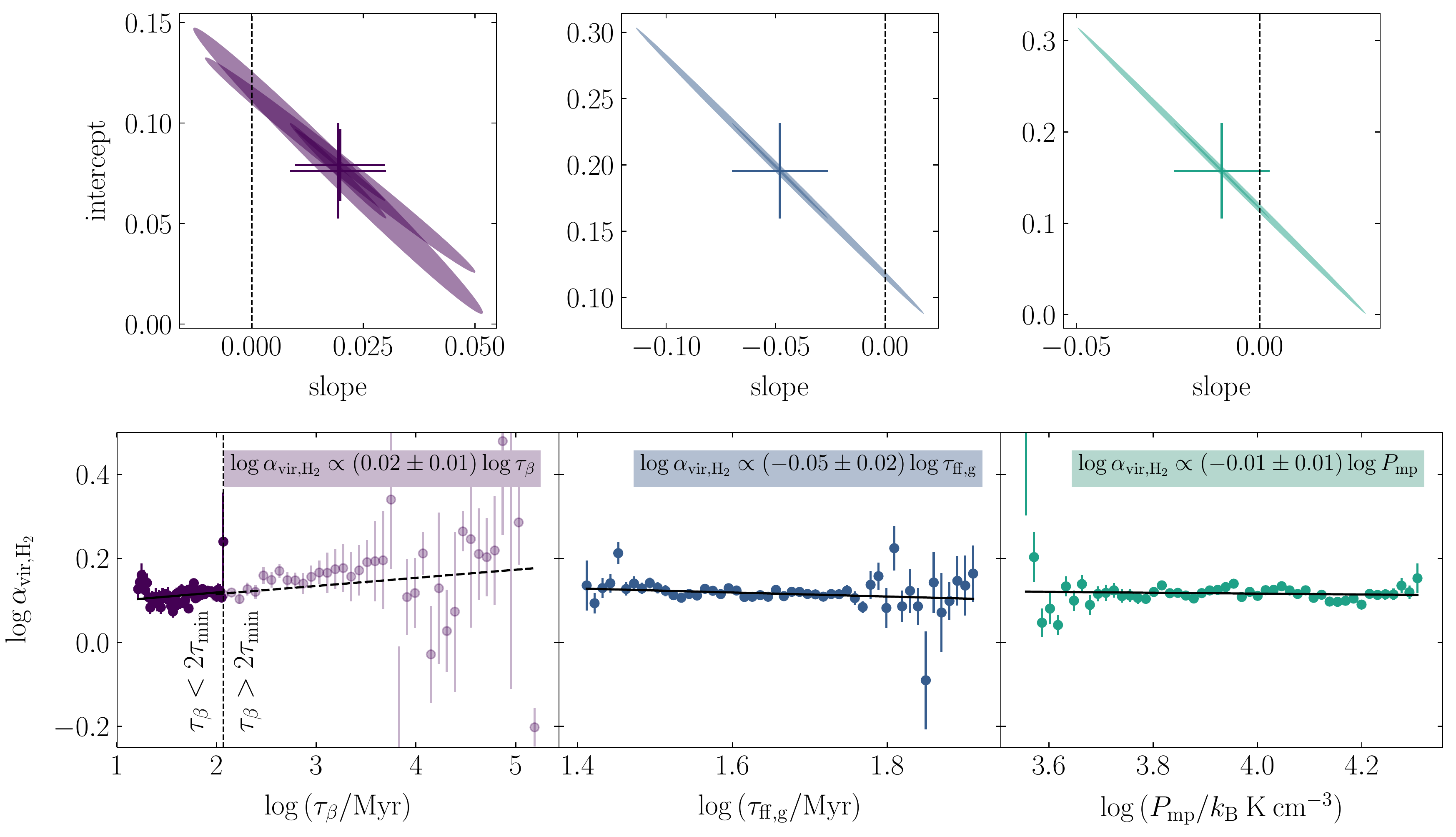}
  \caption{Same as Figure~\protect\ref{Fig::H2-veldisp-stat}, but for the GMC virial parameter, $\alpha_{\rm vir, H_2}$.}
\end{figure*}

\begin{figure*}
  \label{Fig::HI-avir-stat}
  \includegraphics[width=.9\linewidth]{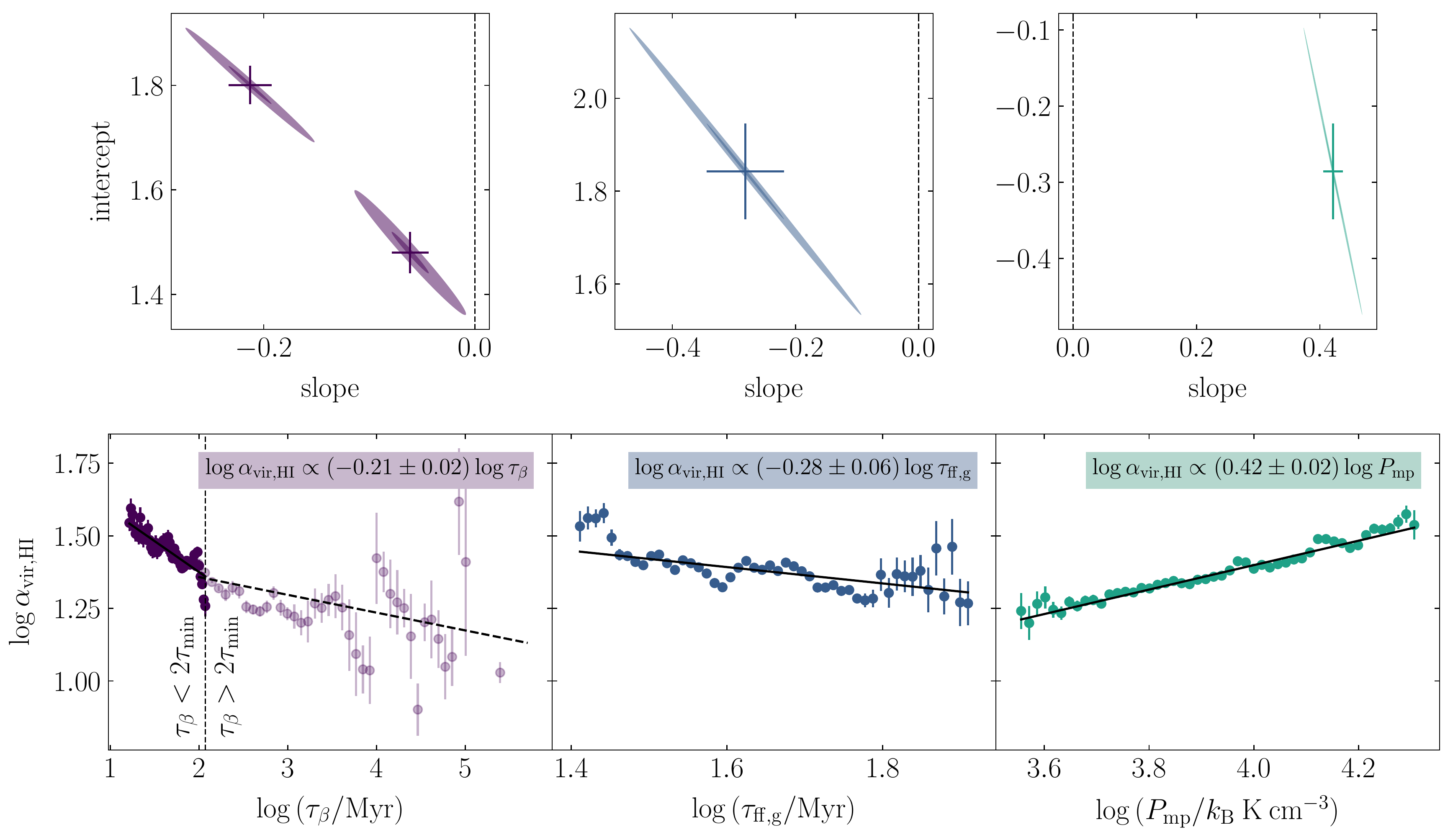}
  \caption{Same as Figure~\protect\ref{Fig::H2-veldisp-stat}, but for the HI cloud virial parameter, $\alpha_{\rm vir, HI}$.}
\end{figure*}

\begin{figure*}
  \label{Fig::H2-Pturb-stat}
  \includegraphics[width=.9\linewidth]{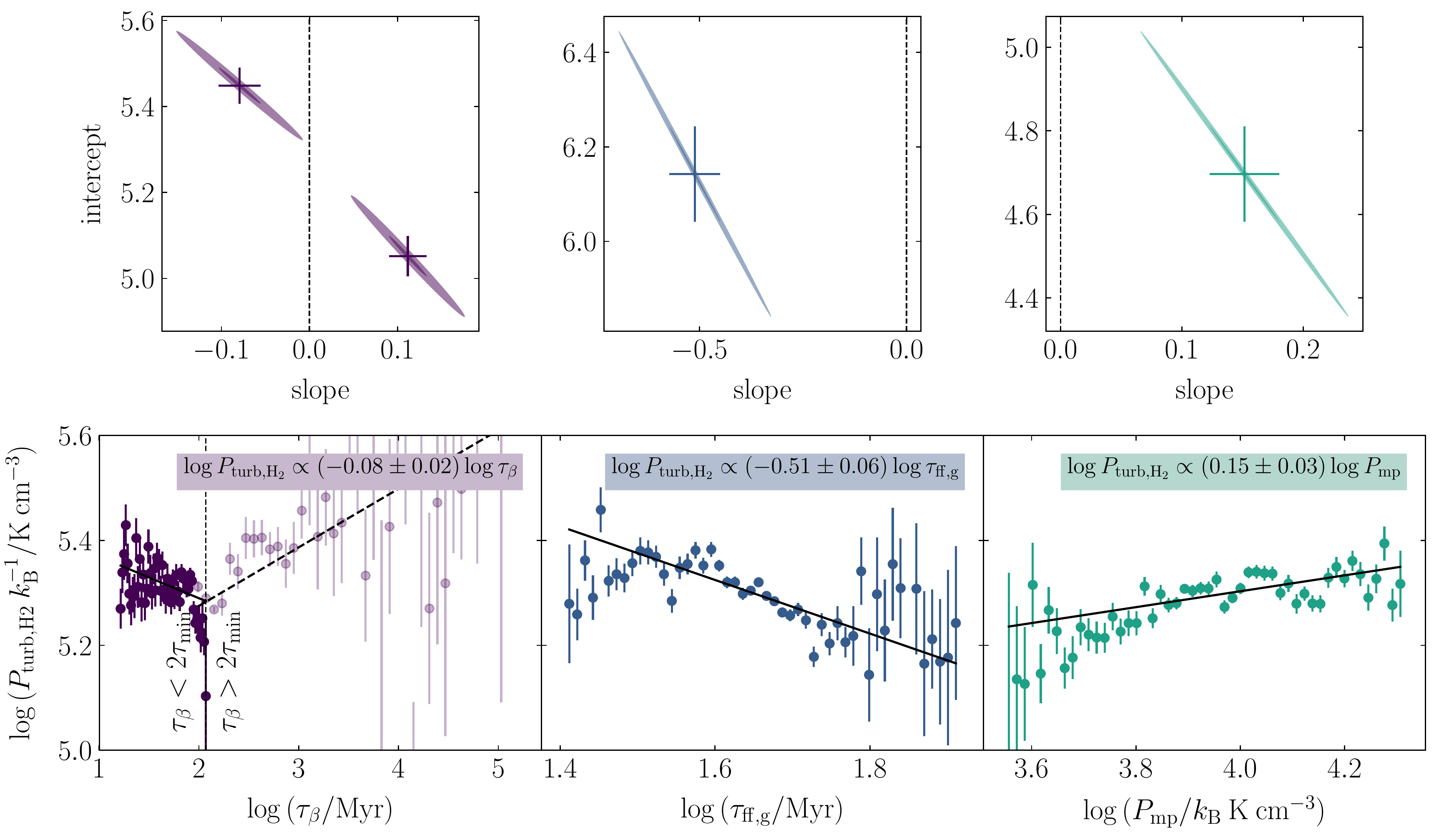}
  \caption{Same as Figure~\protect\ref{Fig::H2-veldisp-stat}, but for the GMC internal turbulent pressure, $P_{\rm turb, H_2}$.}
\end{figure*}

\begin{figure*}
  \label{Fig::HI-Pturb-stat}
  \includegraphics[width=.9\linewidth]{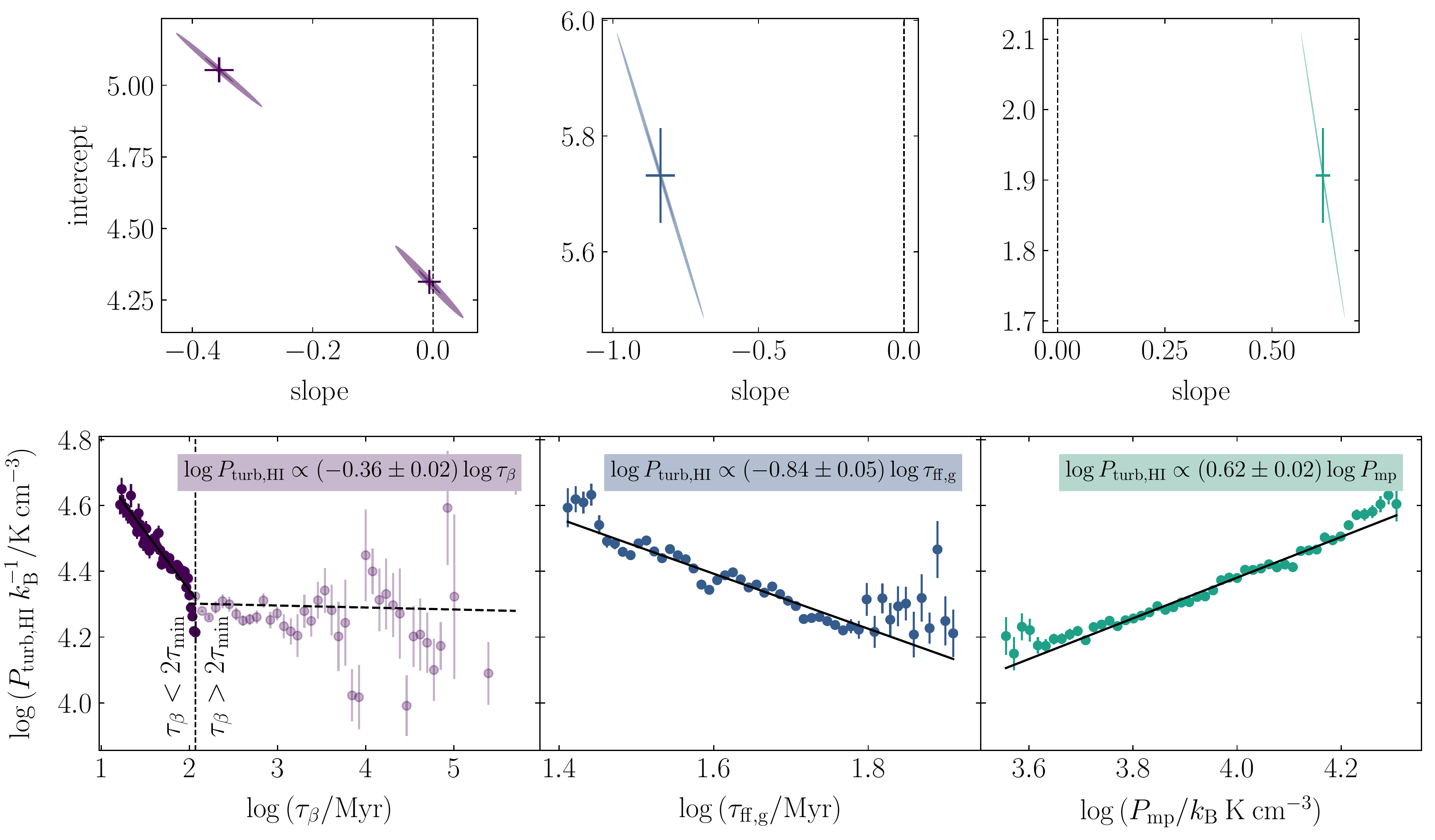}
  \caption{Same as Figure~\protect\ref{Fig::H2-veldisp-stat}, but for the HI cloud internal turbulent pressure, $P_{\rm turb, HI}$.}
\end{figure*}

\begin{figure*}
  \label{Fig::H2-div-stat}
  \includegraphics[width=.9\linewidth]{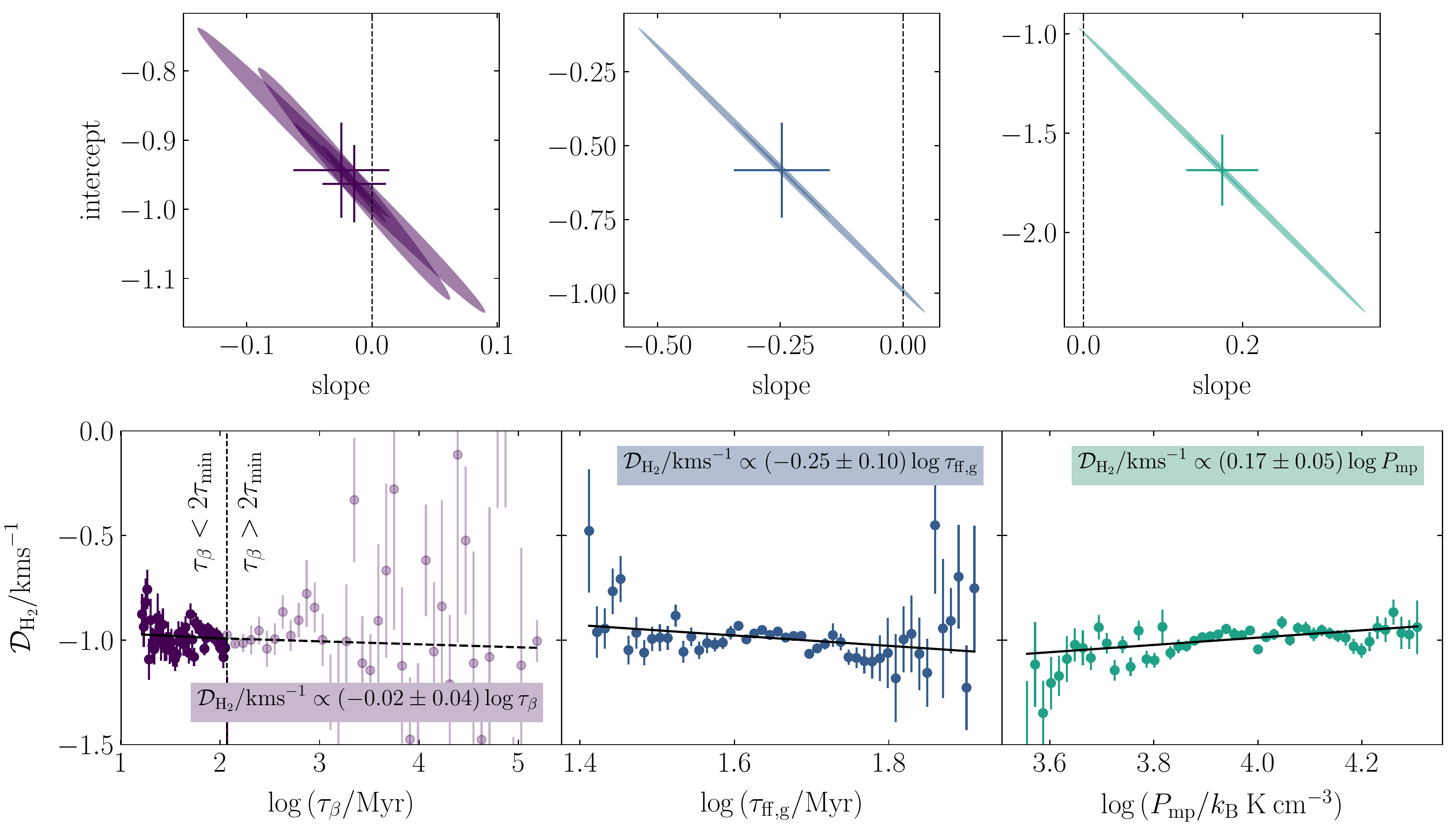}
  \caption{Same as Figure~\protect\ref{Fig::H2-veldisp-stat}, but for the GMC velocity divergence, $\mathcal{D}_{\rm H_2}$.}
\end{figure*}

\begin{figure*}
  \label{Fig::HI-div-stat}
  \includegraphics[width=.9\linewidth]{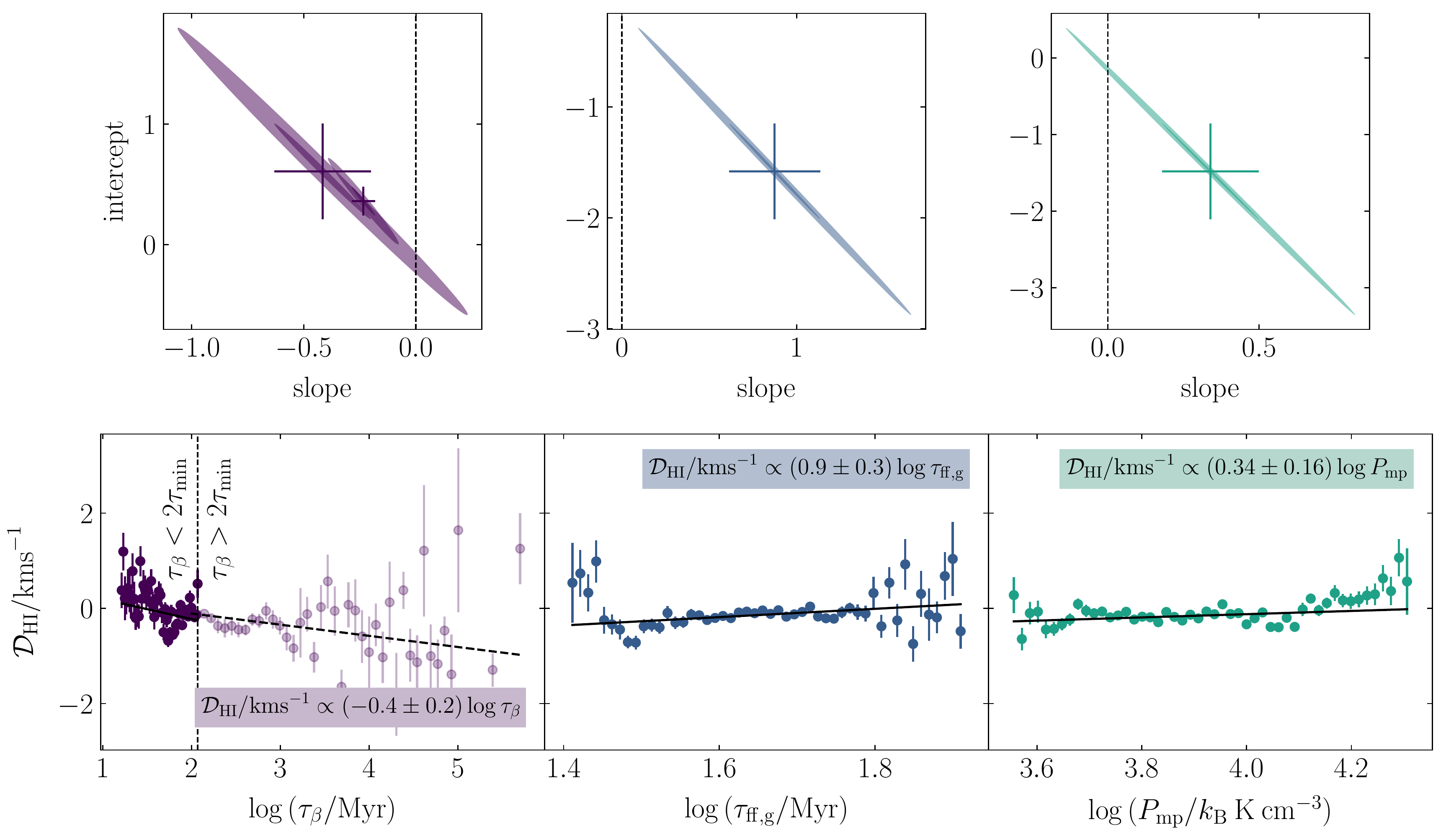}
  \caption{Same as Figure~\protect\ref{Fig::H2-veldisp-stat}, but for the HI cloud velocity divergence, $\mathcal{D}_{\rm HI}$.}
\end{figure*}

\begin{figure*}
  \label{Fig::H2-ellipticity-stat}
  \includegraphics[width=.9\linewidth]{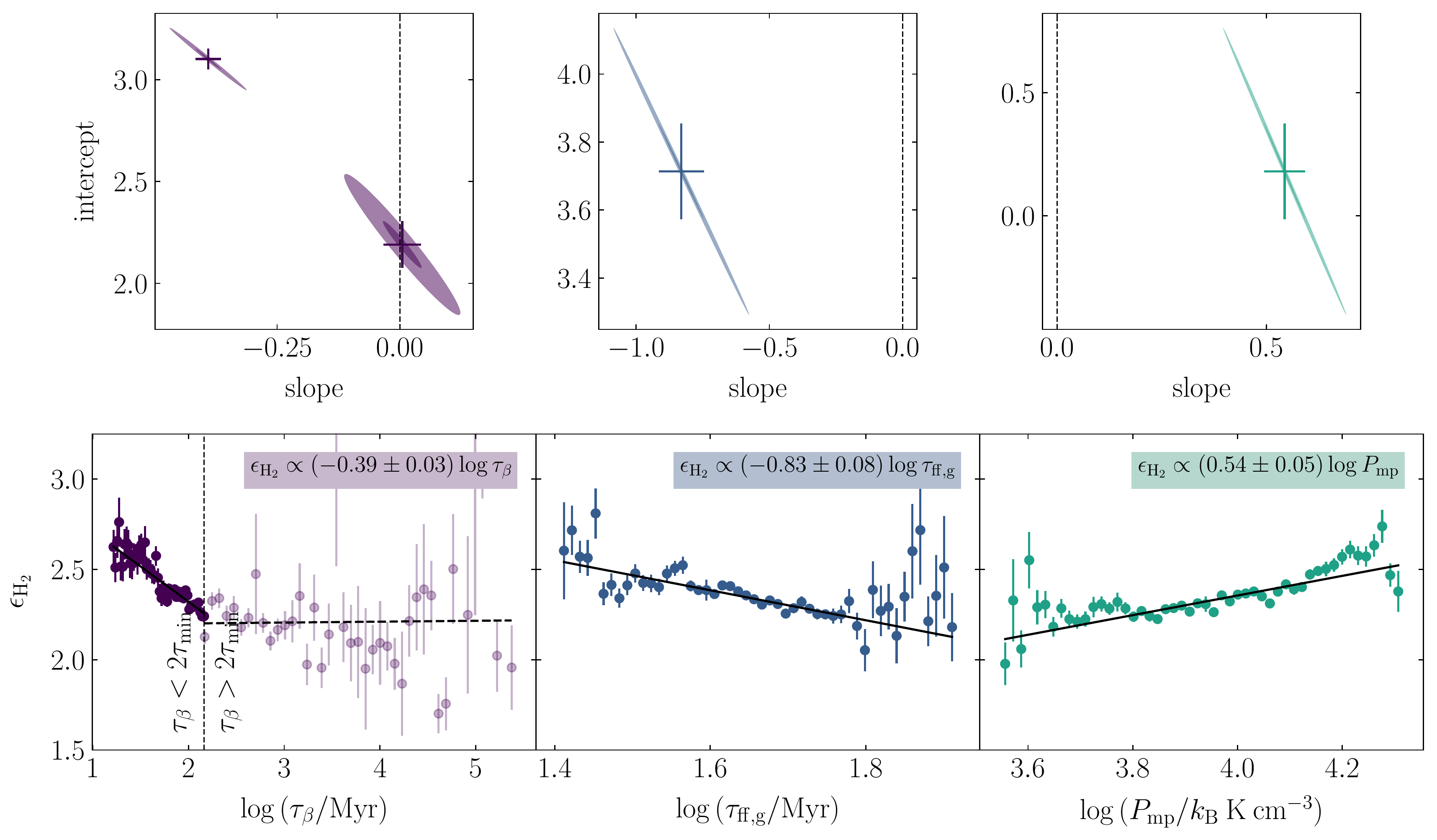}
  \caption{Same as Figure~\protect\ref{Fig::H2-veldisp-stat}, but for the GMC aspect ratio, $\epsilon_{\rm H_2}$.}
\end{figure*}

\begin{figure*}
  \label{Fig::HI-ellipticity-stat}
  \includegraphics[width=.9\linewidth]{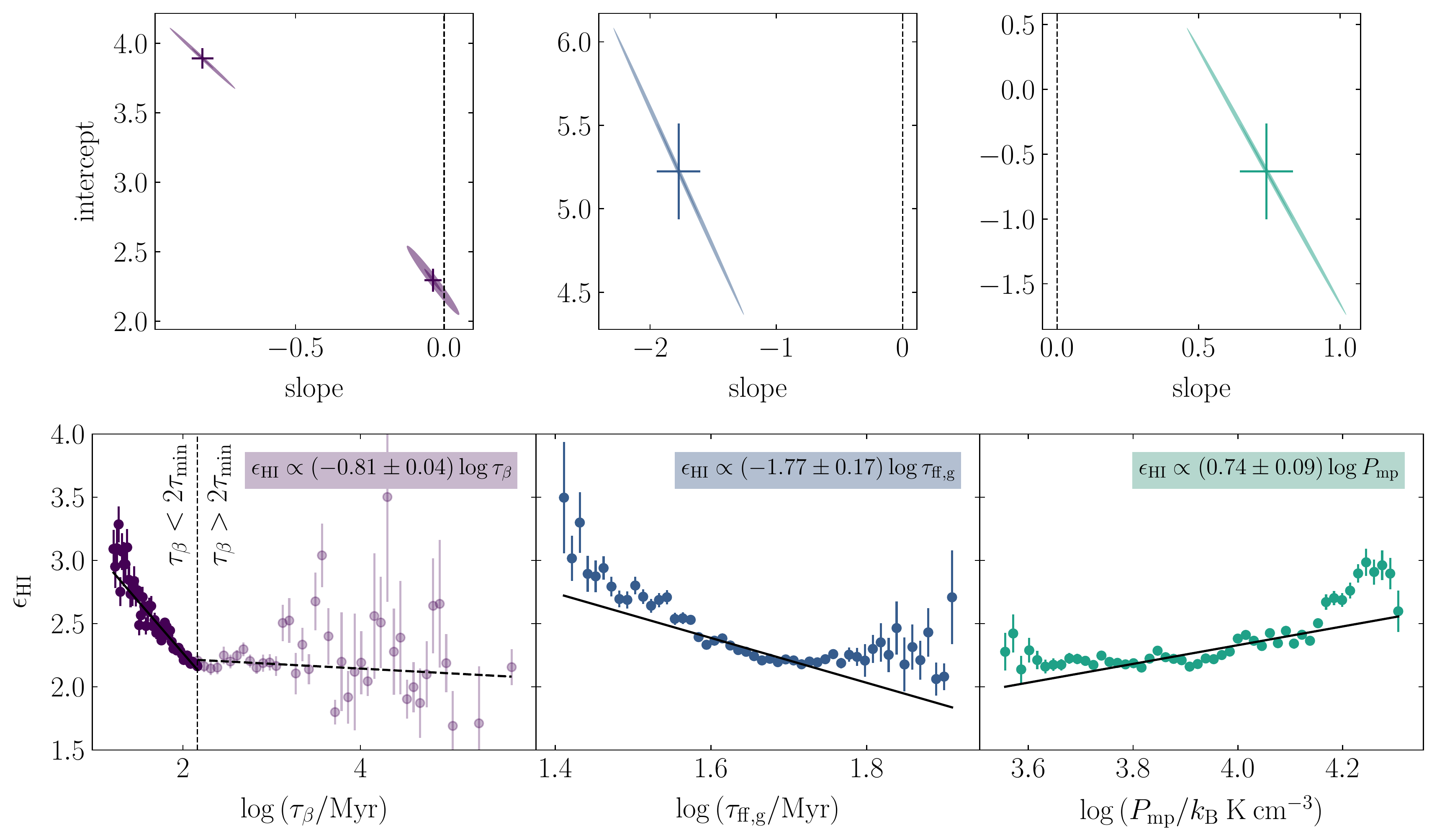}
  \caption{Same as Figure~\protect\ref{Fig::H2-veldisp-stat}, but for the HI cloud aspect ratio, $\epsilon_{\rm HI}$.}
\end{figure*}

\begin{figure*}
  \label{Fig::H2-angmom-stat}
  \includegraphics[width=.9\linewidth]{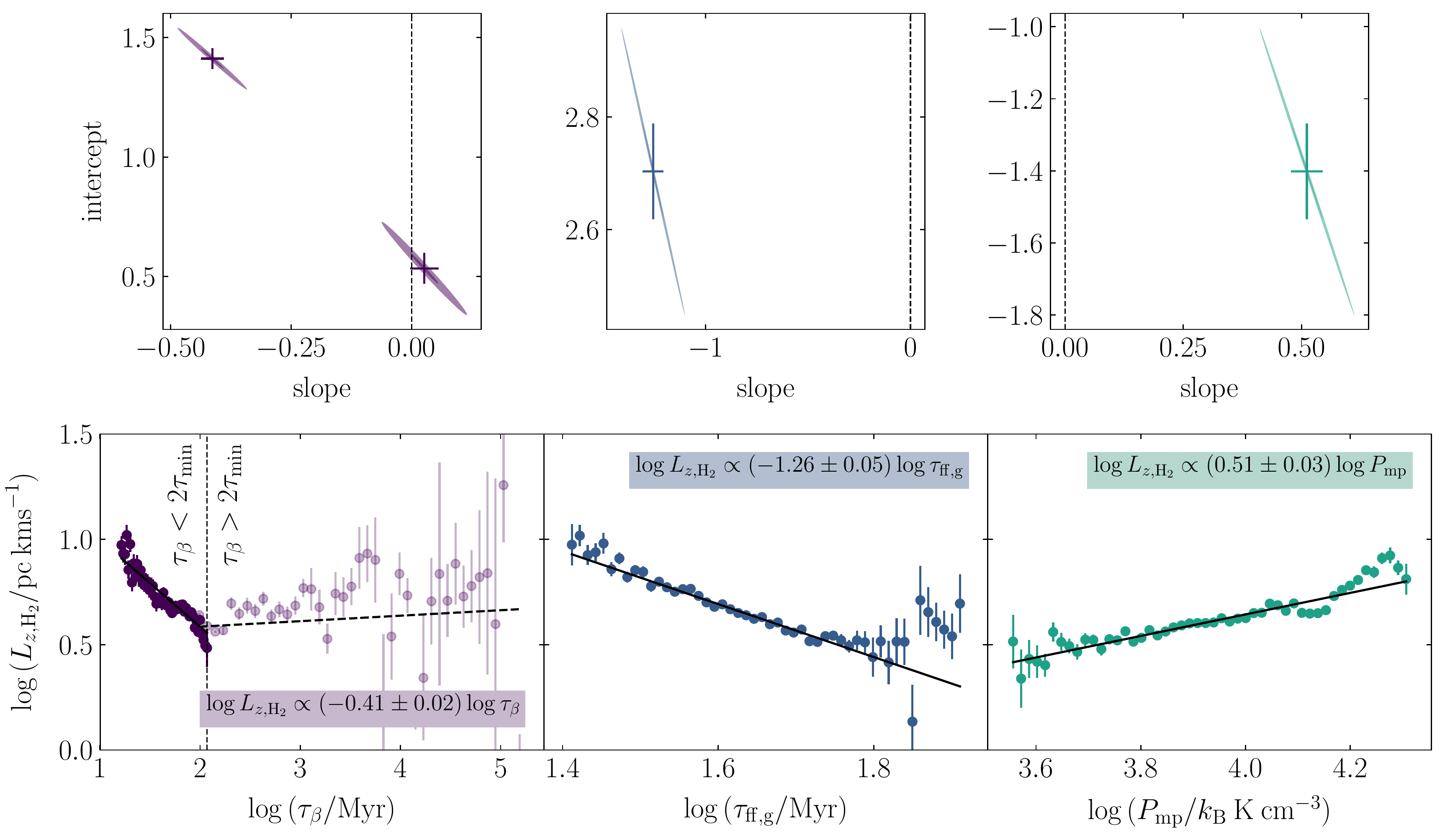}
  \caption{Same as Figure~\protect\ref{Fig::H2-veldisp-stat}, but for the GMC specific angular momentum, $L_{z, {\rm H_2}}$.}
\end{figure*}

\begin{figure*}
  \label{Fig::HI-angmom-stat}
  \includegraphics[width=.9\linewidth]{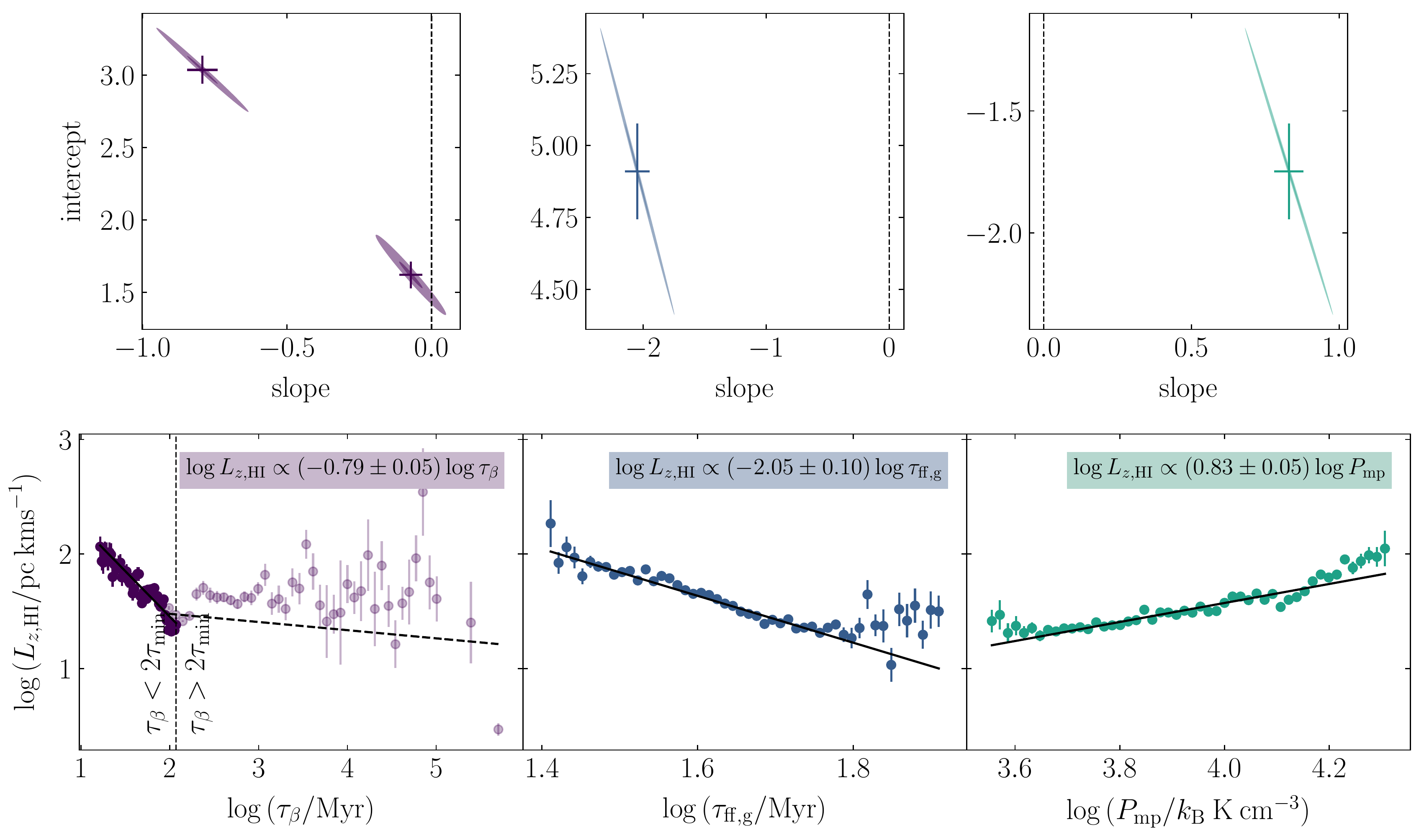}
  \caption{Same as Figure~\protect\ref{Fig::H2-veldisp-stat}, but for the HI cloud specific angular momentum, $L_{z, {\rm HI}}$.}
\end{figure*}

\begin{figure*}
  \label{Fig::H2-anisotropy-stat}
  \includegraphics[width=.9\linewidth]{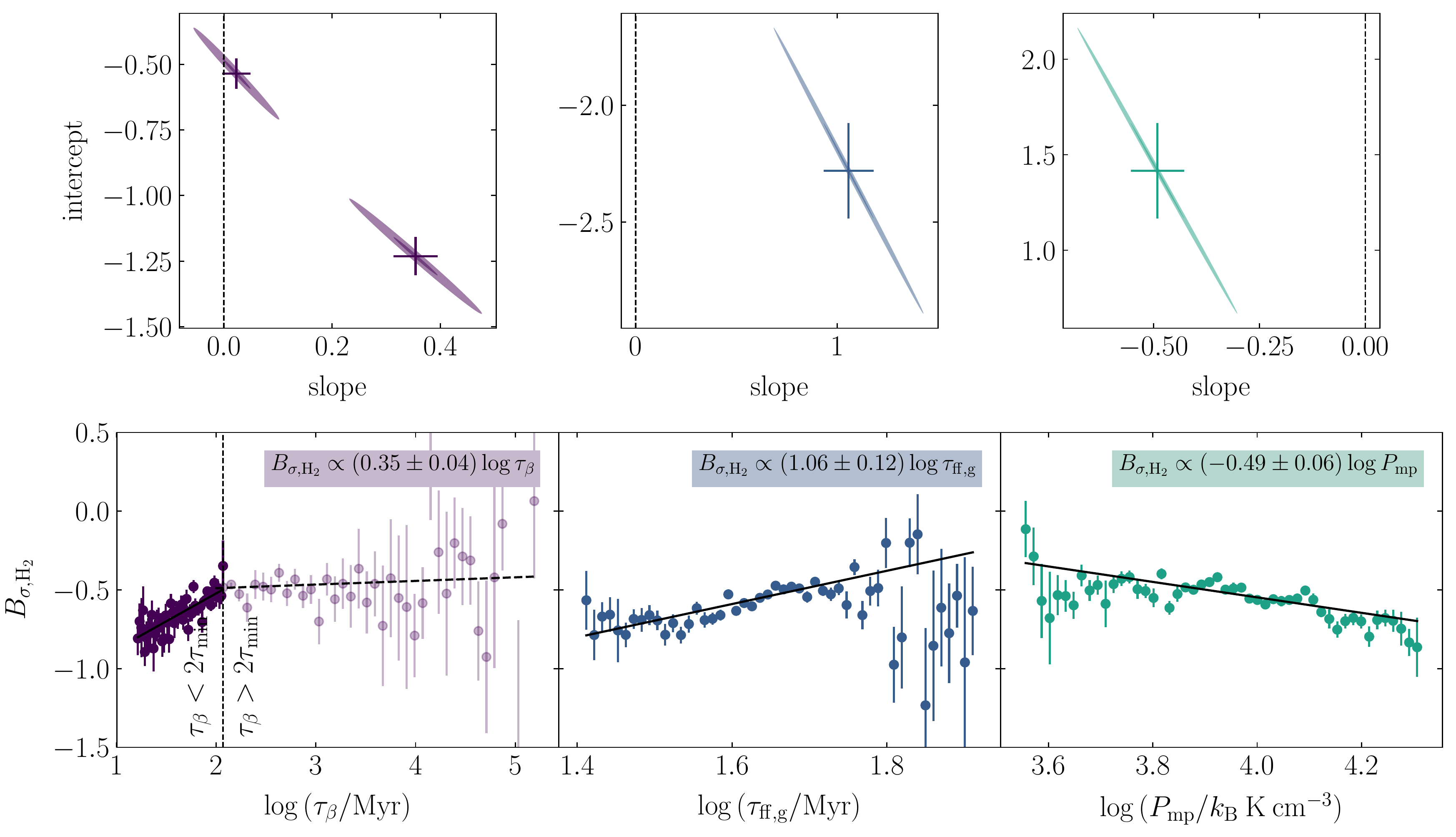}
  \caption{Same as Figure~\protect\ref{Fig::H2-veldisp-stat}, but for the GMC velocity anisotropy, $B_{\sigma, {\rm H_2}}$.}
\end{figure*}

\begin{figure*}
  \label{Fig::HI-anisotropy-stat}
  \includegraphics[width=.9\linewidth]{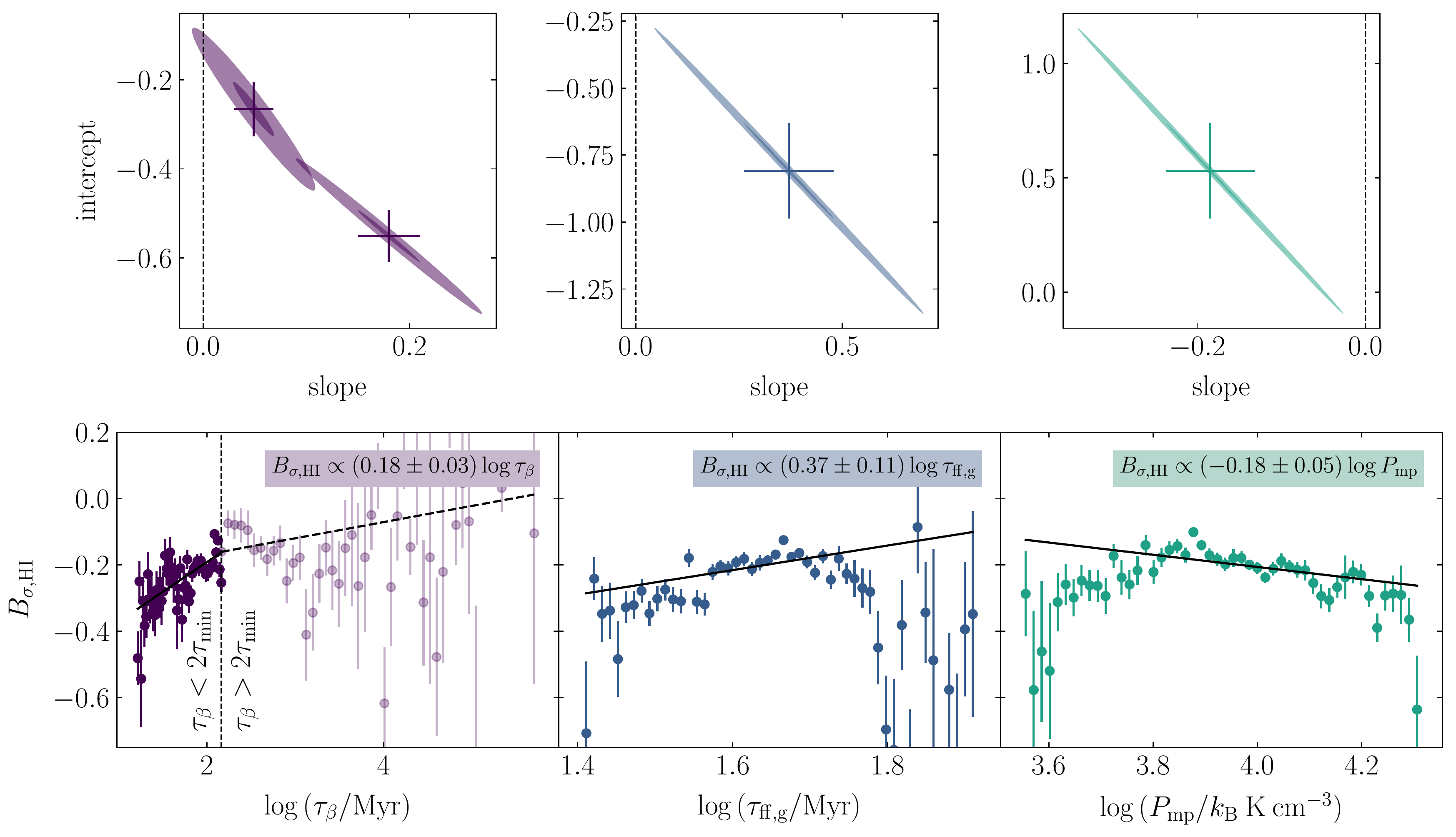}
  \caption{Same as Figure~\protect\ref{Fig::H2-veldisp-stat}, but for the HI cloud velocity anisotropy, $B_{\sigma, {\rm HI}}$.}
\end{figure*}

\begin{figure*}
  \label{Fig::H2-no-density-stat}
  \includegraphics[width=.9\linewidth]{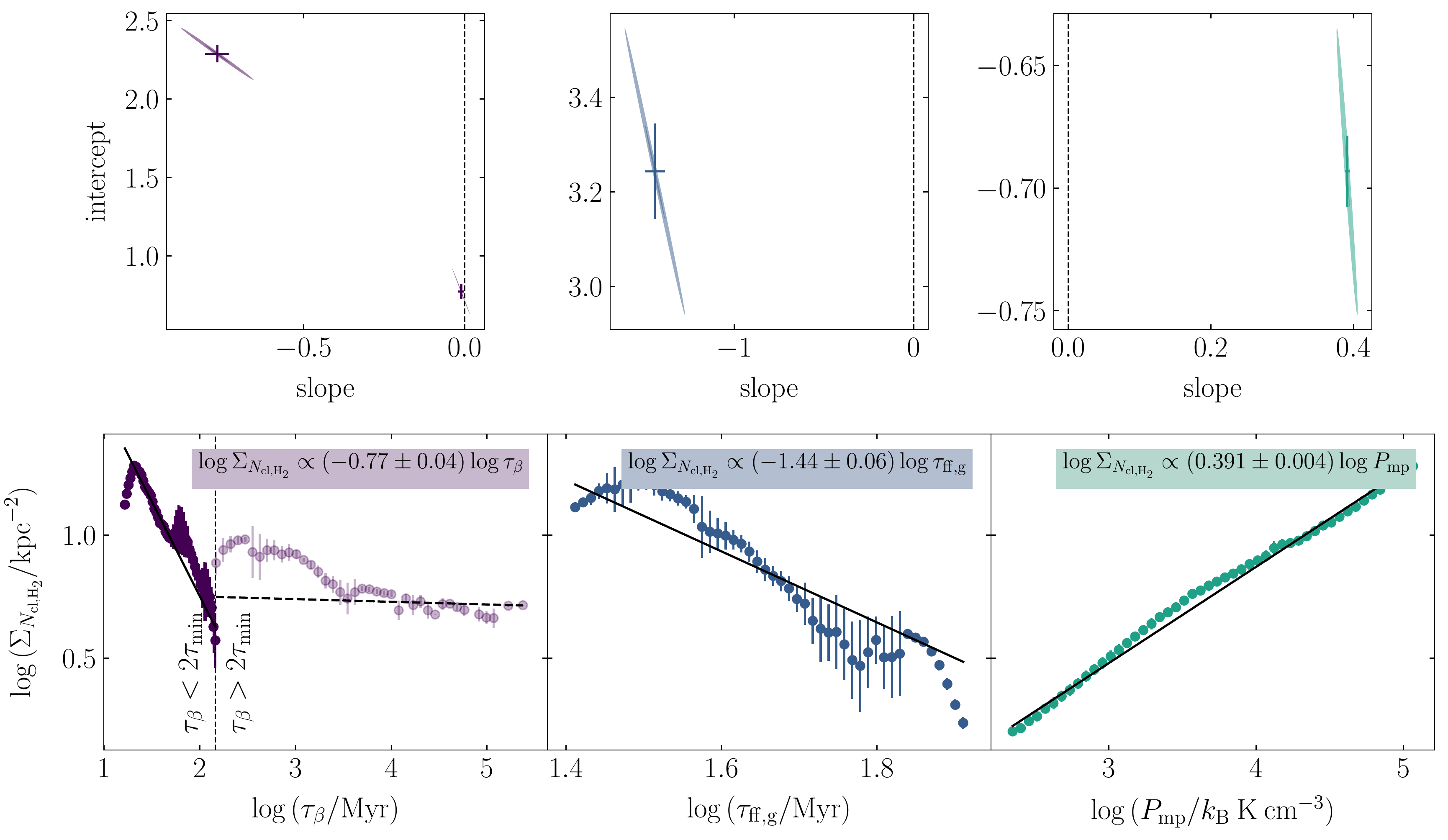}
  \caption{Same as Figure~\protect\ref{Fig::H2-veldisp-stat}, but for the number of GMCs per unit area of the galactic mid-plane, $\Sigma_{N_{\rm cl,H_2}}$.}
\end{figure*}

\begin{figure*}
  \label{Fig::HI-no-density-stat}
  \includegraphics[width=.9\linewidth]{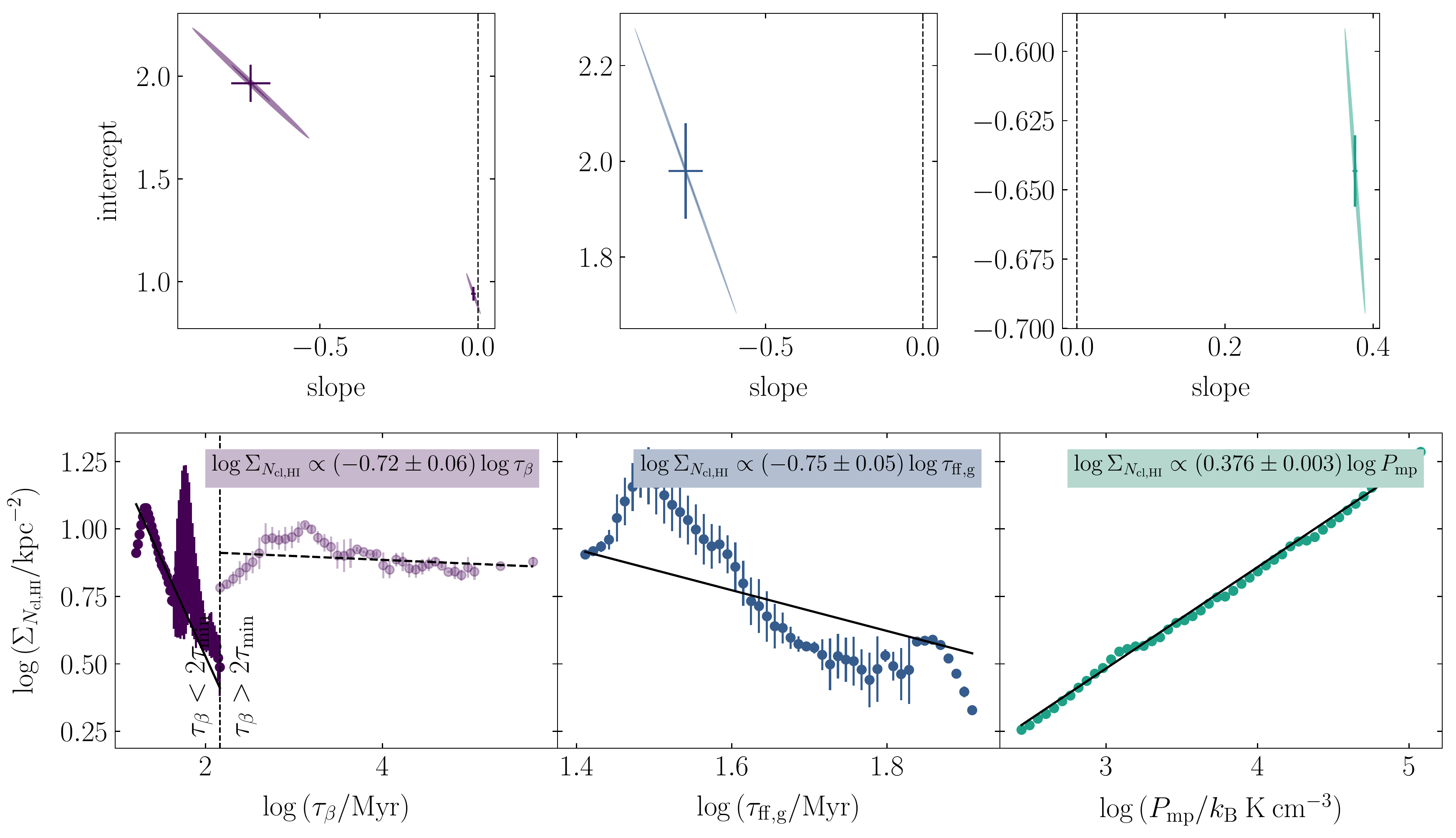}
  \caption{Same as Figure~\protect\ref{Fig::H2-veldisp-stat}, but for the number of HI clouds per unit area of the galactic mid-plane, $\Sigma_{N_{\rm cl,HI}}$.}
\end{figure*}

\begin{figure}
  \label{Fig::HI-div-stat-veldisp}
  \includegraphics[width=.9\linewidth]{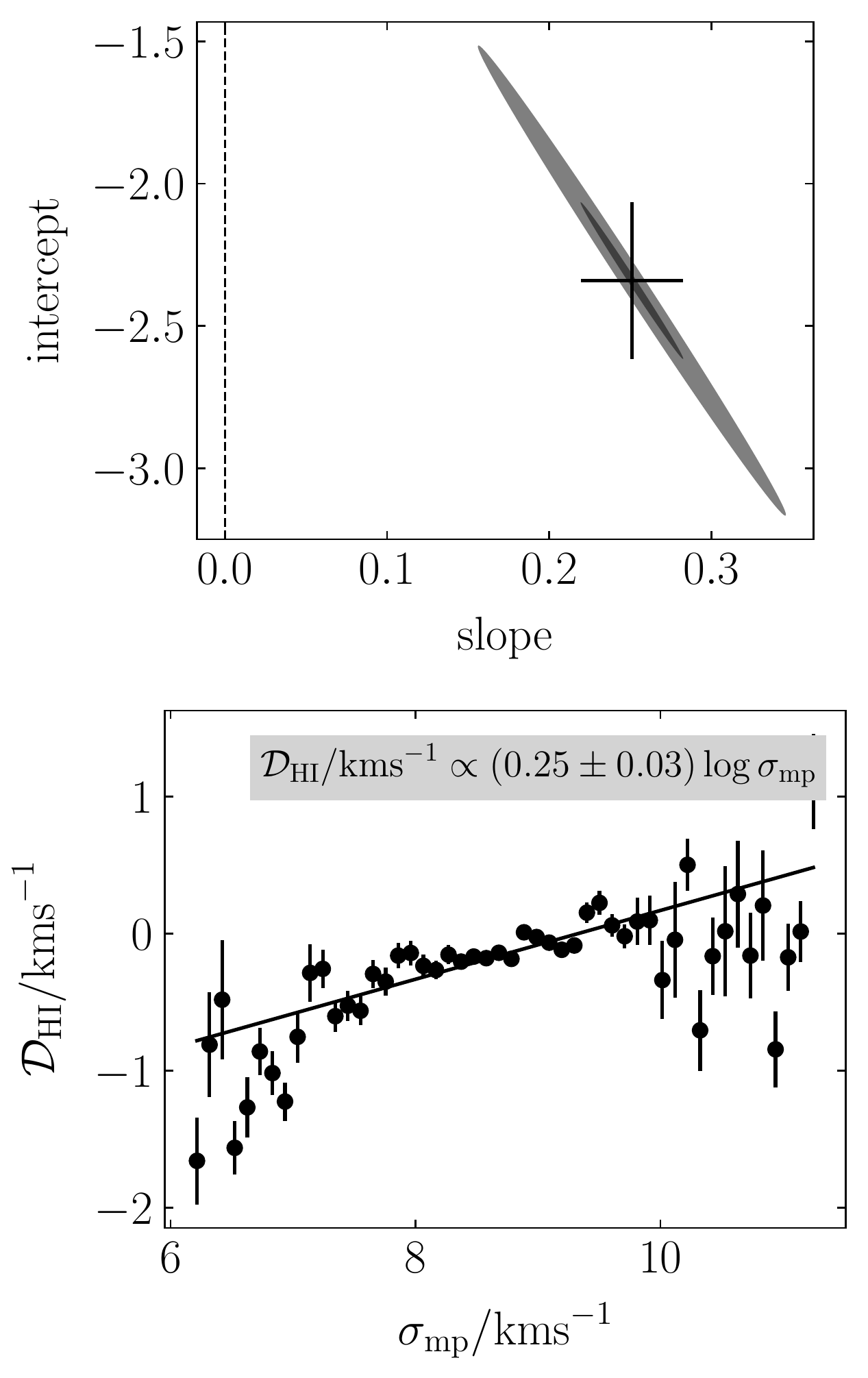}
  \caption{{\it Lower panel:} Mean HI cloud velocity divergence $\mathcal{D}_{\rm HI}$ as a function of the mid-plane velocity dispersion $\sigma_{\rm mp}$. The error-bars correspond to the standard deviation on the mean for the distribution of values in each bin. The equation gives the non-linear least-squares fit to the data. {\it Upper panel:} The $1\sigma$ (dark-coloured) and $3\sigma$ (light-coloured) confidence ellipses for the least-squares fit. The error-bars correspond to the projection of the $1\sigma$ confidence ellipse onto each of the parameter axes.}
\end{figure}

\begin{figure}
  \label{Fig::H2-div-stat-veldisp}
  \includegraphics[width=.94\linewidth]{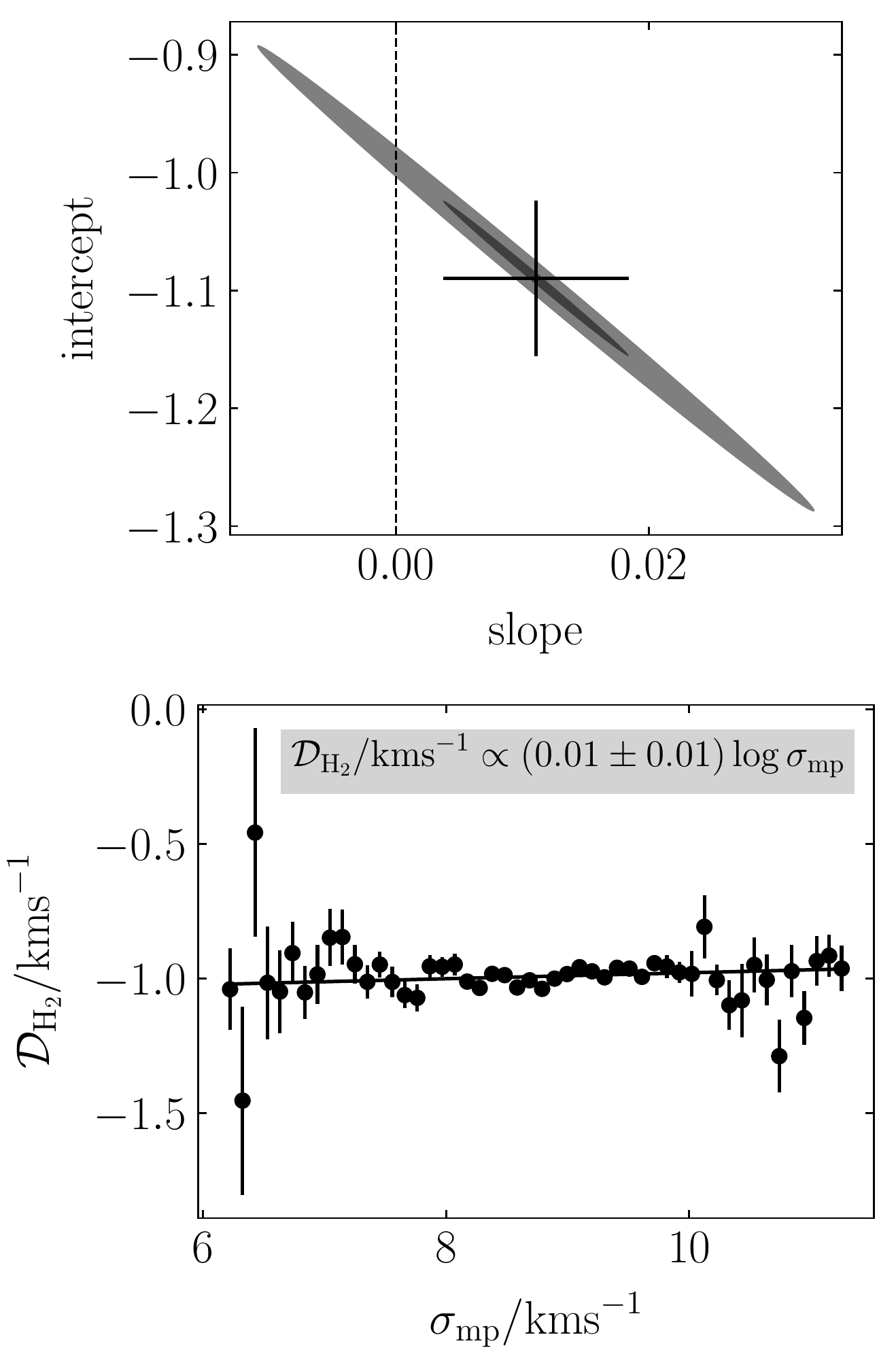}
  \caption{Same as Figure~\protect\ref{Fig::HI-div-stat-veldisp}, but for the GMC velocity divergence, $\mathcal{D}_{\rm H_2}$.}
\end{figure}

\end{document}